\documentclass[12pt,preprint]{aastex}
\pdfoutput=1
\usepackage{emulateapj5}
\received{2007 June 15}
\begin{document}

\newcommand{\kms}{\mbox{km~s$^{-1}$}}
\newcommand{\s}{\mbox{$''$}}
\newcommand{\mloss}{\mbox{$\dot{M}$}}
\newcommand{\my}{\mbox{$M_{\odot}$~yr$^{-1}$}}
\newcommand{\ls}{\mbox{$L_{\odot}$}}
\newcommand{\ms}{\mbox{$M_{\odot}$}}
\newcommand\mdot{$\dot{M}  $}
\title{Preplanetary Nebulae: An HST Imaging
Survey and a New Morphological Classification System}

\author{Raghvendra Sahai\altaffilmark{1},
Mark Morris\altaffilmark{2}, Carmen S\'anchez Contreras\altaffilmark{3}, 
Mark Claussen\altaffilmark{4}
}
                                                                                                                  
\altaffiltext{1}{Jet Propulsion Laboratory, MS\,183-900, California
Institute of Technology, Pasadena, CA 91109}
                                                                                                                                                                                                                                 
\altaffiltext{2}{Department of Physics and Astronomy,
UCLA, Los Angeles, CA 90095-1547}

\altaffiltext{3}{Dpto. de Astrofisica Molecular e
Infraroja, Instituto de 
Estructura de la Materia-CSIC, Serrano 121, 28006 Madrid, Spain}
                                                                                                                  
\altaffiltext{4}{National Radio Astronomy Observatory, 1003 Lopezville Road,
Socorro, NM 87801}

\email{raghvendra.sahai@jpl.nasa.gov}

\begin{abstract}
Using the Hubble Space Telescope (HST), we have carried out a survey of candidate
preplanetary nebulae (PPNs). We report here our discoveries of objects having well-resolved
geometrical structures, and use the large sample of PPNs now imaged with HST (including
previously studied objects in this class) to devise a comprehensive morphological
classification system for this category of objects. The wide variety of aspherical
morphologies which we have found for PPNs are qualitatively similar to those found for
young planetary nebulae in previous surveys. We also find prominent halos surrounding the
central aspherical shapes in many of our objects -- these are direct signatures of the
undisturbed circumstellar envelopes of the progenitor AGB stars. Although the majority of
these have surface-brightness distributions consistent with a constant mass-loss rate with
a constant expansion velocity, there are also examples of objects with varying mass-loss
rates. As in our surveys of young planetary nebulae (PNs), we find no round PPNs. The
similarities in morphologies between our survey objects and young PNs supports the view 
that the former are the progenitors of aspherical planetary nebulae, and that the onset of
aspherical structure begins during the PPN phase (or earlier). Thus, the primary shaping of
a PN clearly does not occur during the PN phase via the fast radiative wind of the hot
central star, but significantly earlier in its evolution.

\end{abstract}


\section{Introduction}
\label{intro}

Pre-planetary nebulae (PPNs), short-lived transition objects between the AGB
and planetary nebula (PN) phases, probably hold the key to understanding how the slowly
expanding
(5--15\,\kms), largely spherical, circumstellar envelopes (CSEs) of AGB stars
(Neri et al. 1998) transform into highly aspherical PNs with fast outflows
($\gtrsim$\,100\,\kms) directed along one or more axes.

We are carrying out a program of multiwavelength imaging and
spectroscopic observations of generally young PPNs, using a large
($\sim$300), morphologically unbiased sample (Sahai \& S\'anchez Contreras
2002, 2004), mostly constructed from catalogs of OH/IR stars (evolved, visually faint,
mass-losing stars with dense circumstellar envelopes, showing, generally
double-peaked, OH maser emission). Interferometric maps of the OH emission in
many of the objects of this class (e.g. Bowers et al. 1983) show that it typically arises
in an extended circular shell of radius $\sim\,10^{16}$cm, indicating that
substantial AGB mass-loss has occurred through a spherical outflow. The IRAS
spectral energy distributions (SEDs) of a large fraction of these objects
indicate a lack of hot dust (25 to 12\micron~flux ratio $F_{25}/F_{12}>1$,
implying a lack of dust hotter than about 450\,K), and therefore
cessation of the dense AGB mass-loss process less than a few hundred years ago.

Here we report on our discoveries of objects with well-resolved geometrical structures from
our imaging surveys of pre-planetary nebulae carried out using the Hubble Space Telescope
(HST) in its SNAPshot mode. A previous SNAPshot survey providing images of resolved
nebulosities in 21 PPNs was carried out by Ueta, Meixner \& Bobrowsky (2000). We use the
large sample ($>$50) of PPNs now available (which includes
previously-imaged objects in this class) to devise a new morphological classification
system for this category of objects. We also derive post-AGB and AGB mass-loss time-scales,
as
well as dust masses, for our objects. The main focus of this paper is the morphology of
PPNs; a more comprehensive statistical analysis which includes results on the unresolved
and undetected objects in our surveys is deferred to a future paper.

%{\it Since we don't have optical spectra of most objects, we need to make the
%case that the IRAS 25/12 ratio is a good proxy that our objects are indeed PPNe; for this
%we need to plot the 25/12 for known PPNe and yPNe (this could be the Sahai \& Trauger
%sample)}

\section{Observations}
\label{obs}

The images presented in this paper were obtained as part of two SNAPshot survey programs
with HST. The target lists for these programs were generated by selecting sources from
published catalogs of OH/IR stars (te Lintel Hekkert et al 1988, 1991, 1992, Chengalur et
al. 1993, te Lintel
Hekkert \& Chapman 1996, Hu et al. 1994, Likkel 1989, Sevenster et al. 2001, 1997a,b) which
had strong emission ($\ge$0.8 Jy) in the OH maser-lines (1665,1667, and/or 1612 MHz) and
with $F_{25}/\,F_{12}\,>\,1.4$. A small number of additional objects were selected from the
compilation of evolved stars with Low-Resolution IRAS
spectra by Kwok, Volk \& Bidelman (1997), subject to
the same IRAS color criterion as
for the OH/IR stars. This subset included carbon-rich
(C-rich) objects and class H objects (showing a red continuum with either a
9.7\micron~silicate absorption feature or a 11.3\micron~PAH emission feature in their LRS
spectra).

A 25\micron~flux threshold of 25\,Jy was then imposed in order to exclude
relatively distant objects. This yielded a sample of about 200 objects. The Digitized Sky
Survey (DSS) optical images, available from the online databases maintained by the Canadian
Astronomy Data Centre, were used to examine a $3\times3$\,arcmin$^2$ field-of-view around
each object in order to check whether any bright field stars were present very close to our
target which would make their imaging difficult. Objects for which no visual detection
could be made on the DSS plates were excluded. The USNO-B.1 and the Guide Star Catalog II
(GSC-II) were queried to determine R magnitudes; when these were unavailable, a rough
assessment of the R magnitude was made from the DSS images. The remaining sample was
divided into 3 magnitude bins (bright: $R<13$, medium: $14<R<17$, and faint: $R>17$), and
roughly equal numbers of objects were randomly selected from each bin in order to meet the
target limitation of 117 objects for the two survey programs, GO\,9101 (17 objects) and
9463 (100 objects), described below. A small number of targets which represent the
evolutionary phases just preceding and following the PPN phase were also included in these
two programs.

Both programs were carried out in the SNAPshot mode of HST in which objects are
selected from the program source list to fill gaps in the general observational schedule,
thus all objects in the list were not observed. 
The first survey program, GO 9101, was a pilot survey with WFPC2 in which 10/17 objects were
imaged. The second, GO 9463, was a more extensive survey in which a total of 58/100 objects
were
imaged with ACS. For GO 9101, we obtained images using the F606W
($\lambda=0.60\micron$, $\Delta\lambda=0.123\micron$) \& F814W ($\lambda=0.80\micron$,
$\Delta\lambda=0.149\micron$ filters, with the 800$\times$800-pixel Planetary Camera (PC)
of Wide-Field \& Planetary Camera 2 (WFPC2), which has a plate scale of 0\farcs0456/pixel.
For GO 9463, we used either the High Resolution Camera (HRC) or the Wide-Field Camera (WFC)
of the Advanced Camera for Surveys (ACS), depending on the optical magnitude of the object.
The HRC has a plate scale of 0\farcs025/pixel, and for objects imaged with this camera, we
used a combination of the F606W filter ($\lambda=0.60\micron$,
$\Delta\lambda=0.123\micron$) and the F814W filter ($\lambda=0.80\micron$,
$\Delta\lambda=0.149\micron$) for the fainter objects, and the F435W ($\lambda=0.43\micron$,
$\Delta\lambda=0.103\micron$) and F606W filter
for the brighter ones. The WFC has a plate scale of 0\farcs050/pixel, and for objects
imaged with this camera, we used a combination of the F606W filter and F814W filters. We
also examined the HST archive for additional PPNs which met our IRAS color and flux
selection criteria (and not published previously). We found one such object,
IRAS\,15553-4230 (GO\,10627, PI: M. Meixner) which was imaged with the ACS/HRC -- we have
included it in this paper.

Some of the objects from GO 9463 were also imaged as part of our GO 9801 program -- a third
survey to image optically-faint PPN candidates with Camera
1 or 2 of the Near Infrared Camera
and Multi-Object Spectrometer (NICMOS) using two or three of the 
F110W ($\lambda=1.1\micron$, $\Delta\lambda=0.55\micron$), F160W ($\lambda=1.6\micron$,
$\Delta\lambda=0.4\micron$) and F205W ($\lambda=2.0\micron$, $\Delta\lambda=0.6\micron$)
filters. The complete results of this survey, which did not yield resolved images of any new
PPN, will be discussed separately; in this paper we utilise the NICMOS images when
available for providing additional details on the morphology.

\section{Results}
\label{results}
\begin{figure}[htb]
\resizebox{0.8\textwidth}{!}{\includegraphics{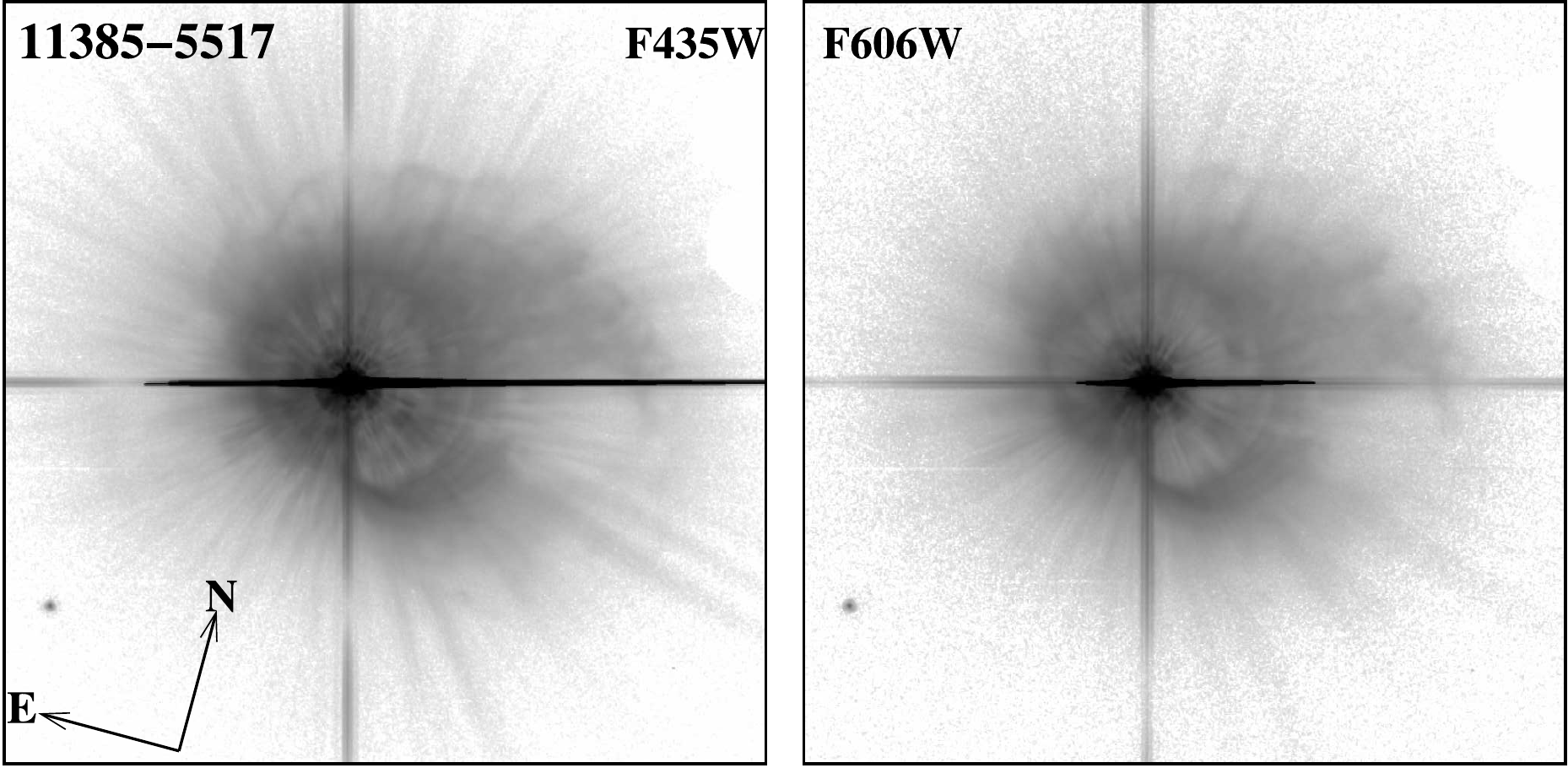}}
%\plotone{f1.eps}
\vskip 0.2in
\caption{HST image ({\it log stretch}) of the young preplanetary 
nebula IRAS\,11385-5517 (12\farcs5$\times$12\farcs5)
}
\label{fig1aa}
\end{figure}

\begin{figure}[htb]
\resizebox{0.8\textwidth}{!}{\includegraphics{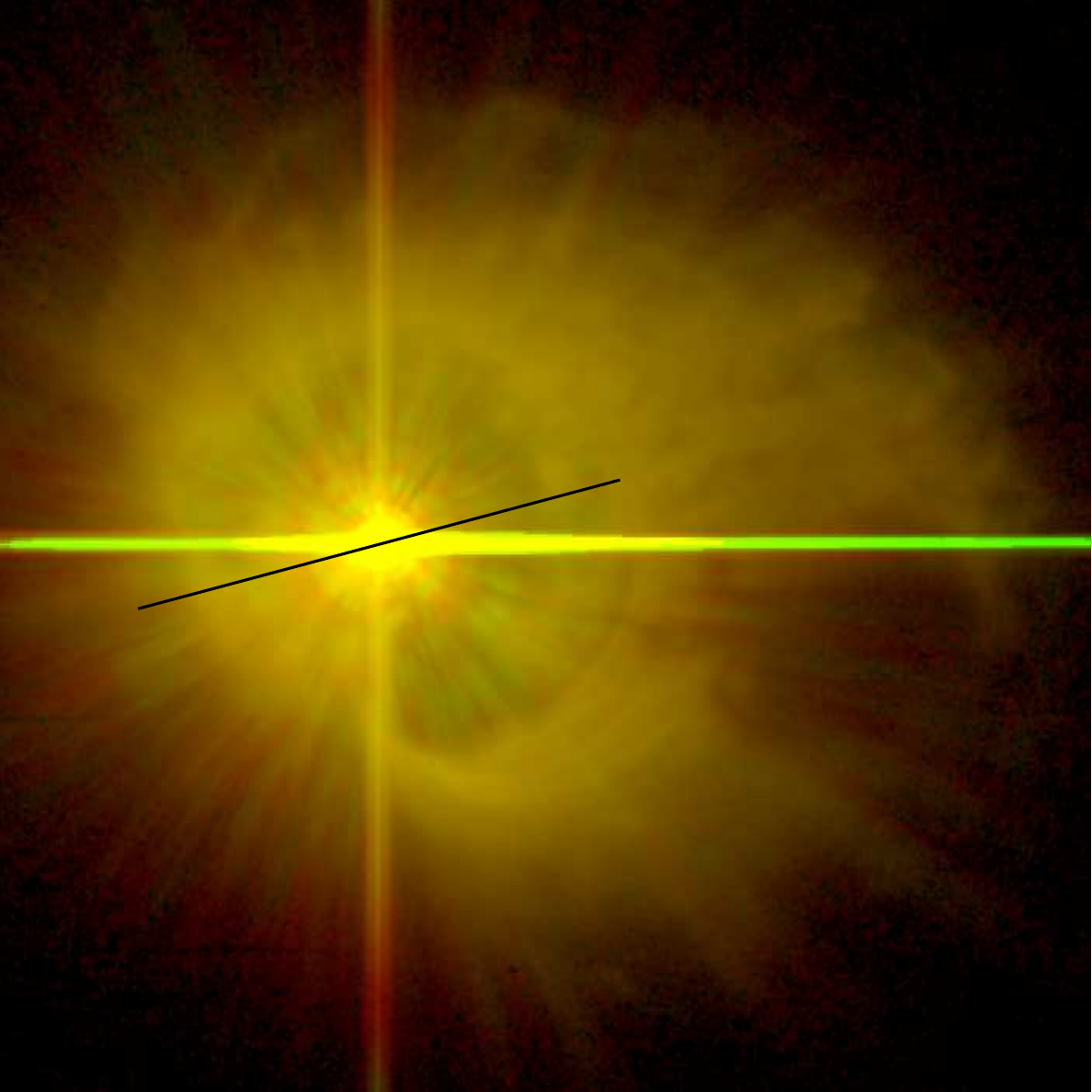}}
%\plotone{f1.eps}
\vskip 0.2in
\caption{HST color-composite image ({\it log stretch, green:F435W, red: F606W}) of
IRAS\,11385-5517 (8\farcs75$\times$8\farcs75). Structures due to the PSF of the bright
central star are readily apparent because corresponding PSF features in the two
wavelength bands occur at slightly different radial distances, whereas real circumstellar
structures overlap. The vector shows the orientation (PA=-60\arcdeg) and extent
($\pm\,2{''}$) of the high-velocity bipolar OH maser outflow. 
}
\label{fig1ab}
\end{figure}

We have listed the sources in our survey in Table\,\ref{t1}, with their 2MASS coordinates.
The images of all resolved objects are shown in Figs.\ref{fig1aa}-\ref{fig11}. 
Optical spectroscopy, allowing spectral types of the central stars to be determined
(either from the
direct starlight if the central star is visible, and/or from the starlight scattered by
nebular dust), is not available for all of our objects. However, we have carried out
spectroscopic surveys of partially-overlapping samples of candidate PPNs selected in the
same manner as for our imaging surveys, using the Palomar 60-inch (Sahai \& S\'anchez
Contreras 2002, 2004) and Keck 10-m telescopes (S\'anchez Contreras et al. 2003, S\'anchez
Contreras et al. 2007, in preparation), and we find that the observed objects do not show
the rich emission-line spectra characteristic of planetary nebulae. For those objects where
spectral types have been determined, either by us or others, the spectral types obtained
typically span the B, A, F,
and G spectral types, which are the typical spectral types for post-AGB central stars. 

The majority of our objects are oxygen-rich (O-rich) due to their selection from OH-maser
catalogs.  Information about the source chemistry is provided in the descriptions of
individual PPNs in \S\,\ref{srcs}.
%{\it Carmen, can you please check the list of PPNs in Table 3 to see for which objects
%spectral types could be obtained from the Keck-ESI data?}
%the available near- to far-infrared broad-band
%photometry using the 2MASS, MSX and IRAS point source catalogs}.
%/data/sahai/ohirdata/spec60/obs_sources.inf, obs_sources2.inf

\subsection{Morphological Classification System}
\label{morph}
The PPNs in our survey show a variety of morphologies. We therefore describe a
systematic method of classifying these. In this morphological classification system, we have
taken care to avoid using
criteria that are based on a 3-dimensional interpretation of the nebular structures whose 
2-dimensional projection is observed in our images.
The main structural components of PPNs in order of
importance are the
lobes, the waist, and the halo. Hence, we classify our objects into various morphological
classes with the
prime discriminant being the lobe structure. The presence or absence of a waist and halo are
included as secondary descriptors. 
The classification system is described below, and the classification codes are summarised in
Table\,\ref{t2}. In Table\,\ref{t3}, we apply these codes to the objects from our survey,
and in Table\,\ref{t4}, to previously imaged PPNs.

First, we divide the observed primary morphologies into 4 major classes -- bipolar (B),
multipolar (M), elongated (E), and irregular (I). There is no class referring to round
shapes, because we have not found any object in which the central post-AGB structure appears
round. The
``bipolar" or B shape is defined by objects which show two primary, diametrically opposed
lobes, centered on the central star (or center of mass if a binary system) or its expected
location (e.g., IRAS\,13557-6442, Fig.\ref{fig2a}; IRAS\,15452-5459, Fig.\ref{fig3}). The
pair
of lobes must have a``pinched in" shape in the region around the center from where they
emanate, and/or the lobes should be visible on both sides of a minimum-light-intensity
central
region (presumably due to an obscuring dust lane, see below). The ``multipolar" or M shape
is defined
by objects having more than one primary lobe on either side of the center
(IRAS\,19024+0044, Fig.\ref{fig6ab}; IRAS\,19475+3119, Fig.\ref{fig8a}). The ``elongated"
or E shape is simply one which is elongated along a specific axis, i.e., is not round (e.g.,
IRAS\,17253-2831, Fig.\ref{fig4a}; IRAS\,18420-0512, Fig.\ref{fig6a}). The ``irregular" or
I shape is defined by objects in which extended circumstellar structure can be seen, but
where no obvious lobe or shell-like structures can be identified, and which therefore do
not fit in any of the previous categories (e.g., IRAS\,11385-5517, Fig.\ref{fig1aa}). As the
name implies, objects in the I category usually do not display axial or point-reflection
symmetry.

Secondary structural features are denoted with lower-case letters following the major
class. First, we add subclassifications which are related to the lobes in the B, M, or E
classes -- these may be open (i.e., like a vase) or closed (i.e., having a bubble-structure)
at their outer ends, and are denoted by $o$ or $c$. For sources where the morphology is not 
sufficiently well-resolved or the images are too noisy to make the $o$ or $c$ determination,
this subclassification is not mentioned. For multipolar (i.e., M) objects, where it is
possible in principle to have some lobes which are open and some which are closed, both $o$
and $c$ should be used.

Some PPNs show bright, compact knots in diametrically-opposed pairs, normally referred to as
ansae (e.g. as in IRAS\,09371+1212, the Frosty Leo Nebula: Morris \& Reipurth 1990, Sahai et
al. 2000a) -- these are
denoted by $an$. 
Sometimes minor lobes can be seen, at low or
intermediate latitudes, as, e.g. in the Egg Nebula (Sahai et al. 1998a), and the Frosty Leo
Nebula -- these are denoted by $ml$. The presence of a skirt-like structure around the
primary lobes (as,
e.g., in Hen\,3-401, Sahai, Bujarrabal, \& Zijlstra 1999a) is denoted by $sk$. This
structure
is an illuminated surface lying outside the primary bipolar lobes but with the same axis of
symmetry; the skirt has a larger opening angle than the primary lobes, and while it
surrounds these lobes it does not have as large a radial extent. Skirt structures are 
most easily discerned when the nebular axis is near the plane of the sky. But if the nebular
axis is sufficiently inclined towards us, it may become difficult to distinguish between a
waist with a sharp edge and a skirt. We discuss this issue in S\,\ref{discus-hw}.

If there is a dark obscuring band that cuts across the
center of the nebula (i.e. a minimum in an intensity cut taken along the primary long axis
of the nebula, and usually described as the ``waist" of the nebula), its presence is denoted
by
$w$. The dusty waist may be due to (i) an opaque dust torus that blocks light that arises or
is scattered from a central region, or (ii) a disk that never allows light from the central
star to emerge in directions near the equatorial plane. However, our morphological
classification does not distinguish between these possibilities.
 
For several PPNs, the waist appears to have a sharp outer (radial) edge or boundary -- we
denote this
by adding $(b)$ after $w$. This outer boundary can show up in two different ways. The
first, and more common, way is when the boundary is seen as a dark, convex (relative to the
center of the nebula) edge cutting across one of the primary lobes\footnote{A detailed
discussion of this issue is given in \S\,\ref{discus-hw}}. A prime example of a
well-studied PPN with such a ``dark-edged" waist is the Egg Nebula (Sahai et al. 1998a); in
our study we also find clear examples in, e.g., IRAS\,19024+0044 (Fig.\ref{fig6ab}),
IRAS\,19292+1806 (Fig.\ref{fig7a}), \& IRAS\,22036+5306 (Fig.\ref{fig8c}). The second way
is when the waist boundary smoothly transitions from a dark feature seen against the bright
lobes to a brightness feature seen against the sky (or faint nebular) background in
equatorial regions on either sides of the lobes. The best examples of this phenomenon are
provided by IRAS\,17106-3046 (Kwok et
al. 2000) and IRAS\,04296+3429 (Sahai 1999, but see Oppenheimer et al. 2005 for a different
interpretation).

If the central star can be observed (in any of the optical filters commonly used with HST
for imaging PPNs), we add the symbol $*$.

Next, if there is evidence for point-symmetry in the nebular structure, we denote it by
$ps$. This classification is not applied to axially symmetric objects, even though axial
symmetry is 
a special case of point-symmetry. The
point-symmetry can be of three general types: (i) the presence of 2 or more pairs of
diametrically-opposed lobes, denoted as $m$ (e.g. IRAS\,19475+3119, Fig.\ref{fig8a}), (ii)
the ansae are distributed point-symmetrically about the center (e.g. as in Frosty Leo), 
denoted as $an$,
and (iii) the overall geometrical shape of the lobes is point-symmetric (e.g. Hen\,3-1475),
denoted as $s$. These three types of point-symmetry are listed in parentheses after the $ps$
symbol, e.g., $ps(m)$ means a nebula which is point-symmetric by virtue of having 2 or more
pairs of diametrically-opposed lobes. An object may possess more than one type of point
symmetry: e.g., IRAS\,19024+0044's point-symmetry would be represented as $ps(m,an)$ since
it shows both the presence of pairs of diametrically-opposed lobes, as well as ansae in one
of the lobe pairs which are point-symmetric (Sahai et al. 2005).

The presence of a halo is denoted with $h$, and an    
$(e)$ is added after it if it has a non-round, or elongated shape. If the halo shape cannot
be determined reasonably (e.g., if the halo is rather tenuous and the source lies in a
field with bright nearby stars, as in the case of IRAS\,17543-3102), we add $(i)$ meaning
that the shape is indeterminate. Thus the default halo
shape is assumed to be round. The presence of arc-like structures in the halo, as for
example seen in the Egg Nebula (Sahai et al. 1998b), is denoted by
$(a)$. Note that a halo may have a smaller visible radial extent than the nebular lobes in 
the images shown. However, this does not necessarily imply that the physical radius to which
the halo can be detected is smaller than the lobes, it is a result of the more rapid radial
decrease of the halo surface brightness compared to the lobes. With azimuthal averaging,
the halo can be traced to radii beyond the radial extents of the lobes.

In a few bipolar PPNs, searchlight-beams which appear to emanate from the center, can be
seen illuminating parts of the halo at high latitudes around the polar axis. Although the
best example of this phenomenon is provided by the Egg Nebula (Sahai et al. 1998b), similar
features have been seen in four other PPNs --  IRAS\,20028+3910 (Hrivnak et al. 2001),
IRAS\,17150-3224 (Su et al. 2003), IRAS\,17245-3951 (Hrivnak et al. 1999) and
IRAS\,18276-1431\footnote{IRAS\,18276 is included in our survey; a detailed study of this
object utilising the HST images shown here and other data has been presented by S\'anchez
Contreras et al. 2007}.

Note that in our system, the M or multipolar classification does not necessarily imply the 
point-symmetric sub-class, denoted by $m$, even though for the two M objects in our survey,
IRAS\,19024+0044 and IRAS\,19475+3119, the lobes appear in diametrically-opposed (i.e.
point-symmetric) pairs. The reason for this is that it may be possible for objects to have
multiple lobes, all of which 
cannot be unambiguously grouped into diametrically-opposed pairs. Specific examples of such
objects are the Starfish Twins (He 2-47
and M1-37), two young planetary nebulae with multiple lobes described by Sahai (2000).

Similarly, the B or bipolar classification does not necessarily imply $w$ (i.e., the
presence of a dusty waist), even though most B objects have dusty waists. Specific examples
of objects which are bipolar, but don't show dusty waists in their images, are,
IRAS\,17440-3310 (Fig.\ref{fig5a}) and IRAS\,19306+1407 (Fig.\ref{fig7b}). It is possible
that these objects do have a dusty waist, but the waist is sufficiently tilted away from an
edge-on orientation, and hence does not appear as an intensity-minimum between the two
lobes. 

We have applied the above system to some of the most well-resolved and best-studied PPNs to
ensure that it captures the important morphological characteristiscs of these objects.
Thus, the Egg Nebula has the
classification Bcw(b),ml,h(e,a,sb); Frosty Leo is Bcw(b)*, an,ml,ps(m,an,s);
Minkowski's Footprint, M\,1-92 (e.g. Bujarrabal et al. 1998) is Bcw(b)*,an;
and Hen\,3-401 is Bow*,sk (Sahai, Bujarrabal \& Zijlstra 1999a).
The water-fountain PPN, IRAS16342-3814 (Sahai et al. 1999b) is classified as Bcw,ps(s).

\subsection{Distances, Sizes, Ages, and Dust Masses}
\label{ages-etc}
% ages to nearest 50 yr
We have estimated the distances to our survey sources (D$_L$) assuming a fixed luminosity of
6000\ls~for each object, rather than using kinematic distances determined from radial
velocities
because these objects can quite often have large and peculiar motions relative to the
general galactic rotation. Both the above choices have been motivated by a VLBA proper
motion study of the young PPN, IRAS\,19134+2131. For this object, Imai, Sahai \& Morris 
(2007) 
determine a trigonometric parallax distance of $8\pm0.8$\,kpc, giving a luminosity of
5500\ls, and
a Galactic rotation velocity of $\sim$125\,\kms, much slower than that given by the Galactic
rotation curve at IRAS\,19134+2131's location. Our choice of a fixed luminosity in our
computations is not meant to imply that all PPNs have this same value of the luminosity,
but simply to provide a convenient normalisation factor for the ages and masses. The values
of these parameters can then be conveniently scaled in future studies when better distance
determinations of individual objects become available.

The bolometric fluxes have been computed for each object by integrating
the SED from the optical to far-infrared wavelengths, using published optical, JHK-band, MSX
and IRAS fluxes. For all objects where the central star is not visible, the total flux for
wavelengths $\lambda\lesssim2$\micron~is quite small. In objects where the central star is
visible, this short-wavelength flux can make a significant contribution to the bolometric
flux, hence our luminosity determination is somewhat affected by the unknown amount of
interstellar extinction. 
%transitional objects, luminosity has declined from AGB peak value says Mark, 
%but evolutionary tracks indicate flat luminosity 

For each PPN, we have estimated the radial extent of the central aspherical
structure (in arcseconds), which
presumably represents the post-AGB mass-loss ($r_{PAGB}$), and the radial extent of
the halo (in arcseconds)
surrounding this structure (when it is present), which presumably represents the AGB
mass-loss($r_{AGB}$). For closed lobes, $r_{PAGB}$ is measured radially outwards from the
geometrical center of the nebula (or
the central star, when visible) out to the most distant lobe structure along the lobe axis.
If the lobes are ``open" at their ends, then $r_{PAGB}$ is defined as
the radial offset of the most distant point on the lobe. If several lobes are present, or
if the lobes on either side of the center have different lengths, the longest one is
selected. For objects classified as I, i.e. when lobe structures are not visible, $r_{PAGB}$
is defined as the radius to the most distant non-spherical structure. We have used azimuthal
averaging to measure radial extent of the halos\footnote{specifically, on a log-log plot of
the averaged-intensity versus radius, the former drops smoothly out to $r_{AGB}$ where it
reaches the background-subtracted sky level and starts showing large fluctuations (either
due to noise and/or the presence of many faint stars)}. The azimuthal averages are done
carefully in order to avoid the angular regions which are occupied by the nebular lobes. If
a halo cannot be detected, we
make the reasonable assumption that $r_{AGB}=r_{PAGB}$, motivated by the fact that the
lobes result from the interaction of a fast AGB wind with a surrounding AGB envelope.

A post-AGB time-scale, $t_{PAGB}$\footnote{rounded to the nearest 5 yr in Table\,\ref{t3}},
is derived by converting $r_{PAGB}$ from angular units to
linear units using the luminosity-based distances, and dividing
it by a nominal outflow speed of 150\,\kms (which is typical of the outflow speeds
derived for the lobes of individual well-studied PPNs), since individual outflow speeds for
the material in the lobes of most objects have not been measured as yet, nor are the
inclination angles of the lobe axes well known. In objects with a
halo, an AGB mass-ejection
time-scale ($t_{AGB}$)\footnote{rounded to the nearest 50 yr in Table\,\ref{t3}} is computed
by dividing the linear radius of the halo ($r_{AGB}\times
D_L$) by the expansion velocity as
derived from molecular line data (either OH 18 cm maser line or CO millimeter-wave line
emission). Since the outer radius of the halo is most likely determined by the steadily
decreasing halo surface-brightness falling below the noise (due to a decrease in density as
well as central illumination), $r_{AGB}$, and correspondingly, $t_{AGB}$, should be
understood as lower limits. By using azimuthal averaging, we can trace the halos out to
significantly larger radial distances than possible with pencil-cuts of the intensity. The
values of $r_{AGB}$ given in Table\,\ref{t3} thus found are typically much larger than can
be seen in the displayed images, where the intensity scales and stretches are chosen in
order to display the central aspherical structures most clearly.

%if Opn Lobe, then Lower Limit on pAGB age, measure size at 5-sigma above bkgrnd? 
%if strong smooth halo present, and Opn lobes superimposed on halo, then use sharpened 
%  image to remove halo, then measure size of aspherical structure

We have fitted the SED of each object from the near-infrared (K-band) to the far-infrared
(as defined by MSX and IRAS fluxes) using a 
multi-component model (Sahai et al. 1991). A 
power-law ($\lambda^{-p}$, with $p$=1.5) dust emissivity with a value of 
150 cm$^2$g$^{-1}$ (per unit {\it dust} mass) at 60 $\mu$m (Jura 1986)
is used to fit the SED with a ``cool", a ``warm", and a ``hot" component. The mass of the
cool
component, which is robustly fitted since the bulk of the emission comes from cool dust, is
significantly larger than the other two components. This mass and the temperature of the 
cool component, are therefore reported in Table\,\ref{t1}. Comparison of the results of our
simple
models with those utilising radiative transfer codes to determine the
temperature distribution in detail (e.g., for IRAS\,19475+3119) indicate that the simple
model gives dust masses that are lower than the more accurate values, by a factor 0.6-0.8.

We have found no obvious correlations between the morphologies and the post-AGB ages, dust
shell masses or temperatures.

%We have derived distances to most of our objects using their radial velocities. In most
%cases these are the near-kinematic distances except if the corresponding luminosity
%derived from integrating the SED is too low for a post-AGB object, i.e. less than
%2500\ls.

%Carmen/ Claussen: construct SEDs, fit dust models?

\section{Individual Sources}
\label{srcs}
\noindent {\it IRAS\,11385-5517}
% 92.6 138.3 193.0 104.0   Irr  pg 74
% current images are 189_435w/3x40sec, and 47a_606w/3x5sec
% 1000 x 1000 0.5 HRC pix
% L=3979(d/1kpc)^2=6000, so d=1.23 kpc, t=180 yr, 1yr=3.156e7 sec
%  rPAGB=4.6", tPAGB=8.49e16/150km/s = 179.3 yr
%  rAGB=assume=rPAGB=4.6, Vexp(AGB)=15, so rAGB=1793 yr=1800 yr
% 0.85E-03, 0.72E-06, 0.24E-08, 73.7 287.1 850.0 Std.Dev.  0.4994E-01
%  Md=(6000/3979)*0.85E-03=1.28e-3
%11(435W) PHOTFLAM=5.4077150E-19 / inverse sensitivity, ergs/cm2/Ang/electron
%ORIENTAT=-74.99 / position angle of image y axis (deg. e of n)
% what axis is indicated by other observations?
This object (HD101584), is well-known for its highly collimated,
high-velocity bipolar outflow seen in OH maser emission (te Lintel
Hekkert, Chapman \& Zijlstra 1992) and
massive
molecular outflows (as seen in CO rotational line emission: Trams et al. 1990). The optical
morphology is irregular (Fig.\ref{fig1aa}).  The central star has been classified as F0Iape
(Hoffleit et al. 1983) and A6Ia (Sivarani et al. 1999, these authors contest Bakker et al.
1996's hotter, i.e., B9II, classification), and believed to be in a close binary (Bakker et
al. 1996). Although the central star is heavily
saturated in both images and produces a significant PSF contribution (e.g.
radial streaks), the nebular structure is easily discernible, since all features
which do not change their radial distance from the center as a function of
wavelength have an astrophysical origin (Fig.\ref{fig1ab}). The red- and blue-shifted 
OH 1667\,MHz emission features define an axis of bipolarity with PA=$-60\arcdeg$ (te Lintel
Hekkert et al. 1992).

\begin{figure}[htb]
\resizebox{0.8\textwidth}{!}{\includegraphics{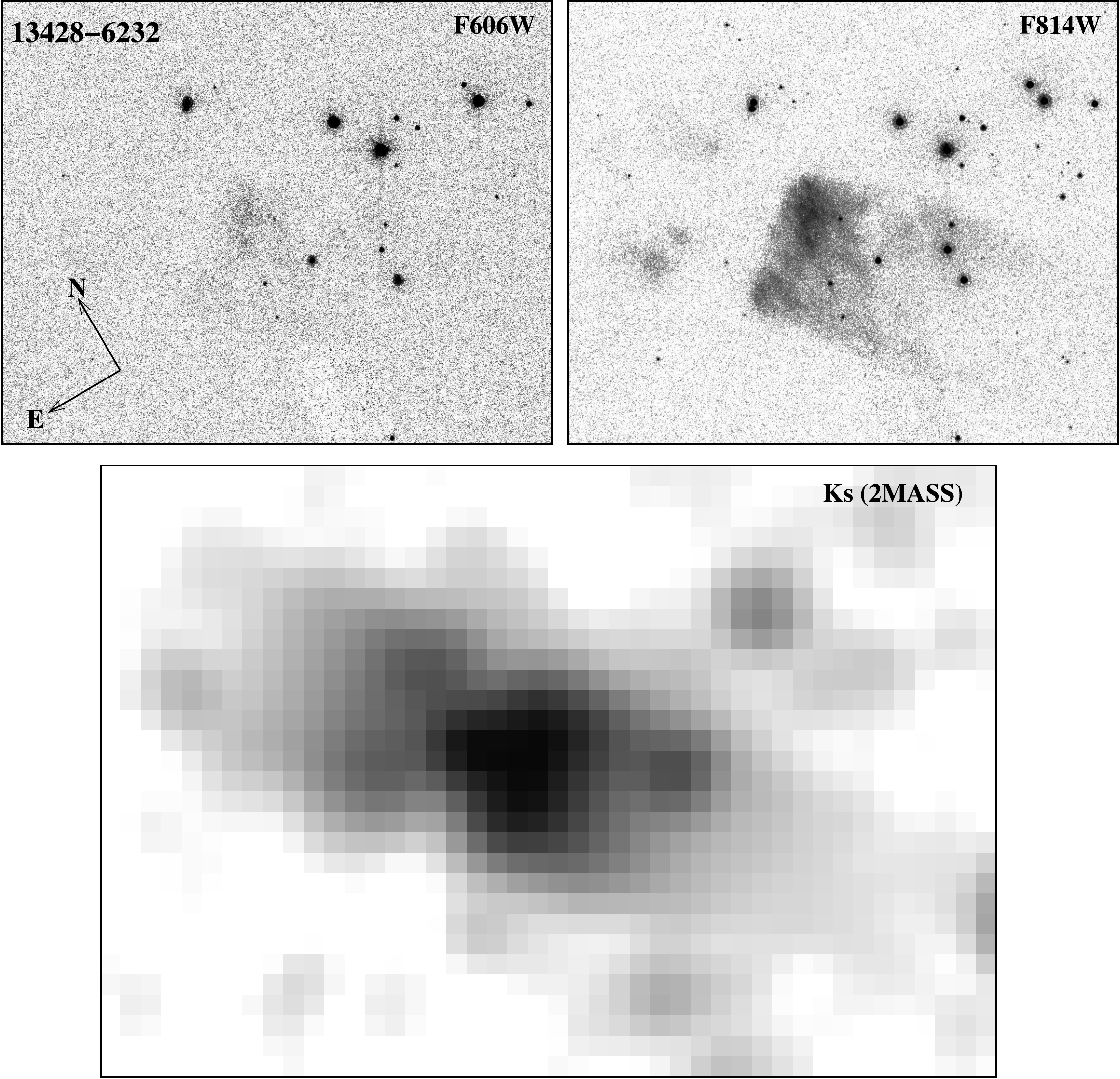}}
%\plotone{f1.eps}
\vskip 0.2in
\caption{{\it top} HST images ({\it log stretch}) of the young preplanetary 
nebula IRAS\,13428-6232 (27\farcs0$\times$21\farcs75); {\it bottom} $K_s$ image from the
2MASS archive (44\farcs0$\times$30\farcs0)
}
\label{fig1b}
\end{figure}

\begin{figure}[htbp]
\vskip -0.6in
\resizebox{1.0\textwidth}{!}{\includegraphics{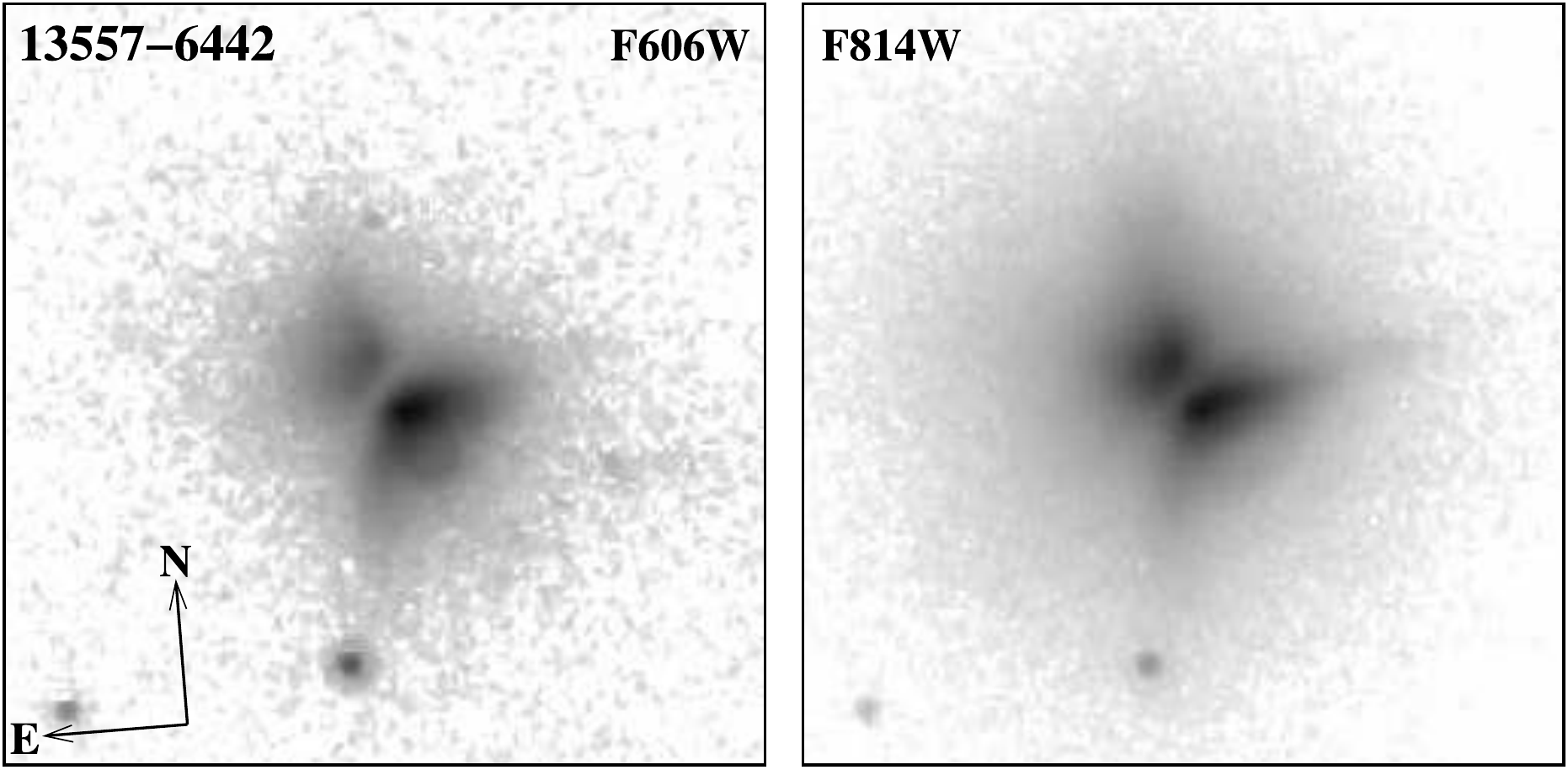}}
%\plotone{f1.eps}
\vskip 0.2in
\caption{HST images ({\it log stretch}) of the young preplanetary 
nebulae IRAS\,13557-6442 (3\farcs125$\times$3\farcs125).
}
\label{fig2a}
\end{figure}

\begin{figure}[htbp]
\vskip -0.6in
\resizebox{1.0\textwidth}{!}{\includegraphics{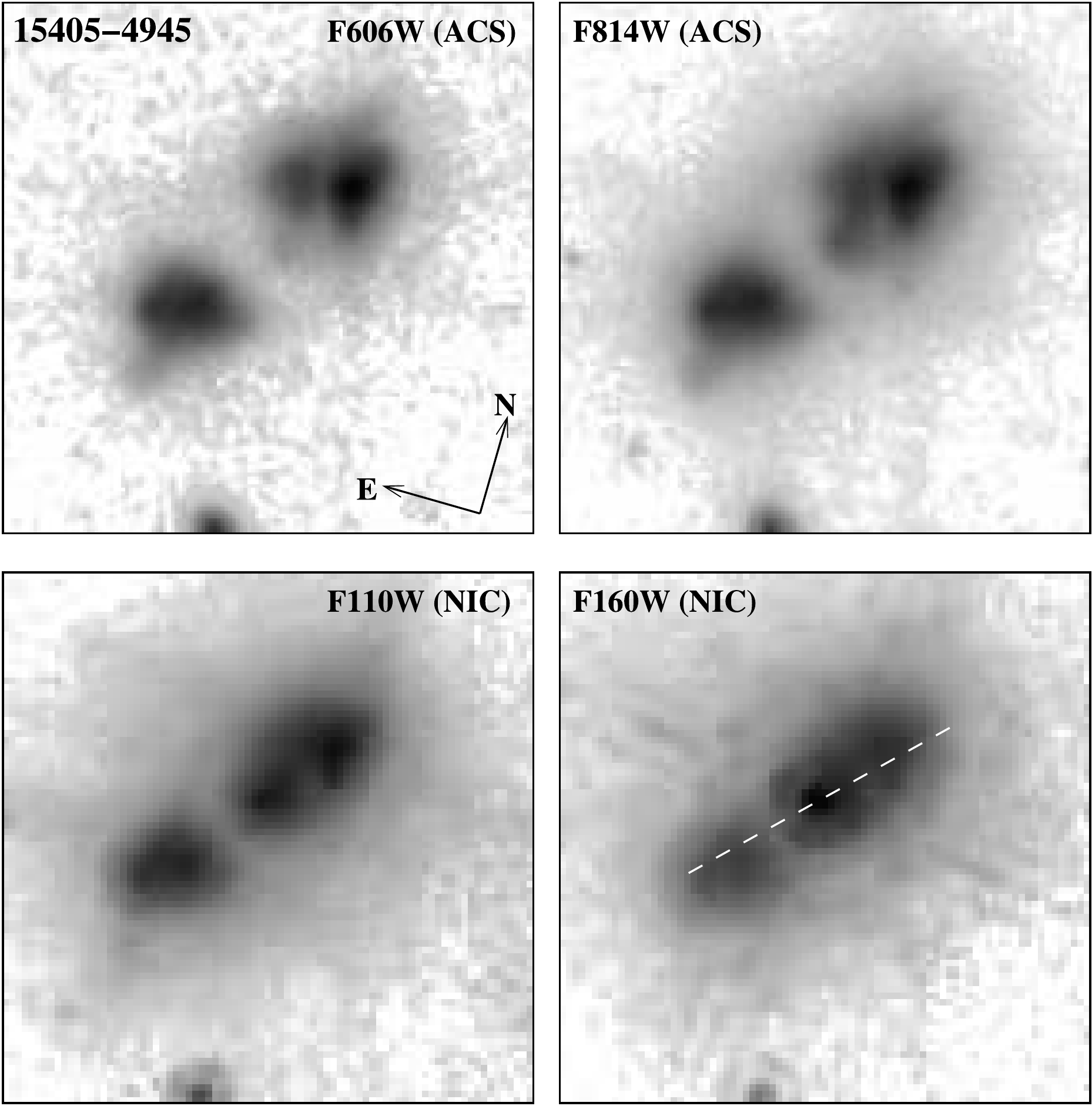}}
%\plotone{f1.eps}
\vskip 0.2in
\caption{HST images ({\it log stretch}) of the young preplanetary 
nebula IRAS\,15405-4945 (3\farcs5$\times$3\farcs5) -- {\it top} ACS, 
{\it bottom} NICMOS. The dashed white vector shown in the F160W image denotes 
the PA (-45\arcdeg) and extent of the bipolar outflow seen in OH maser emission by Zijlstra 
et al. (2001).
}
\label{fig2b}
\end{figure}

\begin{figure}[htbp] \vskip -0.6in
\resizebox{.8\textwidth}{!}{\includegraphics{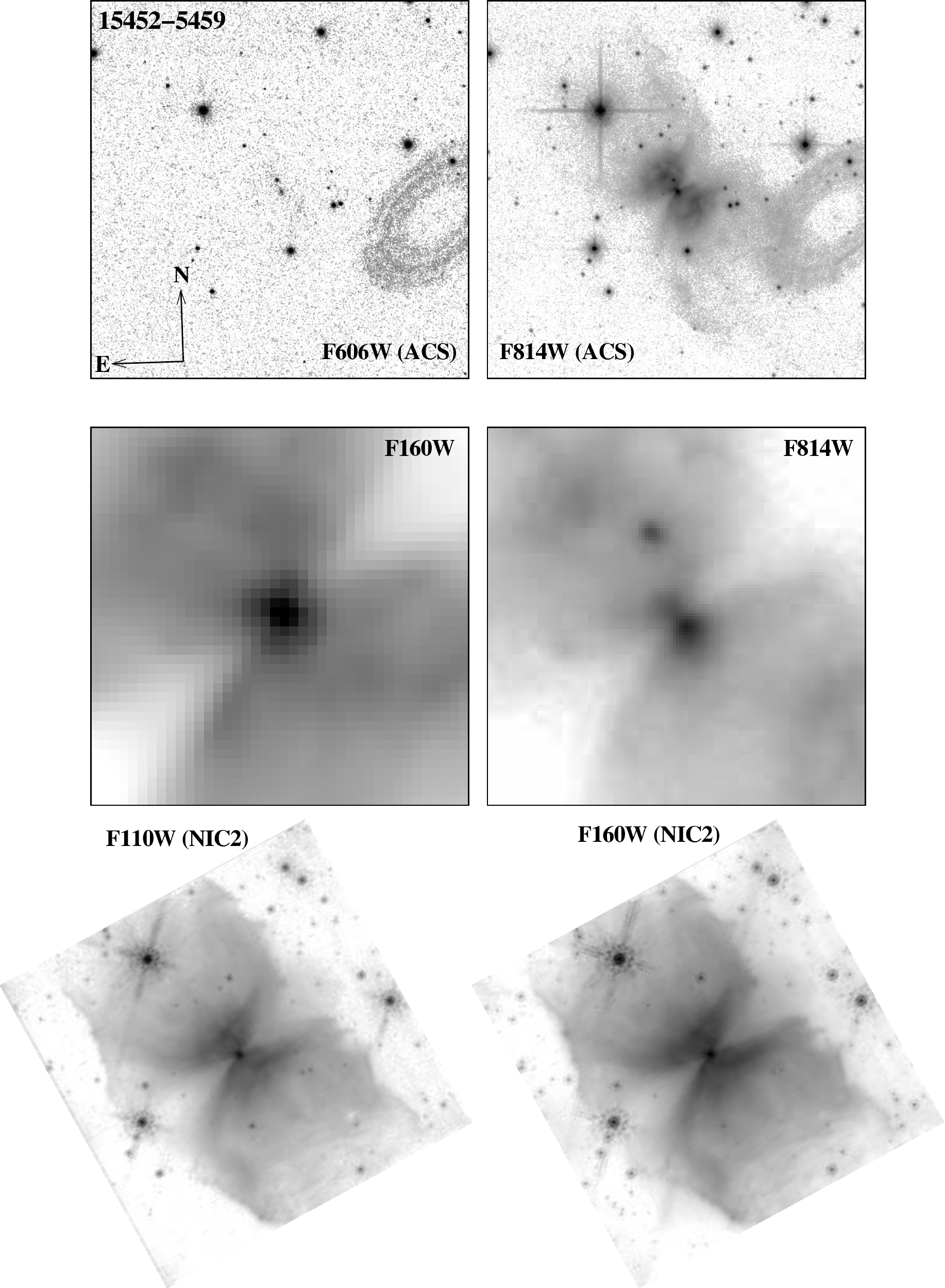}}
%\plotone{f1.eps} 
\vskip 0.2in \caption{HST images ({\it log stretch}) of the young preplanetary nebula
IRAS\,15452-5459 {\it top} -- extended nebula (25\farcs0$\times$25\farcs0 imaged with ACS
(elliptical shape on the right hand edge of image is an instrumental artifact); {\it
bottom} -- extended nebula (19\farcs3$\times$19\farcs3) imaged with NICMOS, {\it middle} --
central region (3\farcs0$\times$3\farcs0)
}
\label{fig3}
\end{figure}

\begin{figure}[htbp] \vskip -0.6in
\resizebox{1.0\textwidth}{!}{\includegraphics{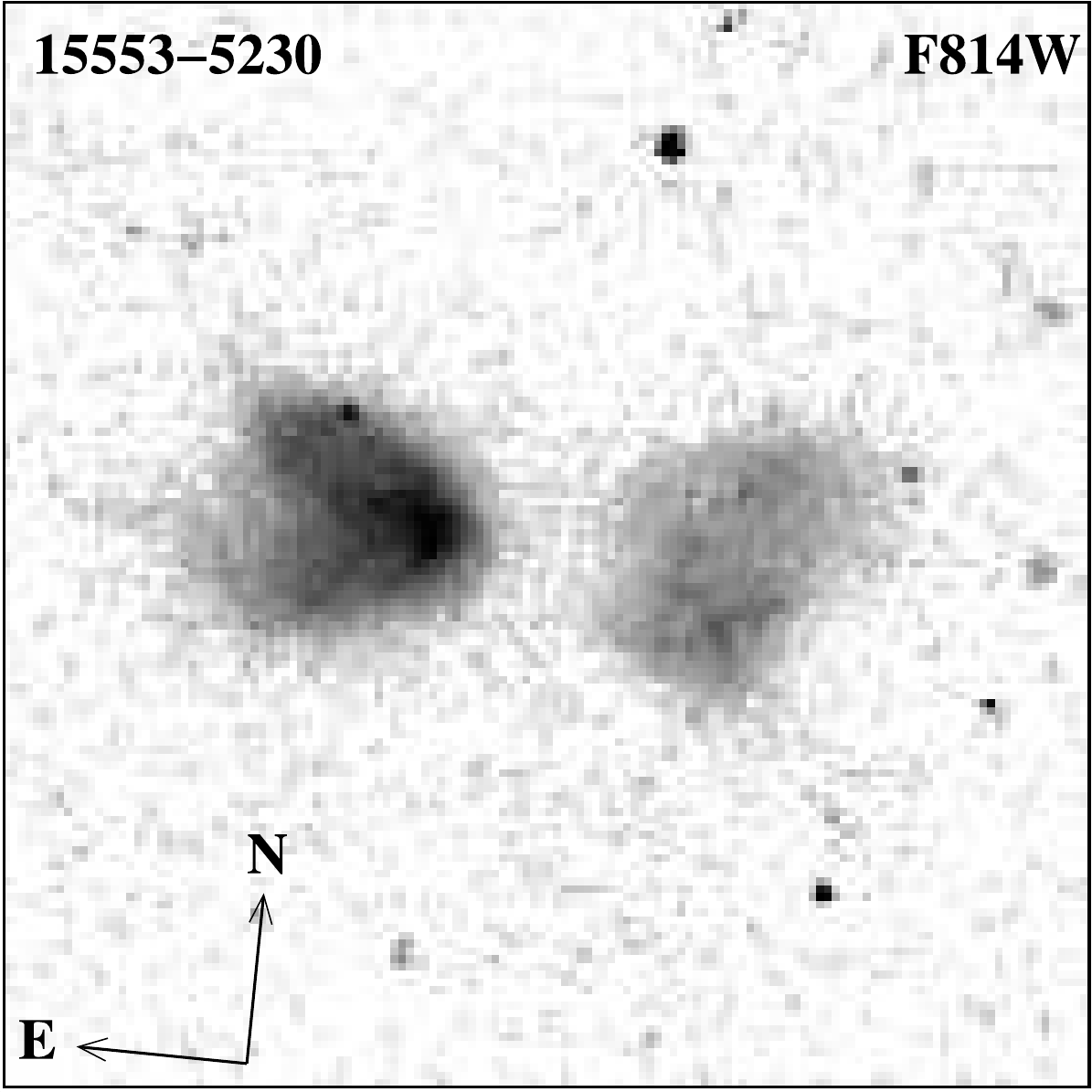}}
%\plotone{f1.eps} 
\vskip 0.2in \caption{HST image ({\it log stretch}) of the young preplanetary nebula
IRAS\,15553-5230 (3\farcs5$\times$3\farcs5)
}
\label{fig3a}
\end{figure}

\begin{figure}[htbp] \vskip -0.6in
\resizebox{1.0\textwidth}{!}{\includegraphics{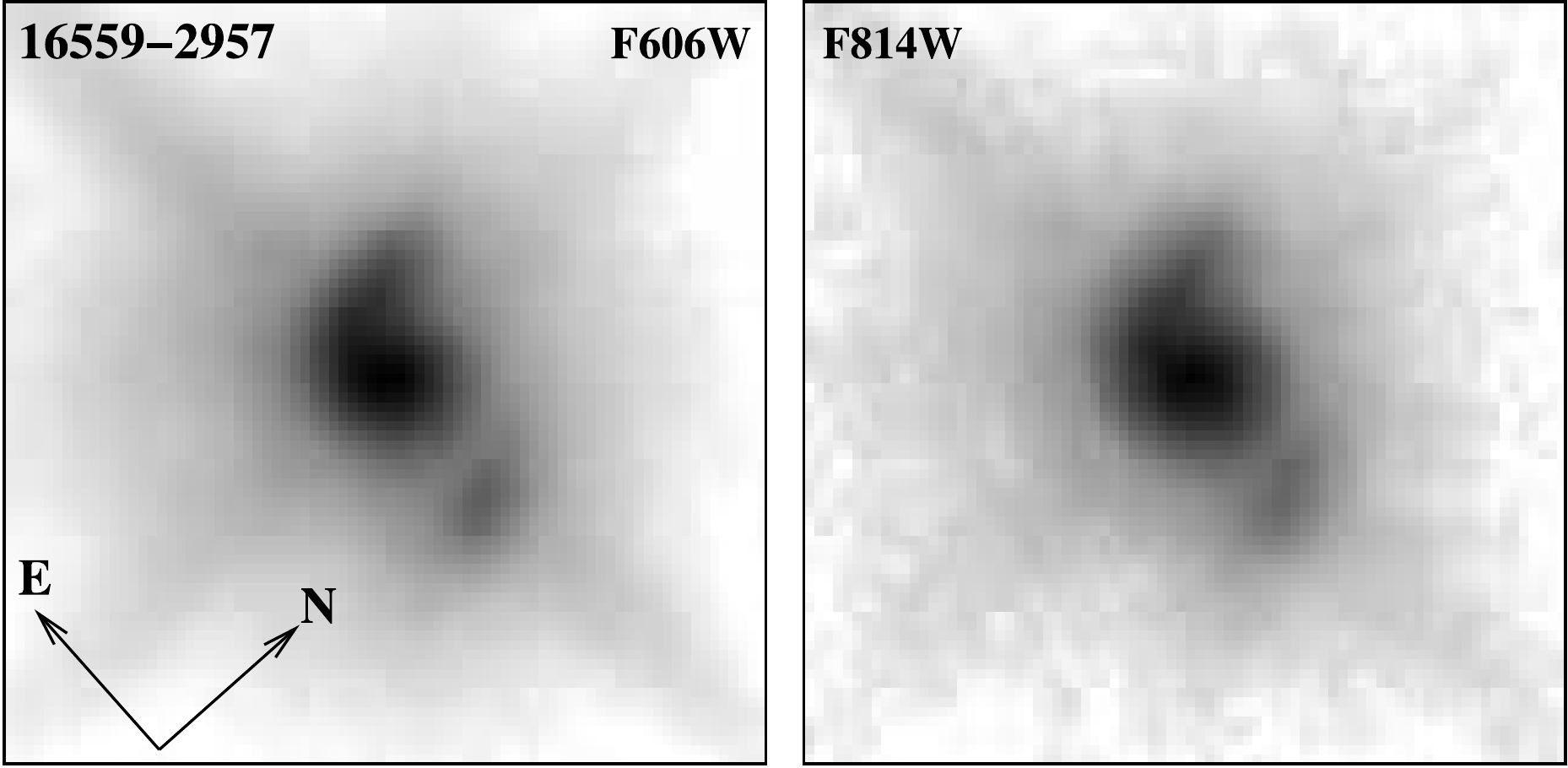}}
%\plotone{f1.eps} 
\vskip 0.2in \caption{HST images ({\it log stretch}) of the young preplanetary nebula
IRAS\,16559-2957 (1\farcs82$\times$1\farcs82)
}
\label{fig3b}
\end{figure}

\begin{figure}[htbp]
\vskip -0.6in
\resizebox{1.0\textwidth}{!}{\includegraphics{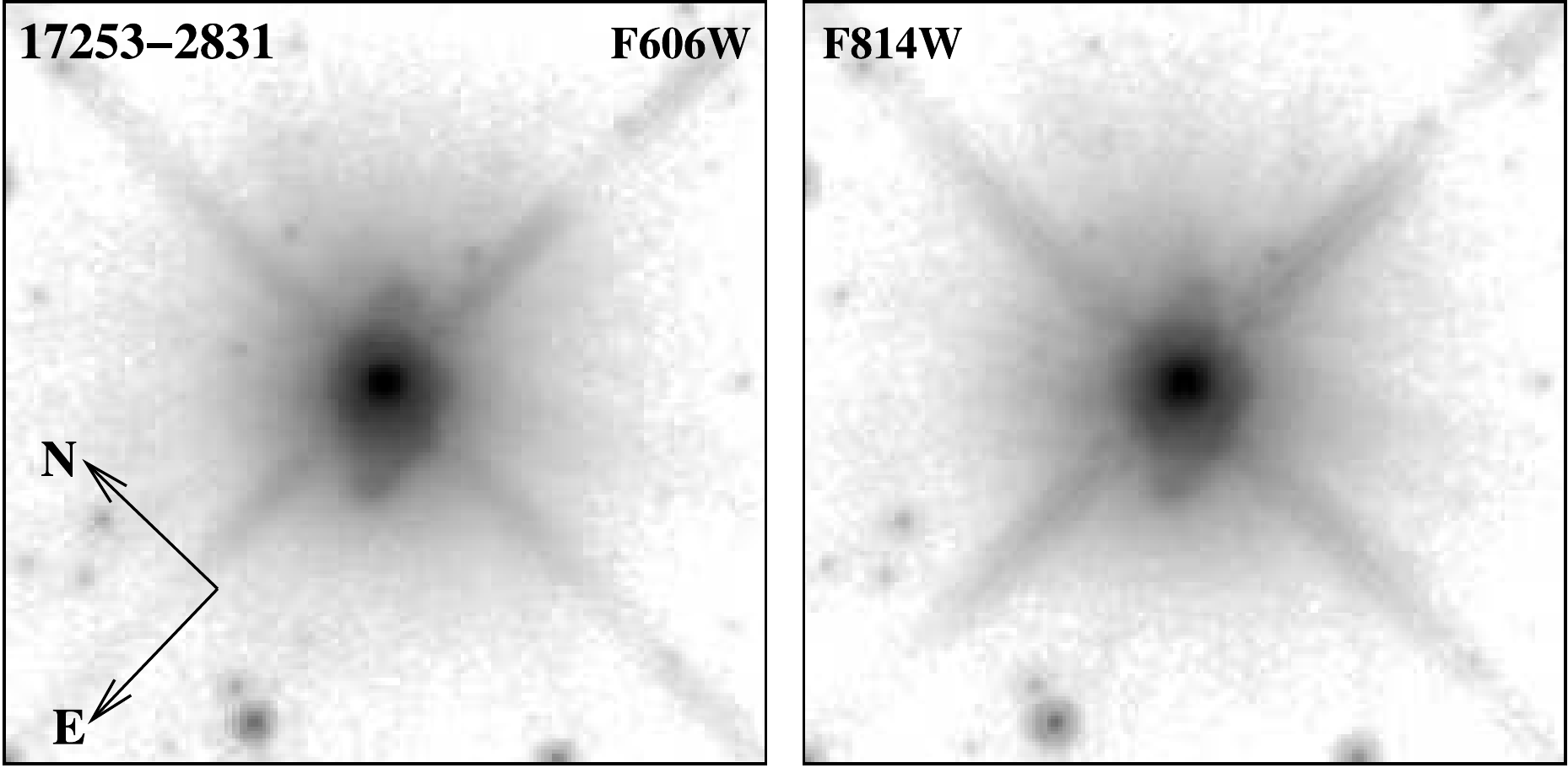}}
%\plotone{f1.eps}
\vskip 0.2in
\caption{HST images ({\it log stretch}) of young preplanetary 
nebula IRAS\,17253-2831 (4\farcs57$\times$4\farcs57)
}
\label{fig4a}
\end{figure}

\noindent {\it IRAS\,13428-6232}
% 21.88 337.4 394.6 185.8L vLrg-Bip  pg68
% 11_41_606/2x400sec 21_31_814/2x200sec 1080x870 HRC pix
%11(606w) PHOTFLAM=1.2746497E-19 / inverse sensitivity, ergs/cm2/Ang/electron
%ORIENTAT=149.53 / position angle of image y axis (deg. e of n)
% given its relatively small R(12/25)=0.065
% L=1689(d/1kpc)^2=6000, so d=1.88 kpc, 1yr=3.156e7 sec
%  SED integration gives 1487 Lsun
% 0.81E-03,  0.42E-06, 0.84E-09,  88.0 219.3 511.8 Std.Dev.  0.2646E-01
%  Md=(6000/1689)*0.81E-03=2.88e-3
% rPAGB=15.7", tPAGB=935 yr (HST), 24.7"  from 2MASS,tPAGB=1470
% assume Ve=15 km/s, rAGB=rPAGB=24.7, tAGB=
This object appears to be rather extended, with a bright cylindrical-shaped lobe seen in the
F814W image on the southwest side; the inner parts of a very faint counterpart can be seen
on the northeast side (Fig.\ref{fig1b}). The geometry is intrinsically bipolar, as can be
seen in JHK images from the 2MASS archive and a K-band image by van de Steene, van Hoof \&
Wood (2000). The near-IR images show both lobes of the bipolar nebula very clearly. The
cylindrical lobes are significantly more extended as seen in the 2MASS images: e.g., the
K-band image (Fig.\ref{fig1b}) shows that the brighter (southern) lobe extends to about
24\farcs7 along the nebular axis, compared to about 15\farcs7 in the HST F814W image. The
dusty waist has a flared shape in the K-band image, compared to the F814W image, where its
edge bordering the southwest lobe is quite straight. This change in shape is easily
understood as resulting from a decrease in the disk optical depth with wavelength from 0.8
to 2\micron. The central star of the nebula is clearly visible in the K-band
image in van de Steene et al (2000) which has smaller pixels (0\farcs25) than the 2MASS
image (1\farcs0). The nebula is much fainter in the F606W image. Therefore, although it has
been listed as a possible young PN by Preite-Martinez (1988), the lack of nebulosity in the
F606W filter (which covers the H$\alpha$ line) shows that any H$\alpha$ emission, if
present, is very weak, indicating that the circumstellar material has not been ionised
substantially. In the compilation by Kwok et al.
(1997), the LRS spectrum of IRAS\,13428 is classified as ``H", i.e. having a red continuum
with either a
9.7\micron~silicate absorption feature or a 11.3\micron~PAH emission feature, so its
chemistry is unknown.

\noindent {\it IRAS\,13557-6442}
% 33.73 63.58 24.92 13.4L  Smal-Bip  pg 76
%  MSX fluxes are too high compared to IRAS
%  1.465e+01, 1; 3.573e+01, 4; 5.766e+01, 4; 6.960e+01, 4; 8.359e+01, 4: MSX6C F,Qual
%   above are B2[4.35micr], A,C,D,E
% current images are 11_41_606/2x400sec, and 21_31_814/2x200sec
% in del-cosmic script, introduce code to correctly treat Fastie Finger
% 250 x 250 0.5 HRC pix
% ORIENTAT=-94.58 / position angle of image y axis (deg. e of n)
% (OH)Vr=-0.4,Vb=-3.4
% 11(606w): PHOTFLAM=1.2746497E-19 / inverse sensitivity, ergs/cm2/Ang/electron
% L=536(d/1kpc)^2=6000, so d=3.34 kpc, 1yr=3.156e7 sec
%  t=3175, if Ve=(15) km/s
%  0.47E-04, 0.33E-06, 0.31E-10 103.2 270.0 850.0 Std.Dev. 0.5093E-01
%   Md=(6000/536)*0.47E-04=5.26e-4
% rPAGB=0.8, tPAGB=84.7yr
This object is bipolar, with an hourglass shape and a dense waist seen almost 
edge-on (Fig.\ref{fig2a}). The hourglass nebula is surrounded by an extended round
halo. Using azimuthal averaging, we can trace the halo out 
to a radius of $\sim$3$''$; the radial distribution roughly follows a power-law,
r$^{-\beta}$ with $\beta\sim$2.3. In this source, the MSX6C fluxes are roughly a factor
$1.7-1.5$ higher in the 12-21\micron~range compared to the corresponding IRAS fluxes
(either due to variability or poor calibration). We
have therefore scaled down the MSX6C fluxes by a factor 1.6 in generating its SED. Since the
blue and red peaks of the OH maser emission line are separated by only 3\,\kms, it is
unlikely that the standard interpretation of this separation representing twice the recent 
AGB CSE expansion velocity is applicable. We have therefore assumed V$_{exp}$=15~\kms,
which gives an age $t_{AGB}$=3200($D_L$/3.3\,kpc)\,yr for the halo.
%{\it what does OH profile look like -- classical double peak, supporting AGB
%interpretation, or different?}

%{\it add intensity cuts of halo, but then have to do this for all objects with halo?}

\noindent {\it IRAS\,15405-4945}
% 2.38L 26.51 82.99 34.22: Medm-Bip  pg 48, NIC pg 82
% current images are 606W/2x338sec (hm_606.ps), and 814W/2x338sec (ik_814.ps)
% 140 x 140 images, 0.5 * WFC/ACS pix
%ORIENTAT=-74.0623208268169 / position angle of image y axis (deg. e of n)
%hq(606w):PHOTFLAM=7.8697440E-20 / inverse sensitivity, ergs/cm2/Ang/electron
% L=249(d/1kpc)^2=6000, so d=4.91 kpc, t=xx yr, 1yr=3.156e7 sec
% dust model: 0.29E-03  0.93E-07 0.31E-09  71.9 219.3 600.8 Std.Dev.  0.7239E-01
%  Md=(6000/249)*0.29e-3=6.98e-3
% rPAGB=1" to tips of main lobes; tPAGB=155 yr (using Ve=150 km/s)
% rAGB=3", tAGB=5180 yr
This bipolar nebula shows complex, highly structured lobes (Fig.\ref{fig2b}). A faint
elliptical halo can be seen in the F814W image. If the halo is due to a spherical AGB
mass-loss envelope, the elliptical shape indicates that the AGB envelope is being
indirectly illuminated by the lobes, and not by a central source, or that the light from the
central source cannot get out to the larger distances at low latitudes. We also obtained
NICMOS images with NIC1 in the F110W and F160W filters. Although the overall shape of the
bipolar lobes in these near-IR filters is roughly similar to that in the optical ones,
there are distinct differences. The central star is visible in the NIC1 images, lying
midway on the line joining the tips of the lobes. The OH masers
in this object span a very large range (80\,\kms~in the 1612\,MHz line, 105\,\kms~in 
the 1665\,MHz line and 155\,\kms~in the 1667\,MHz line), and lie along a PA of -45\arcdeg,
covering an elongated region with major axis of $\sim2{''}$ (Zijlstra et al 2001).  Thus
the extent of the fast outflow as traced by the OH masers matches the size of the bipolar
nebula seen in the HST images. We estimate an expansion velocity of the AGB envelope of
13.5\kms~from the separation of the prominent peaks in the OH 1665 MHz line profile; and
given the 3$''$ extent of the halo, we find $t_{AGB}$=5200($D_L$/4.9 kpc)\,yr.
%{\it what about water masers?)}

\noindent {\it IRAS\,15452-5459}
% 87.05 242.7 273.6 401.6L Larg-Bip  pgs 43 & 44 & 78
% v1_606/2x338, wz_814/2x338 1000x1000, 0.5 WFC pix
% v1_cen, wz_cen 120x120
%ORIENTAT=-88.0821779590589 / position angle of image y axis (deg. e of n)
%(vw: 814W) PHOTFLAM=7.1726731E-20 / inverse sensitivity, ergs/cm2/Ang/electron
% L=2117(d/1kpc)^2=6000, so d=1.68 kpc, t= yr, 1yr=3.156e7 sec
% dust model fit L=1907; 0.63E-03, 0.14E-05, 0.27E-09  87.0 240.0 800.8 Std.Dev. 0.4532E-01
%  Md=0.63E-03(6000/2117)=1.78e-3
% rPAGB=11.7 to tips of main lobes; tPAGB=622.8yr; TAKE rAGB=11.7, get tAGB=9833.7
This object shows a very extended hourglass shaped nebula in the F814W image; almost no
nebulosity is visible in the F606W image. 
The two lobes are of roughly
equal brightness and are separated by a flared, edge-on waist
(Fig.\ref{fig3}). We also obtained NICMOS images with NIC2 in the F110W, F160W and F205W
filters -- the shape of the extended nebula appears quite similar in all three near-IR
filters compared to that in the F814W filter. However, the central star, which is not
visible in the F814W image, appears visible at the geometrical center of the dust lane in
the near-IR images. The waist, which appears as a minimum in intensity between the bright
inner regions of the bipolar lobes, appears as a bright feature against the sky background
beyond the western lateral periphery of the lobes, and shows a distinct, curved, outer
edge. The OH maser
emission in this object shows blue and red peaks at -67.2 and -48.3\,\kms, implying an
outflow speed of 9.5\kms. 
Since a halo is not observed directly, we have assumed $r_{AGB}$=$r_{PAGB}$ to compute  an
estimated AGB mass-loss time-scale, $t_{AGB}$=9850($D_L$/1.7 kpc)\,yr.

\noindent {\it IRAS\,15553-5230}
% SED integration: L=307.7
% IRAS: 9.99, 70, 49.6, <282.7
% L=329.9(d/1kpc)^2=6000, so d=4.26 kpc, tAGB= yr, 1yr=3.156e7 sec
%   dust model L=329.9 (use this as it account sfor in cool component beyond 60 micr)
% dust: 0.60E-04 0.90E-07 0.70E-10 104.0 230.0 580.0 Std.Dev.  0.5755E-01
% Md=1.09e-3
% rPAGB=1.06 asec=1430 yr
% not detected in CO 1-0 by Loup et al AA 227, L29 (1990)
This object is bipolar, with an hourglass shape and a dense waist seen almost 
edge-on (Fig.\ref{fig3a}), very similar to IRAS\,13557. It was only imaged in F814W. No halo
can be seen. No line emission (either CO or OH) has been detected from the object, and the
AGB expansion velocity is unknown. Assuming a typical AGB outflow velocity of
V$_{exp}$=15\,\kms, we find $t_{AGB}$=1430($D_L$/4.3 kpc) yr. We do not know if IRAS\,15553
is oxygen- or carbon-rich; although it is included in the compilation by Kwok et al.
(1997), its LRS spectrum is classified as ``H", i.e. having a red continuum with either a
9.7\micron~silicate absorption feature or a 11.3\micron~PAH emission feature.

\noindent {\it IRAS\,16559-2957}
% Smal  old notebook, no MSX, but Spitzer
% 2MASS	16 59 08.22, -30 01 40.3: 11.596 0.023	10.713 	0.023 	9.347 	0.021
% IRAS  9.17 32.37 16.38  4.18L %
% 80 x 80 1/2 PC images 2m_am, 4m_8m 
% L=222(d/1kpc)^2=6000, so d=5.2 kpc, tAGB= yr, 1yr=3.156e7 sec
%  dust model L=195
%  0.14E-04     0.62E-08     0.15E-10 118.1 401.3 850.0 Std.Dev.  0.4938E-01
%   Md=(6000/222)*0.14E-04=3.78e-4
% OH Bonn data from telintel, Vr=85.6, Vb=57, Ve=14.75
% rPAGB=0.7/2=0.35"
This object (with an F5Iab:e central star: SIMBAD database) shows a compact nebula with
filamentary structures on either side of a central,
round core (Fig.\ref{fig3b}). The eastern filament emanates from the central core, whereas
the
western one is separated from the latter by a dark band, which most likely represents a
tilted, dusty equatorial structure. Using azimuthal averaging, we can trace a
tenuous halo out to a radius of about 3\farcs5. The radial distribution follows a
power-law, r$^{-\beta}$
with $\beta\sim$2.8. Using an AGB envelope expansion velocity of 14.8\kms~derived from OH
maser
data, we find that the AGB envelope was characterised by a 
roughly constant mass-loss rate over a period $t_{AGB}$=5850($D_L$/5.2 kpc)\,yr.

\noindent {\it IRAS\,17253-2831}
% IRAS 1.35  20.39 15.89 20L   Smal-Bip   pg 2
% image 3m_9m_606.ps, 200sec, 200x200 0.5xPC pixels
% image 5m_6m_814.ps, 100sec, 200x200 0.5xPC pixels
% no ISO, MSX+IRAS, 12/25=1.35/20.39
% OH masers, Ve=9.1 (te Lintel Hekkert et al 1991, Zijlstra et al 2001)
% L=101(d/1kpc)^2=6000, so d=7.7 kpc, tAGB=10816 yr, 1yr=3.156e7 sec
%  0.22E-04, 0.16E-07, 0.31E-10  97.5 192.9 850.0 Std.Dev. 0.4305E-01
%  Md=(6000/101)*0.22E-04=1.31e-3
% rPAGB=0.66, tPAGB=161.0
This object has point-symmetrically shaped lobes, and is surrounded by an extended round
halo (Fig.\ref{fig4a}). The OH maser 
1612\,MHz profile shows 2 prominent peaks at $V_{lsr}$ velocities of -71.4 and -53.2\kms
(implying an AGB envelope expansion velocity of 9.1\kms) 
the emission at velocities within these peaks is confined to a poorly resolved, roughly
elliptical region of size about 0\farcs35 (Zijlstra et al. 2001). Using azimuthal averaging,
we can trace the halo out 
to a radius of $\sim$2\farcs7; the radial distribution follows a power-law, r$^{-\beta}$
with $\beta\sim$2.9, signifying a 
roughly constant mass-loss rate over a period $t_{AGB}$=10800($D_L$/7.7 kpc) yr. Zijlstra et
al.
(2001) believe
that there is a fast bipolar outflow based on the locations of maser features at velocities
outside the main profile peaks, but do not provide a position angle which can be compared
with that of the optical nebula in our HST images. 

\begin{figure}[htbp]
%vskip -0.6in
\resizebox{1.0\textwidth}{!}{\includegraphics{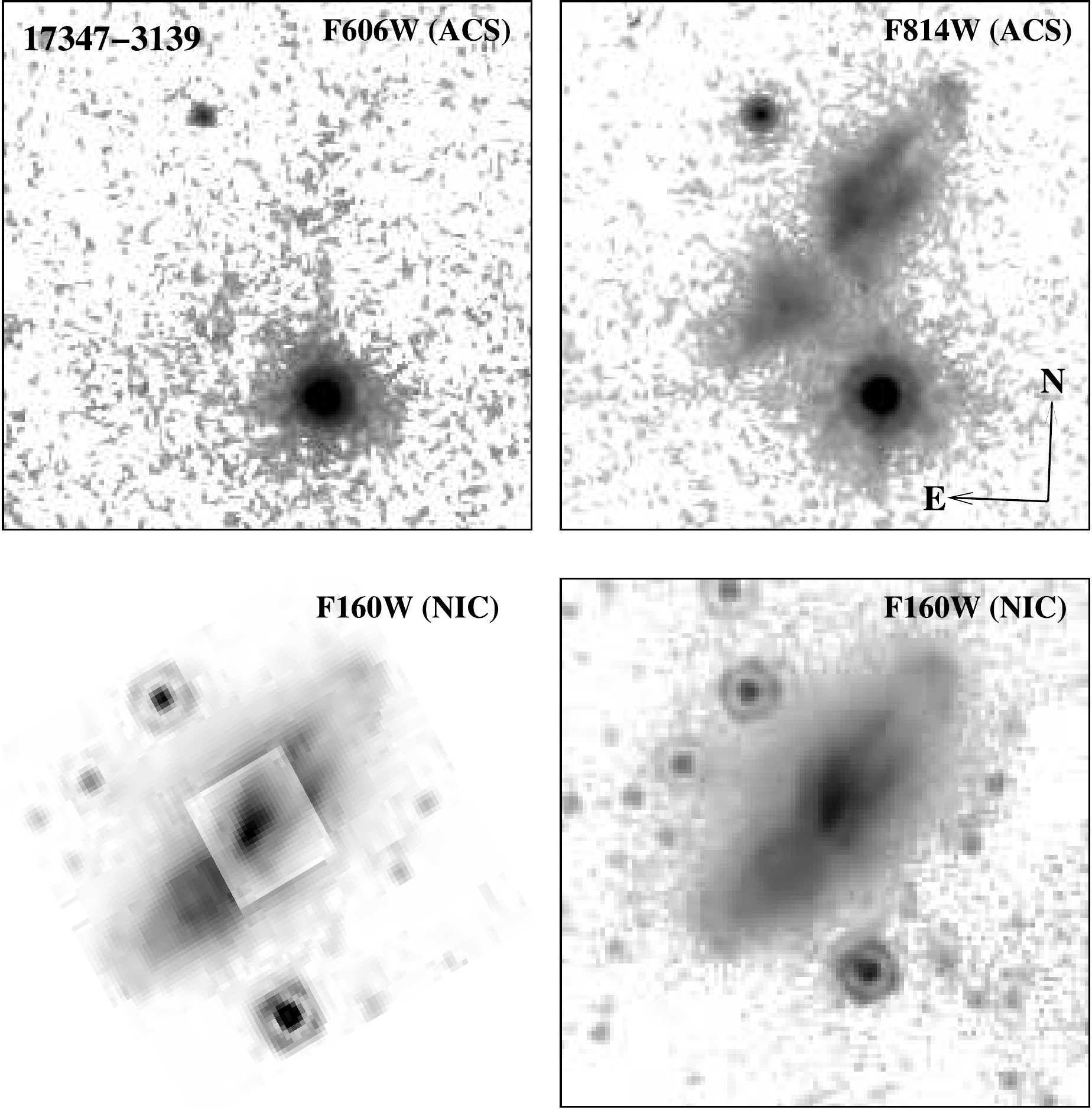}}
%\plotone{f1.eps}
\vskip 0.2in
\caption{HST images of the young preplanetary 
nebula IRAS\,17347-3139 (5\farcs0$\times$5\farcs0). The {\it top} panel shows ACS images
using a {\it log stretch}. The {\it bottom} panel shows NICMOS (NIC1) images -- on the {\it
right} is the F160W image using a  
{\it log stretch}; on the {\it left} the same image is shown using a linear stretch, with
the intensity reduced by a factor 5 in a central 0\farcs85$\times$0\farcs85 patch, in order
to better display the structure within the lobe. The faint patchy structure seen to the
right of, and below the middle of, the nebula, in
the F160W log-stretch image is an instrumental artifact.
}
\label{fig4b}
\end{figure}

\begin{figure}[htbp]
\vskip -0.6in
\resizebox{1.0\textwidth}{!}{\includegraphics{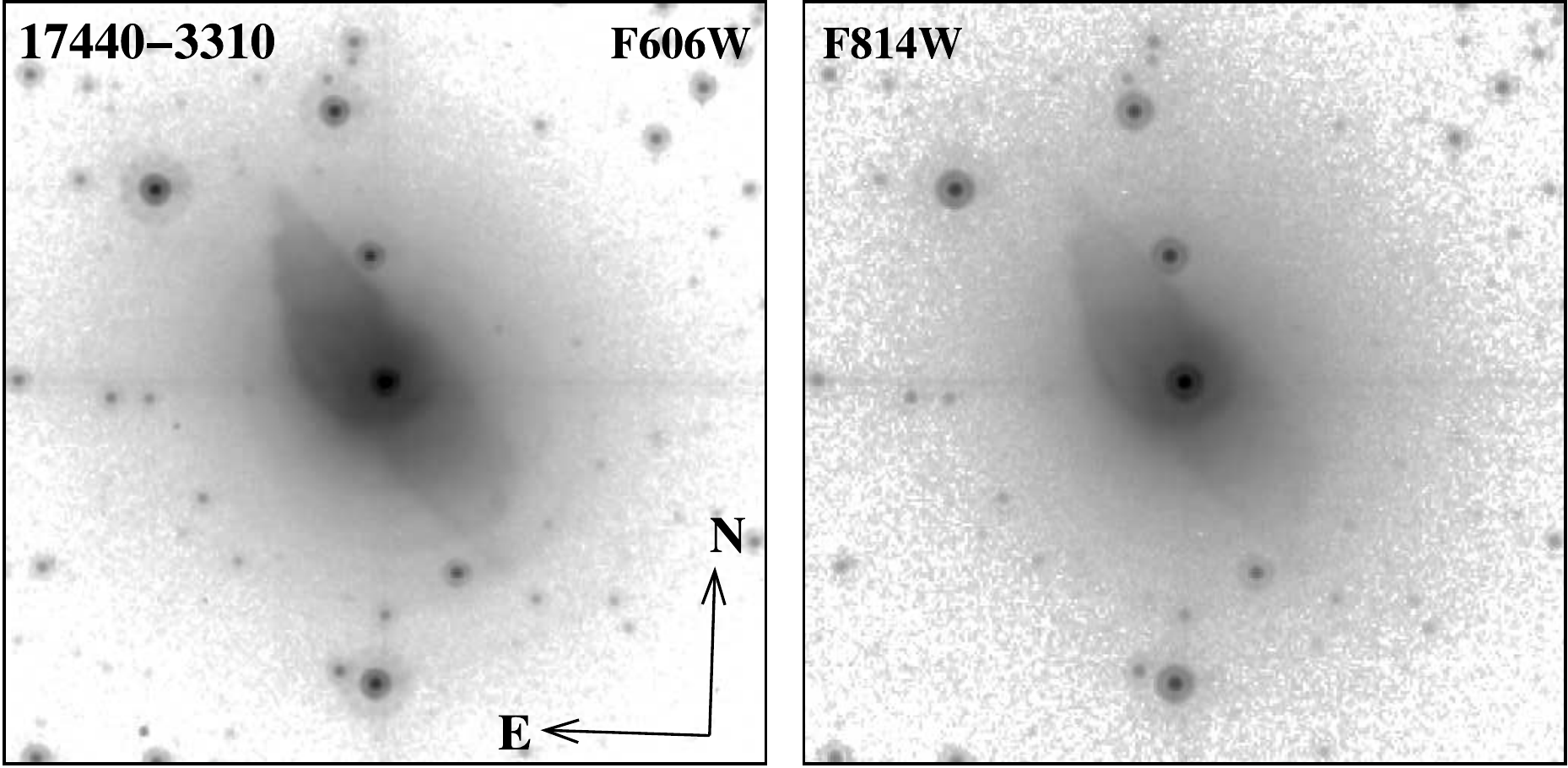}}
%\plotone{f1.eps}
\vskip 0.2in
\caption{HST images ({\it log stretch}) of the young preplanetary 
nebulae IRAS\,17440-3310 (5\farcs0$\times$5\farcs0).
}
\label{fig5a}
\end{figure}

\begin{figure}[htbp]
\vskip -0.6in
\resizebox{1.0\textwidth}{!}{\includegraphics{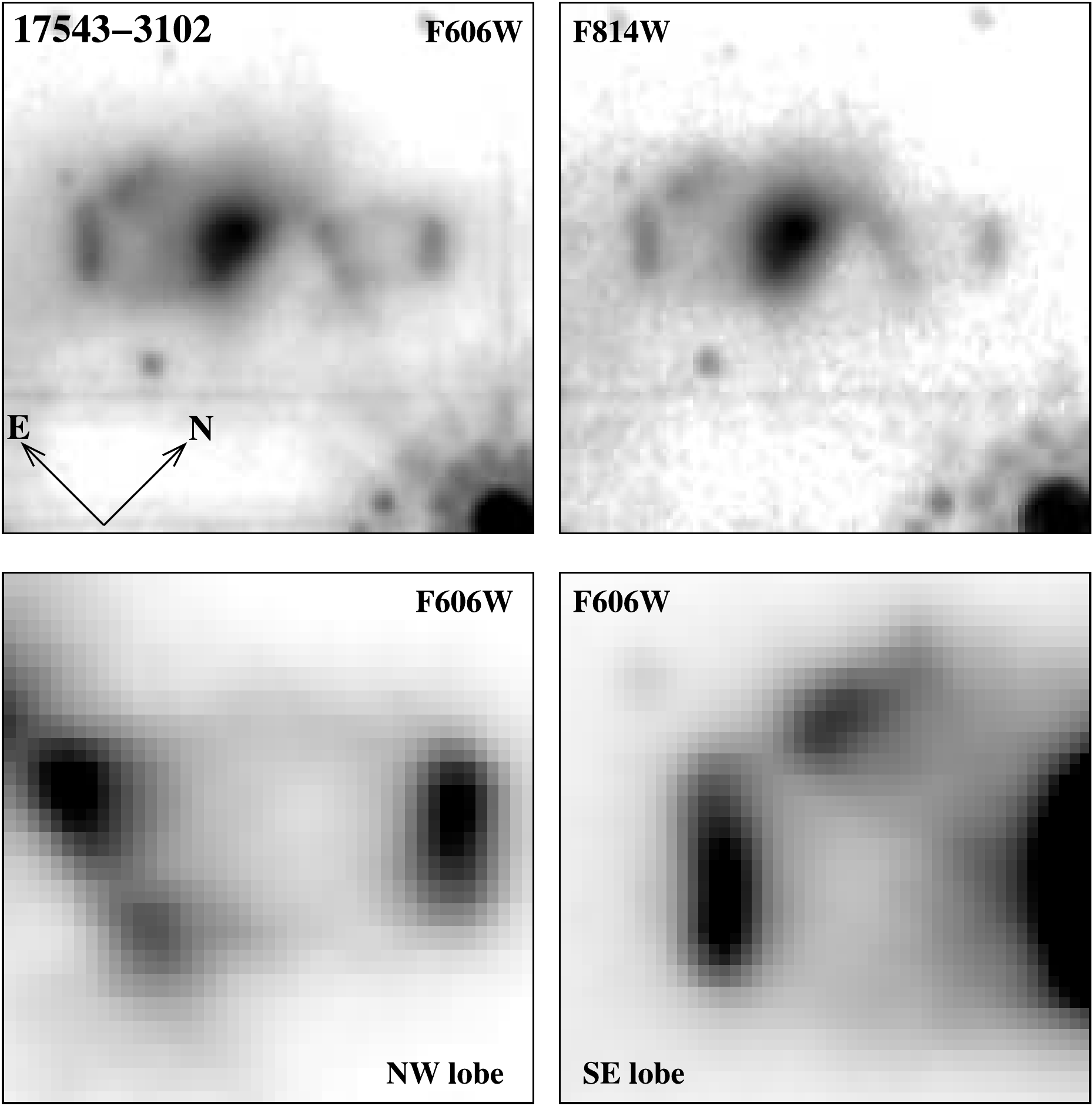}}
%\plotone{f1.eps}
\vskip 0.2in
\caption{HST images of the young preplanetary 
nebulae IRAS\,17543-3102 (2\farcs0$\times$2\farcs0). {\it Top} panel shows the F606W and
F814 images on a {\it log stretch}, {\it bottom} panel shows magnified views
(0\farcs56$\times$0\farcs56) of the SE and NW parts of the nebula in the F606W image on a
linear
stretch, with the central region saturated in order to bring out the 
shape and structure of
the lobes
}
\label{fig5ab}
\end{figure}

\begin{figure}[htbp]
\vskip -0.6in
\resizebox{1.0\textwidth}{!}{\includegraphics{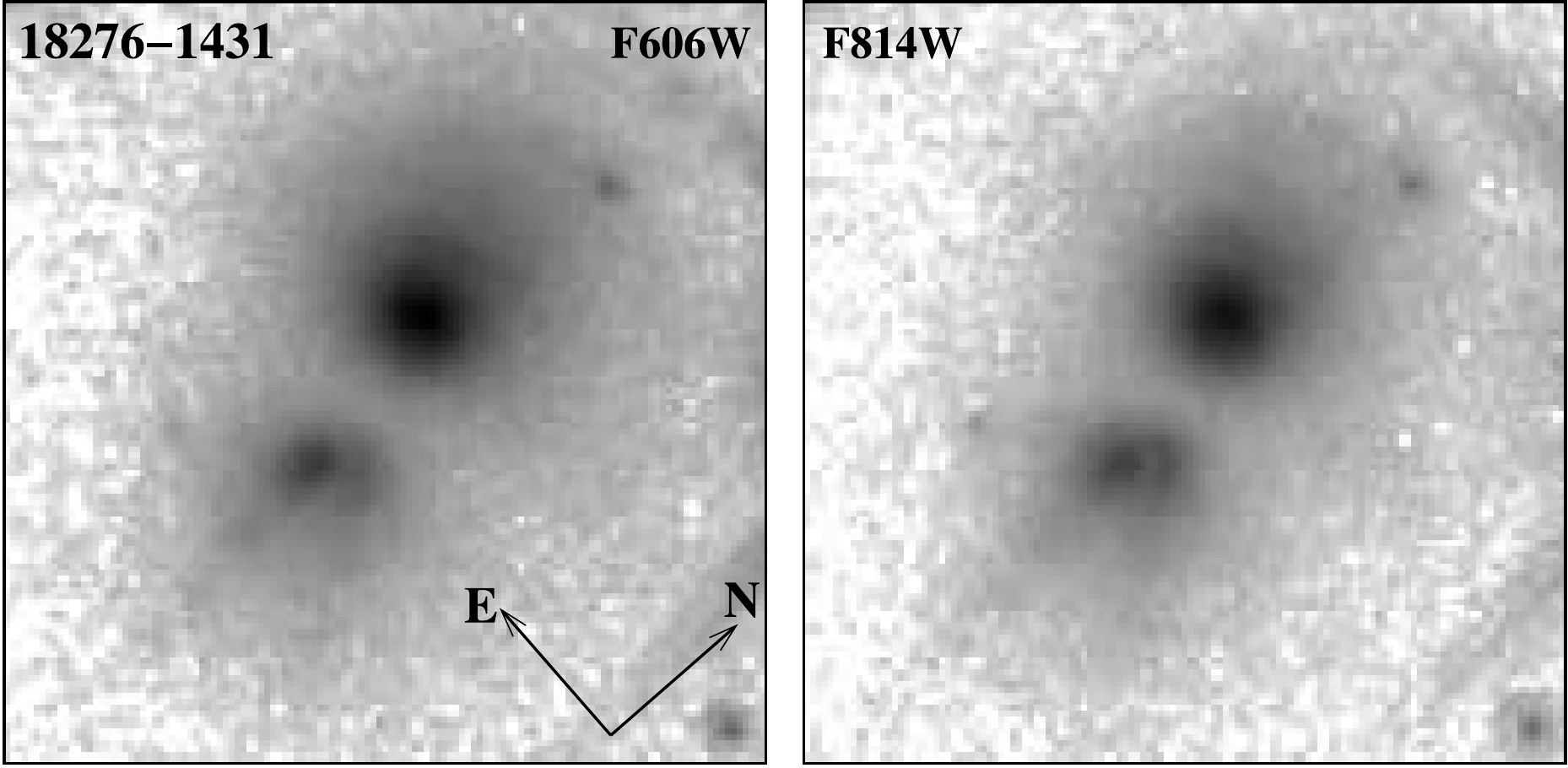}}
%\plotone{f1.eps}
\vskip 0.2in
\caption{HST images ({\it log stretch}) of the young preplanetary 
nebula IRAS\,18276-1431 (2\farcs96$\times$2\farcs96). The linear structures on the right
edge of the image are diffraction spikes due to a bright field star.
}
\label{fig5b}
\end{figure}

\begin{figure}[htbp]
%\vskip -0.6in
\resizebox{1.0\textwidth}{!}{\includegraphics{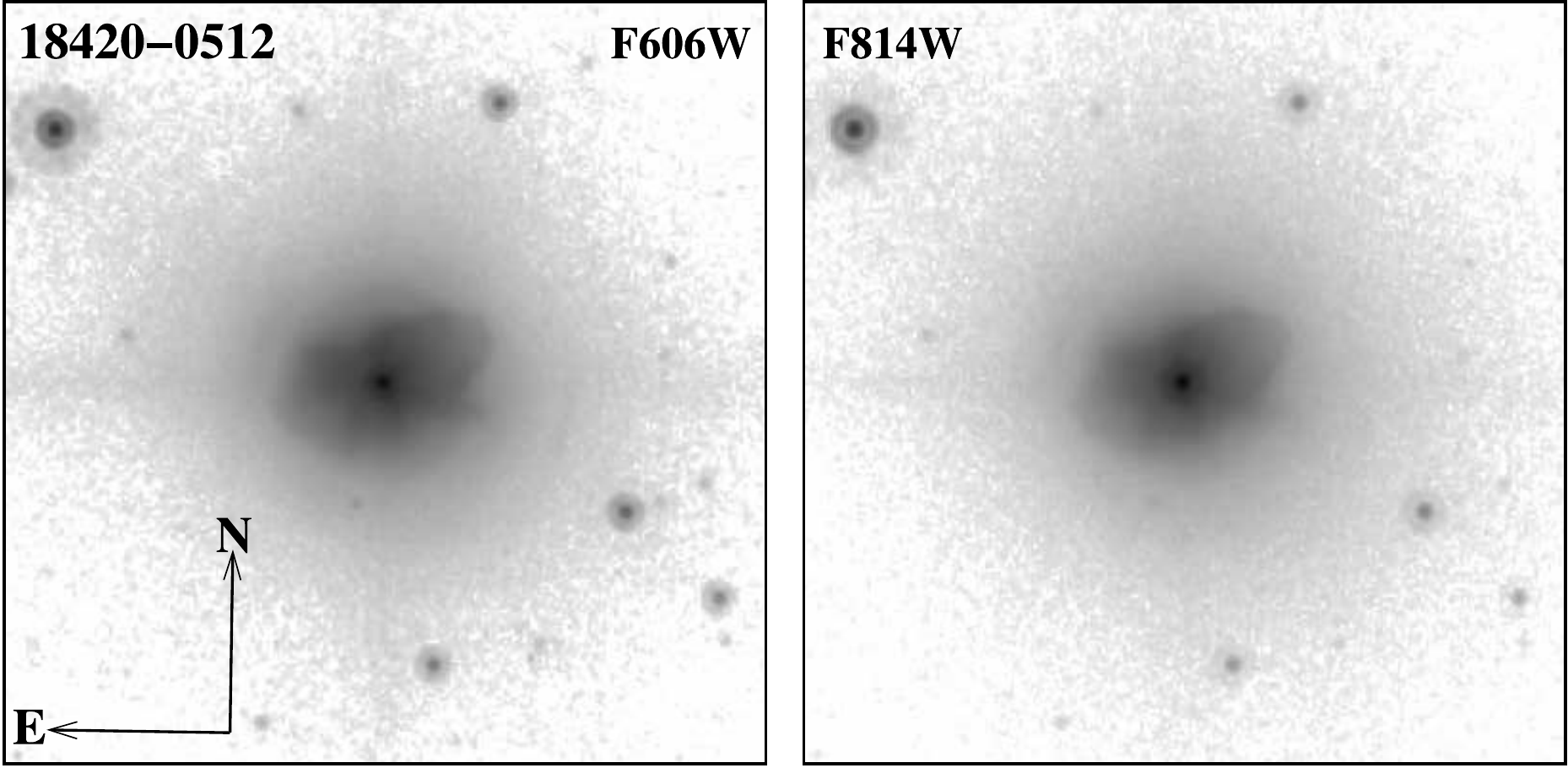}}
%\plotone{f1.eps}
\vskip 0.2in
\caption{HST images ({\it log stretch}) of young preplanetary 
nebula IRAS\,18420-0512 (5\farcs0$\times$5\farcs0)
}
\label{fig6a}
\end{figure}

\begin{figure}[htbp]
\vskip -0.6in
\resizebox{.9\textwidth}{!}{\includegraphics{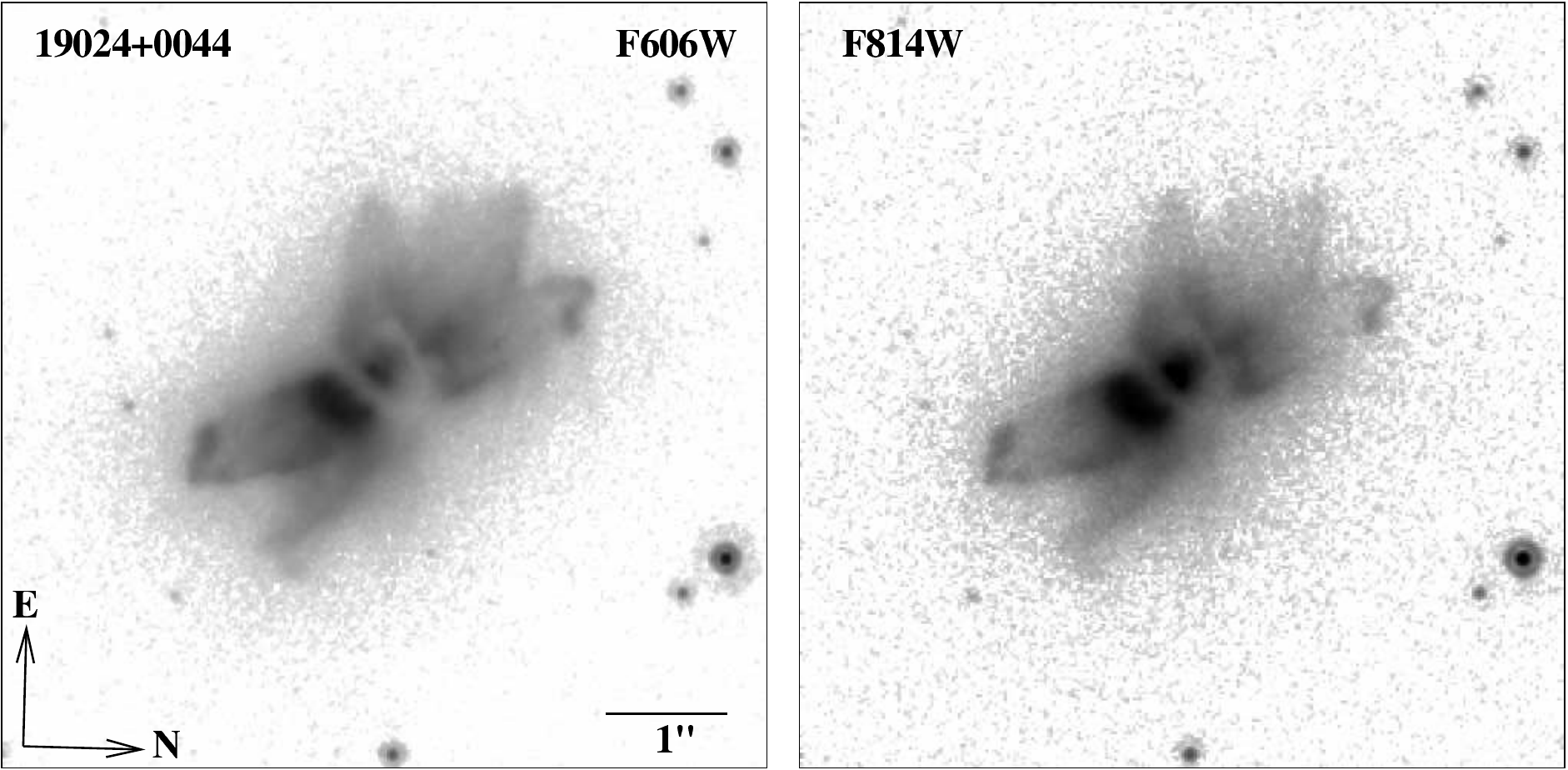}}
%\plotone{f1.eps}
\vskip 0.2in
\caption{HST images ({\it log stretch}) of the young preplanetary 
nebula IRAS\,19024+0044 (6\farcs25$\times$6\farcs25)
}
\label{fig6ab}
\end{figure}

\begin{figure}[htbp]
\vskip -0.6in
\resizebox{1.0\textwidth}{!}{\includegraphics{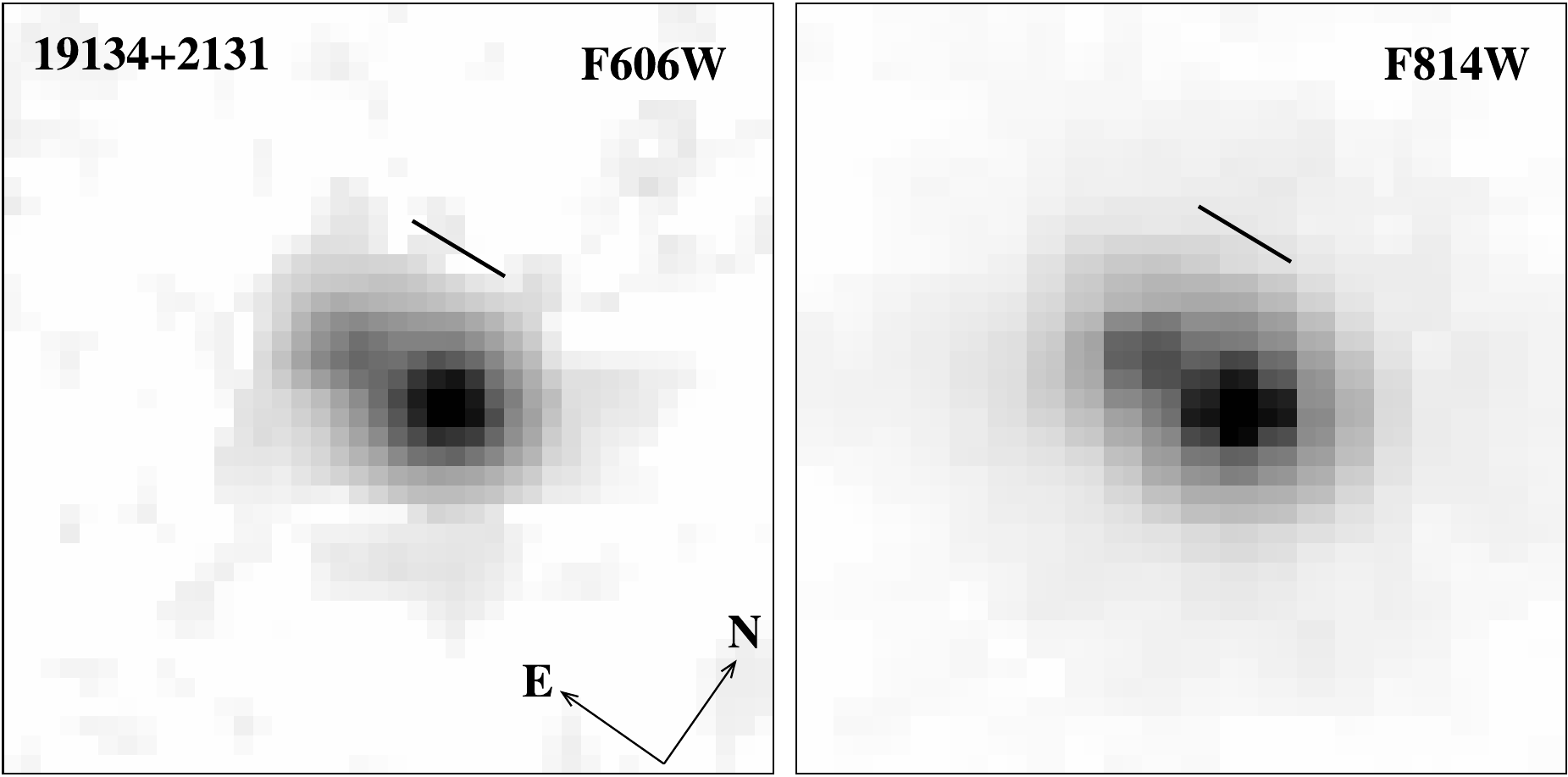}}
%\plotone{f1.eps}
\vskip 0.2in
\caption{HST images ({\it square-root stretch}) of the young preplanetary 
nebula IRAS\,19134+2131 (1\farcs0$\times$1\farcs0). The straight line segment shows the PA
and separation of the blue- and red-shifted H$_2$O maser-jet spot clusters in this object
observed by Imai et al. (2007). 
%An unrelated field star is seen below the nebula in the image.
}
\label{fig6b}
\end{figure}

\begin{figure}[htbp]
\vskip -0.6in
\resizebox{1.0\textwidth}{!}{\includegraphics{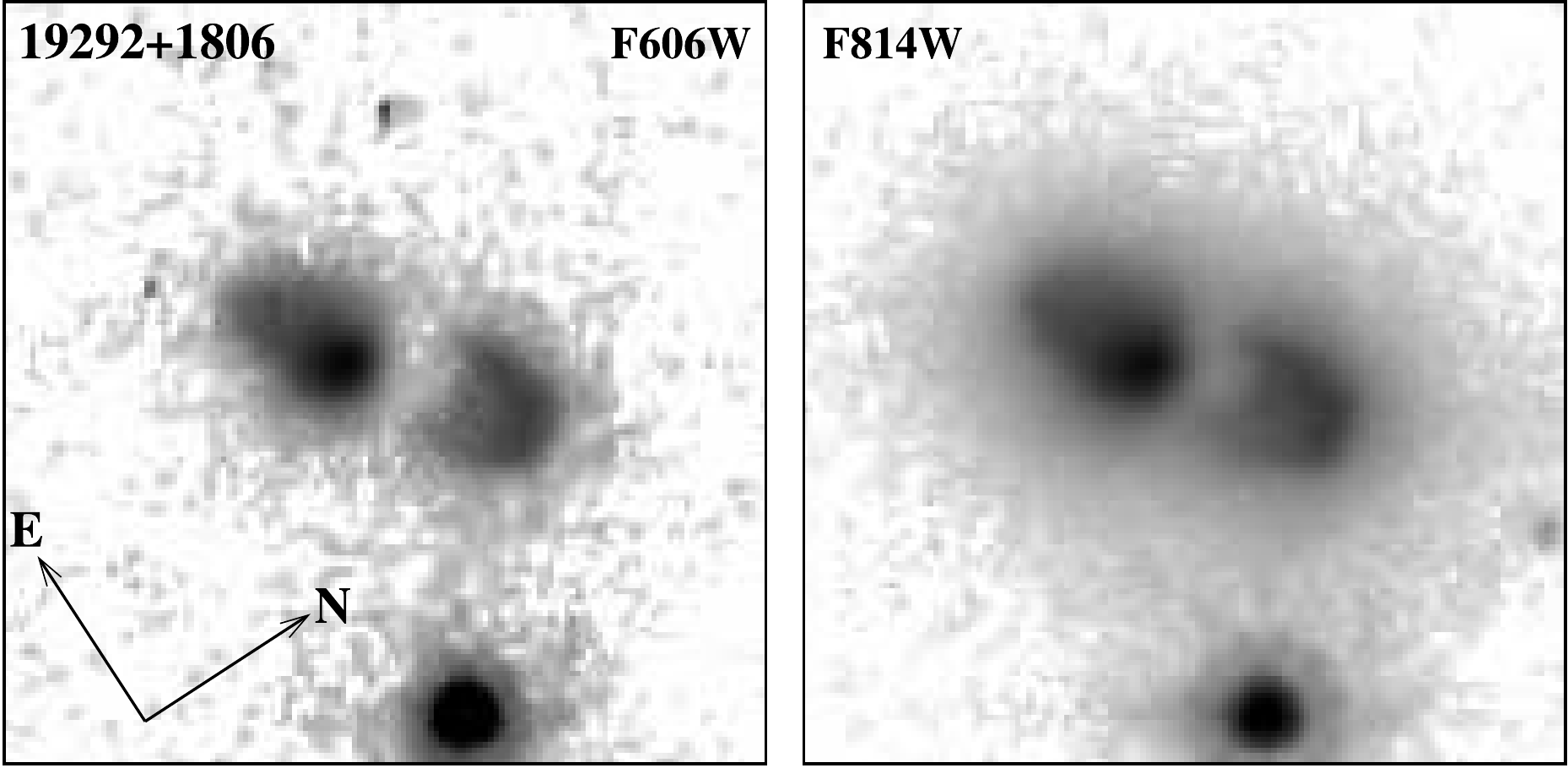}}
%\plotone{f1.eps}
\vskip 0.2in
\caption{HST images ({\it log stretch}) of young preplanetary 
nebula IRAS\,19292+1806 (3\farcs5$\times$3\farcs5).
}
\label{fig7a}
\end{figure}

\begin{figure}[htbp]
\vskip -0.6in
\resizebox{1.0\textwidth}{!}{\includegraphics{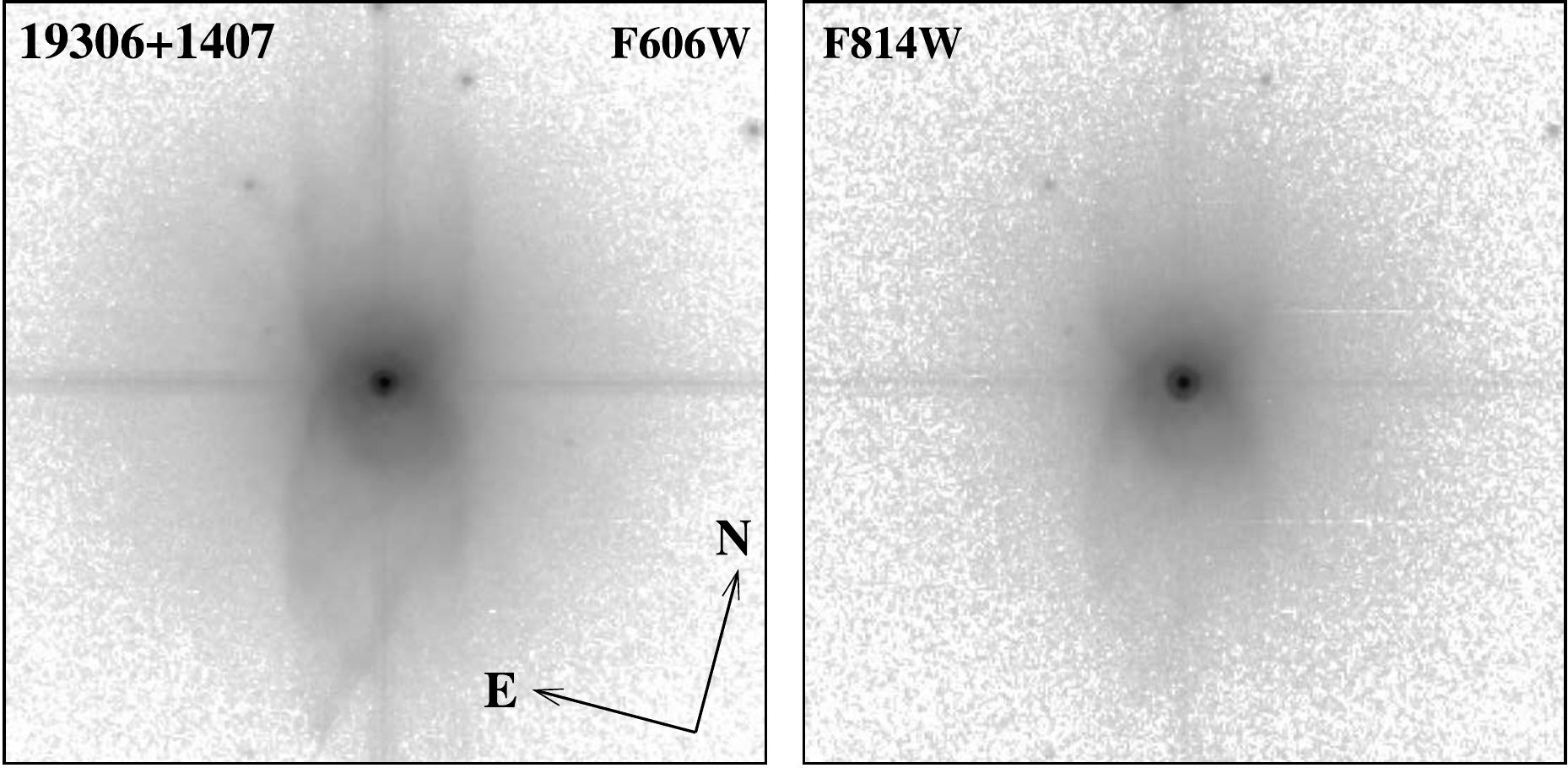}}
%\plotone{f1.eps}
\vskip 0.2in
\caption{HST images ({\it log stretch}) of young preplanetary 
nebula IRAS\,19306+1407 (6\farcs75$\times$6\farcs75).
}
\label{fig7b}
\end{figure}

\begin{figure}[htbp]
\vskip -0.6in
\resizebox{1.0\textwidth}{!}{\includegraphics{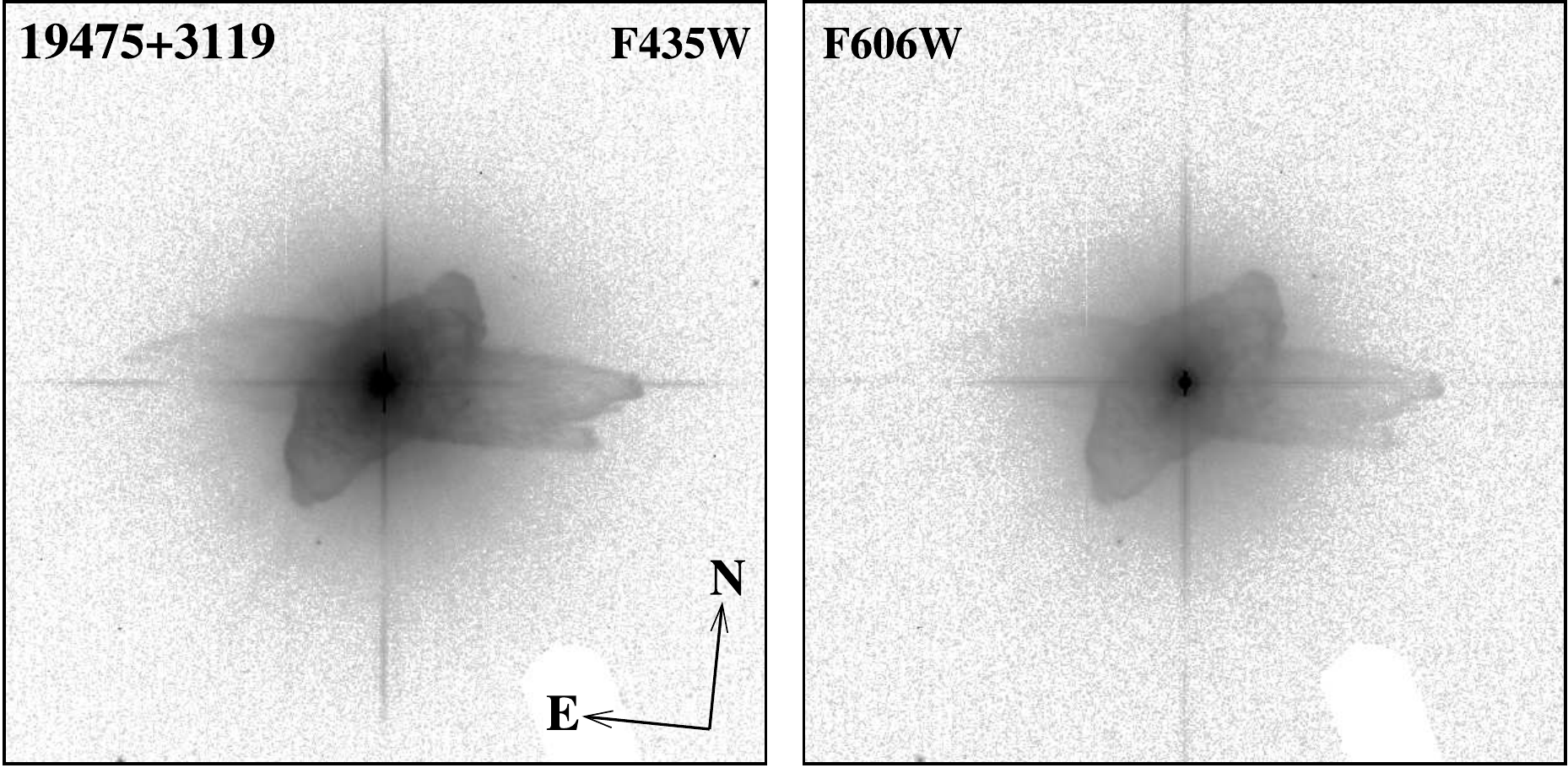}}
%\plotone{f1.eps}
\vskip 0.2in
\caption{HST images ({\it log stretch}) of young preplanetary 
nebula IRAS\,19475+3119 (15\farcs0$\times$15\farcs0)
}
\label{fig8a}
\end{figure}

\begin{figure}[htbp]
\vskip -0.6in
\resizebox{1.0\textwidth}{!}{\includegraphics{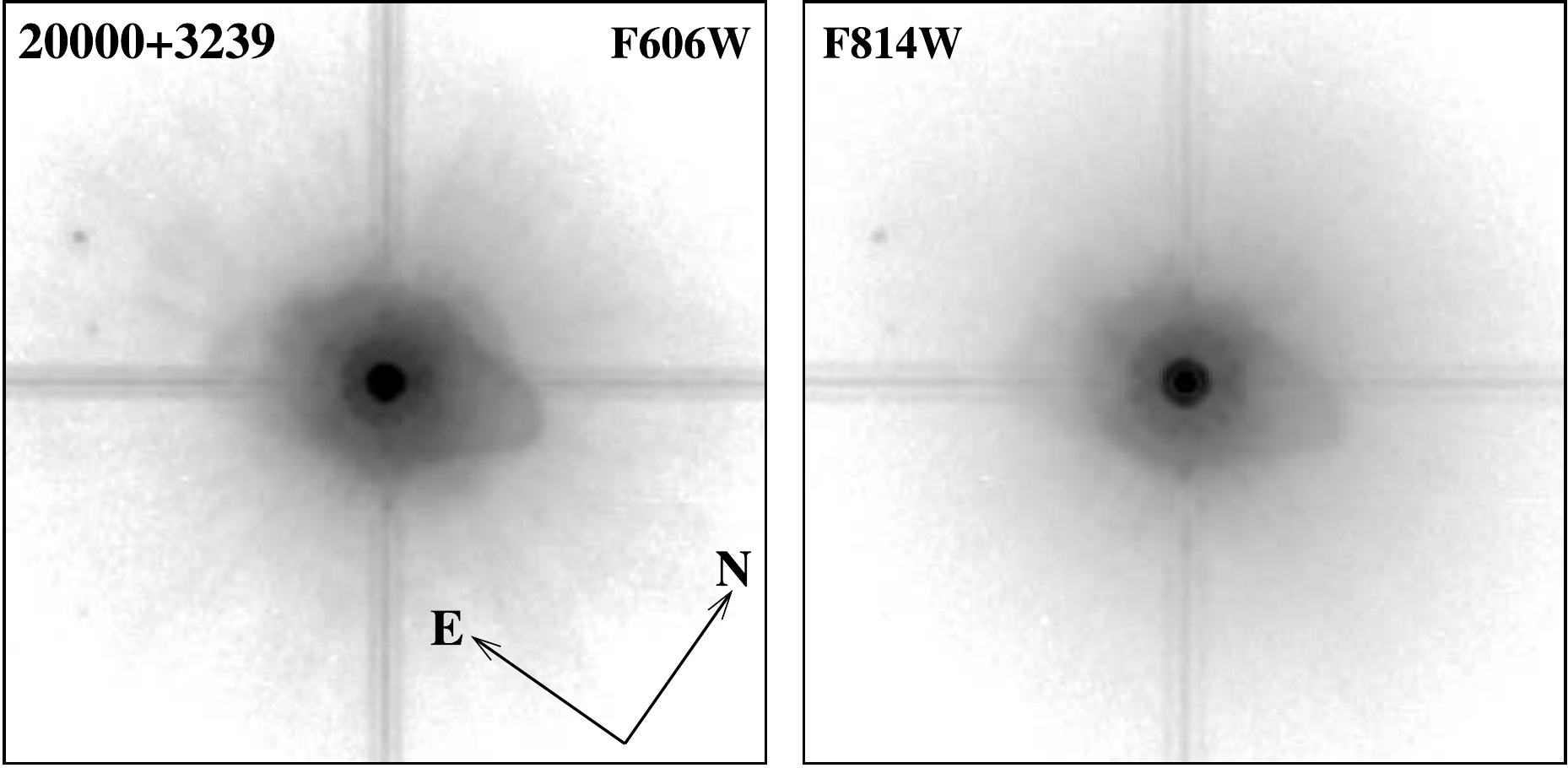}}
%\plotone{f1.eps}
\vskip 0.2in
\caption{HST images ({\it log stretch}) of the young preplanetary 
nebula IRAS\,20000+3239 (5\farcs0$\times$5\farcs0) 
}
\label{fig8b}
\end{figure}

\begin{figure}[htbp]
\vskip -0.6in
\resizebox{1.0\textwidth}{!}{\includegraphics{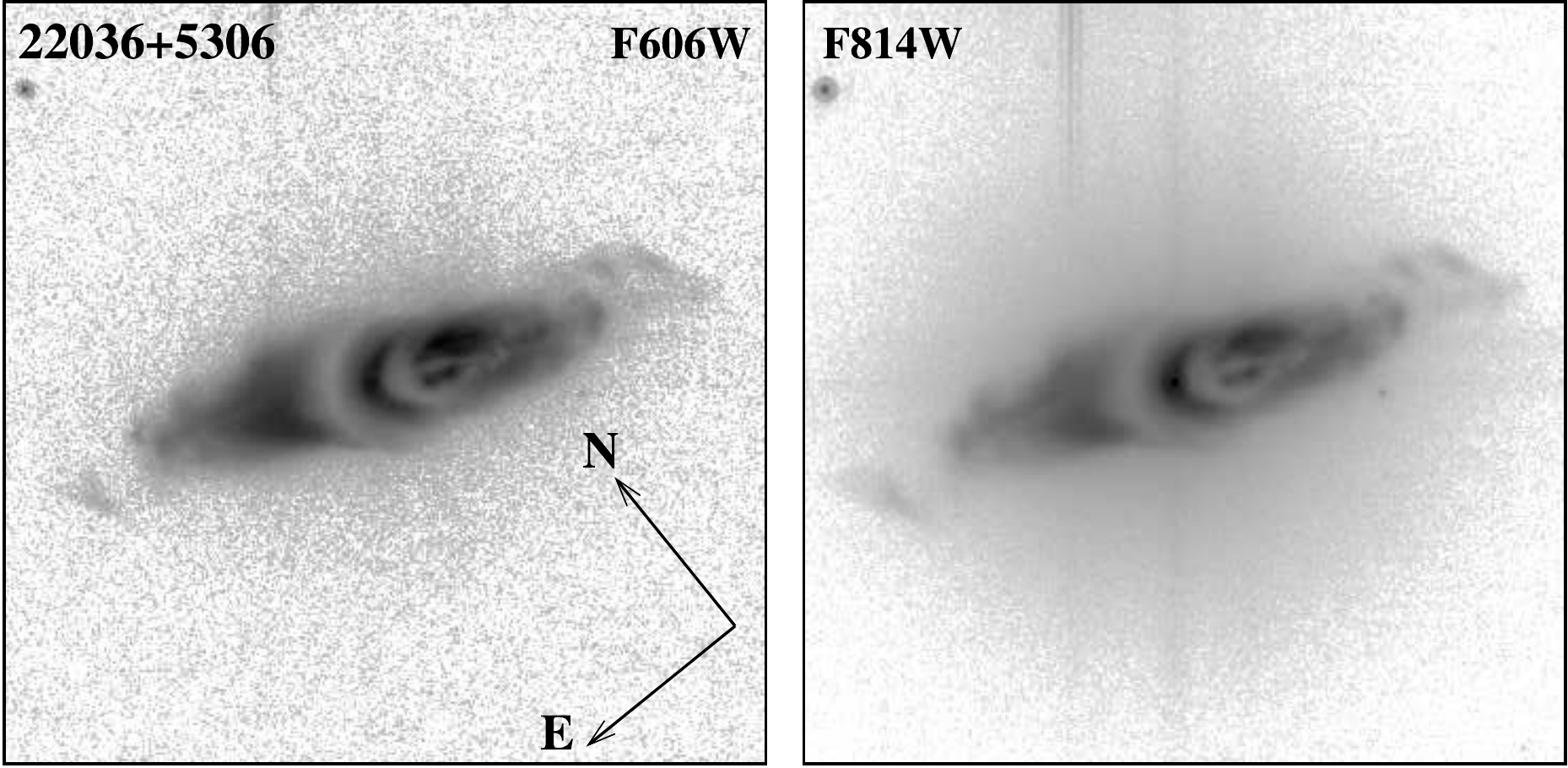}}
%\plotone{f1.eps}
\vskip 0.2in
\caption{HST images ({\it log stretch}) of young preplanetary 
nebula IRAS\,22036+5306 (9\farcs0$\times$9\farcs0). The vertical stripes are optical
artifacts.
}
\label{fig8c}
\end{figure}

\begin{figure}[htbp]
\vskip -0.6in
\resizebox{1.0\textwidth}{!}{\includegraphics{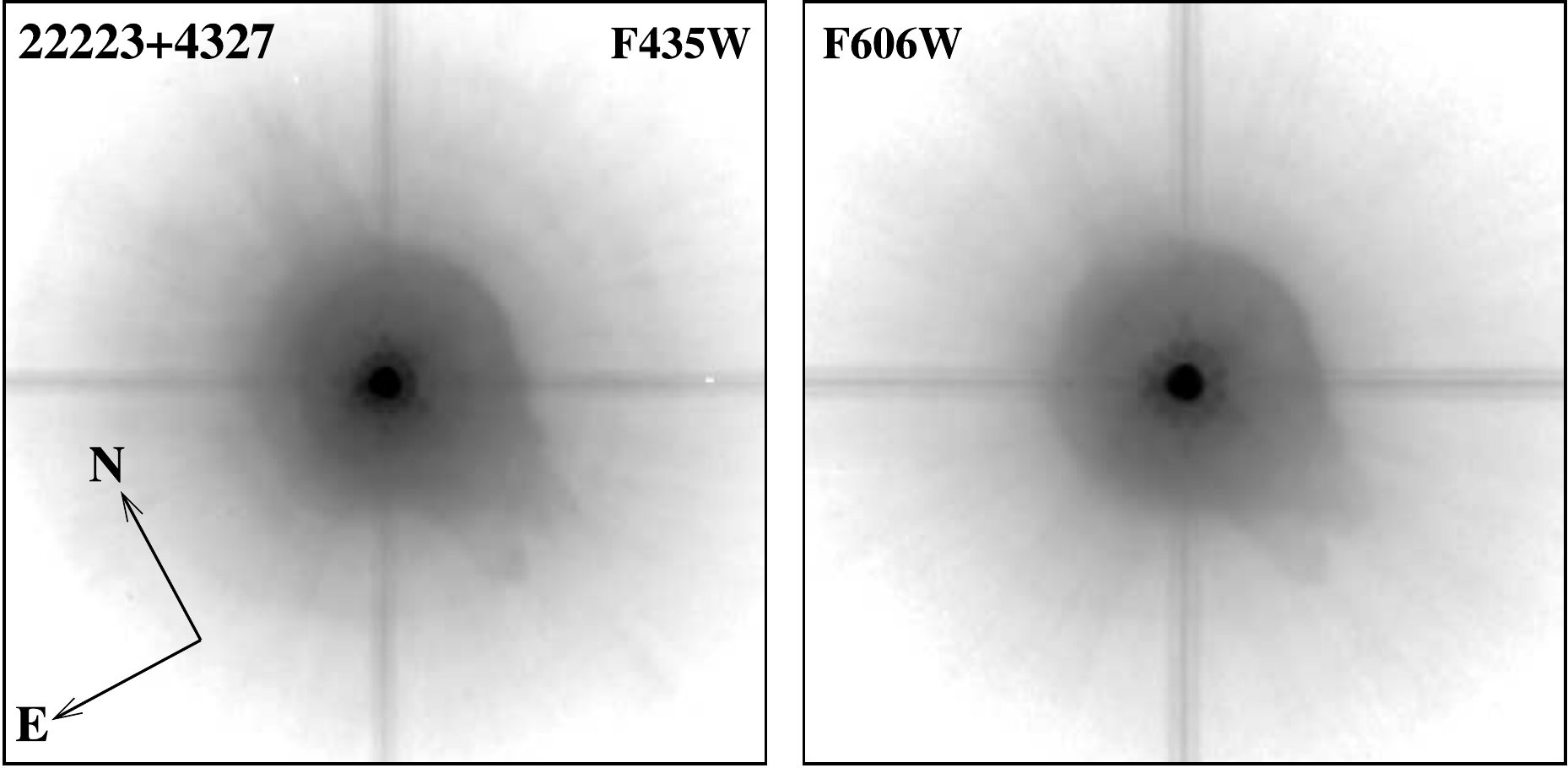}}
%\plotone{f1.eps}
\vskip 0.2in
\caption{HST images ({\it log stretch}) of young preplanetary 
nebula IRAS\,22223+4327 (5\farcs0$\times$5\farcs0)
}
\label{fig9a}
\end{figure}

\begin{figure}[htbp]
%\vskip -0.6in
\resizebox{1.0\textwidth}{!}{\includegraphics{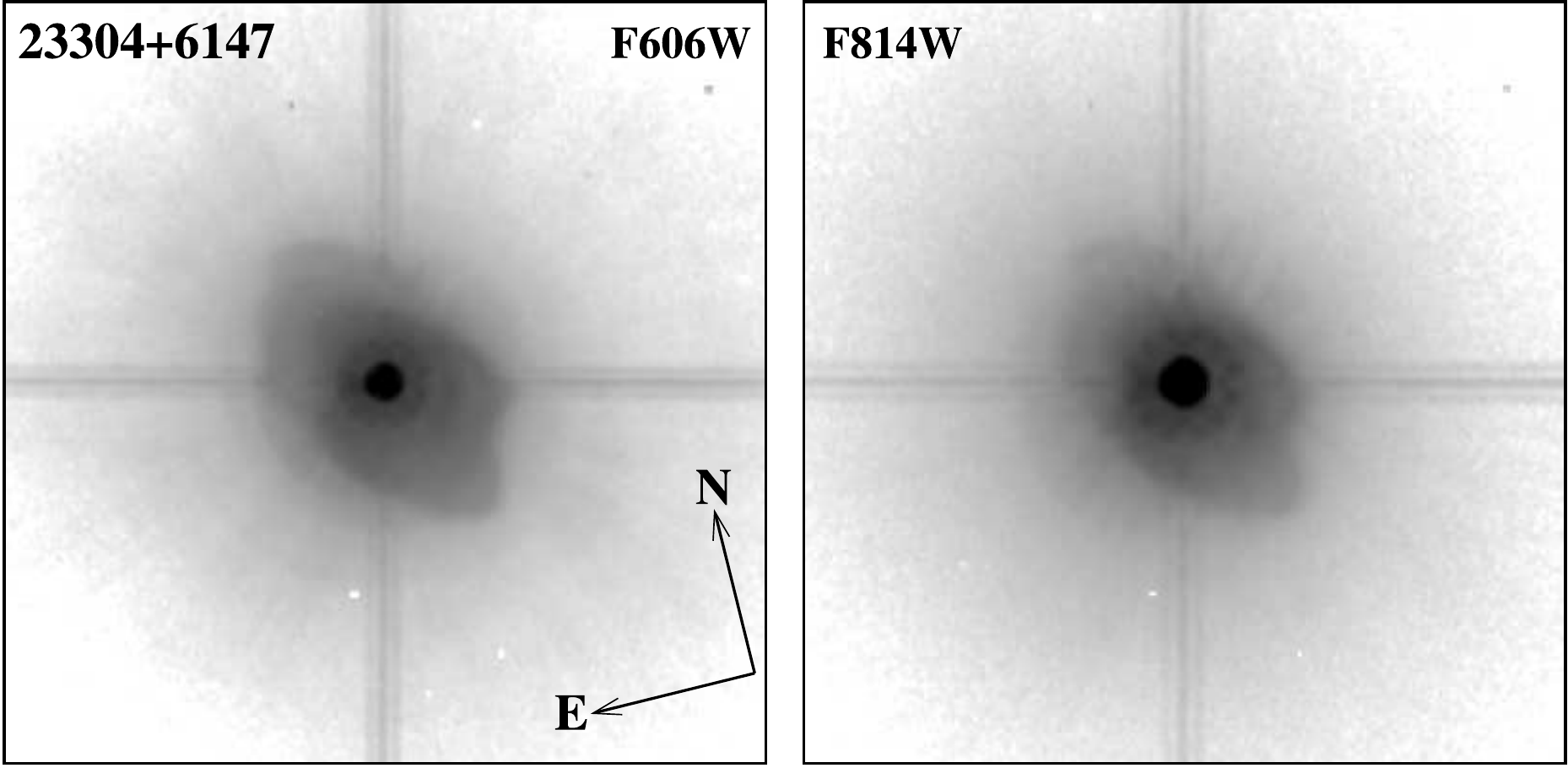}}
%\plotone{f1.eps}
\vskip 0.2in
\caption{HST images ({\it log stretch}) of young preplanetary 
nebula IRAS\,23304+6147 (5\farcs0$\times$5\farcs0)
}
\label{fig9b}
\end{figure}

\begin{figure}[htbp]
\vskip -0.6in
\resizebox{1.0\textwidth}{!}{\includegraphics{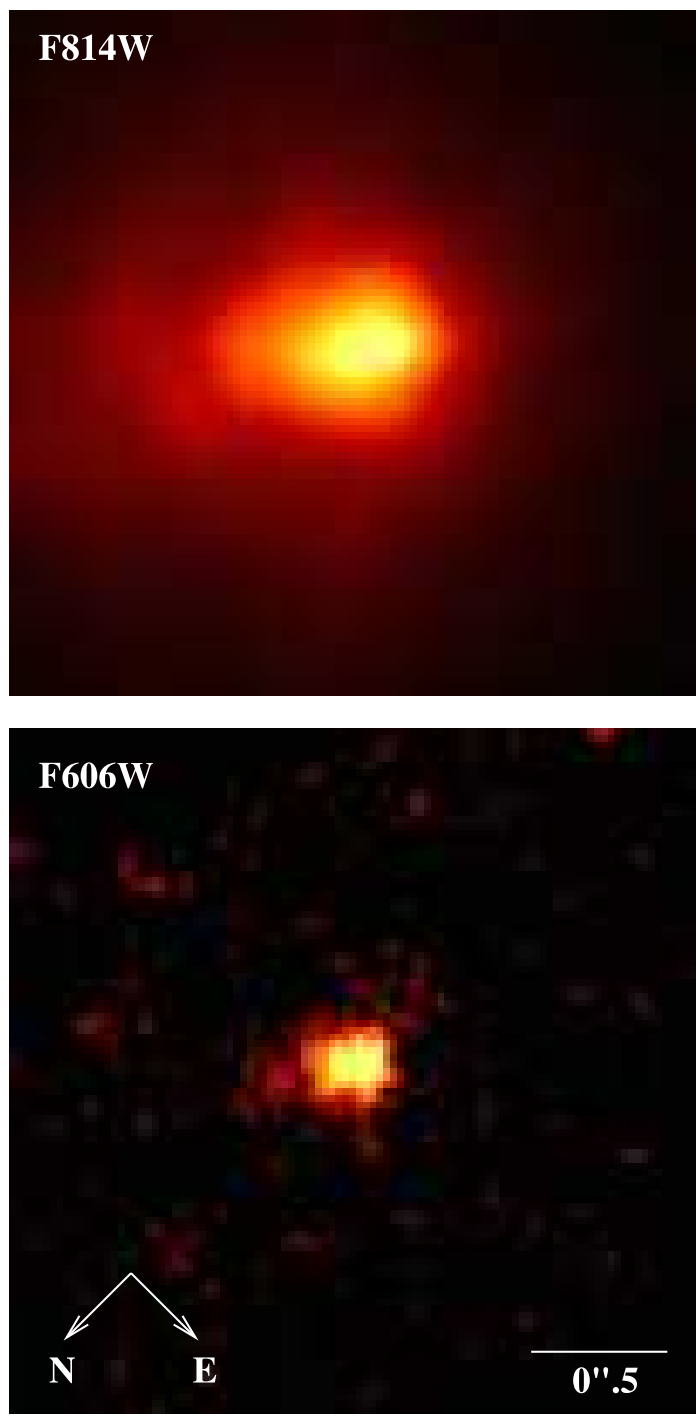}}
%\plotone{f1.eps}
\vskip 0.2in
\caption{False-color HST images of the nascent preplanetary 
nebula IRAS\,01037+1219 ({\it with sharp features enhanced})}
\label{figm1}
\end{figure}

\begin{figure}[htb]
\vskip -0.6in
\resizebox{0.6\textwidth}{!}{\includegraphics{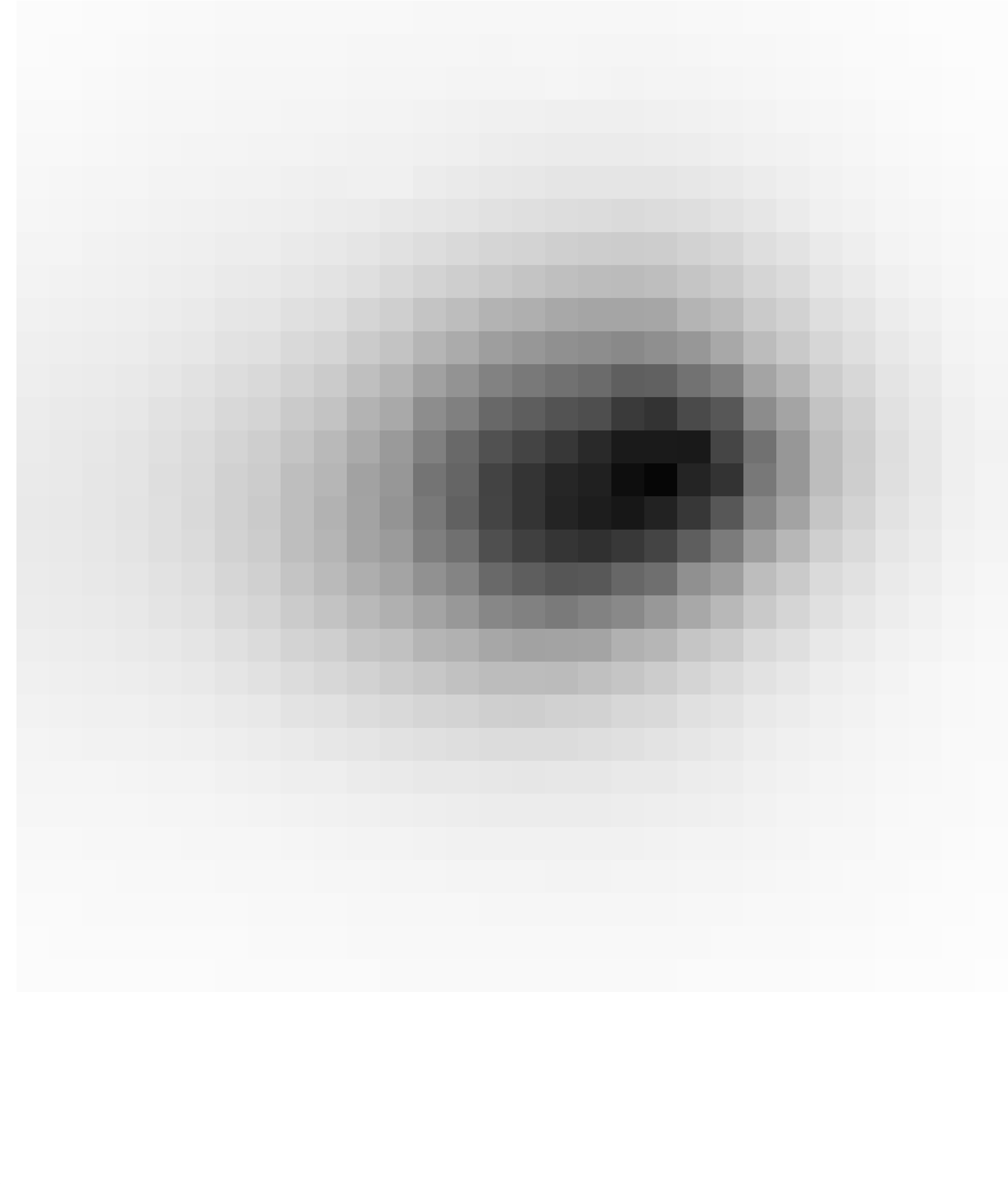}}
%\plotone{f1.eps}
\vskip 0.2in
\caption{Expanded view of the central region (0\farcs75$\times$0\farcs75) of the nascent
preplanetary nebula IRAS\,01037+1219 ({\it linear stretch})}
\label{figm1a}
\end{figure}

\begin{figure}[htb]
\resizebox{0.5\textwidth}{!}{\includegraphics{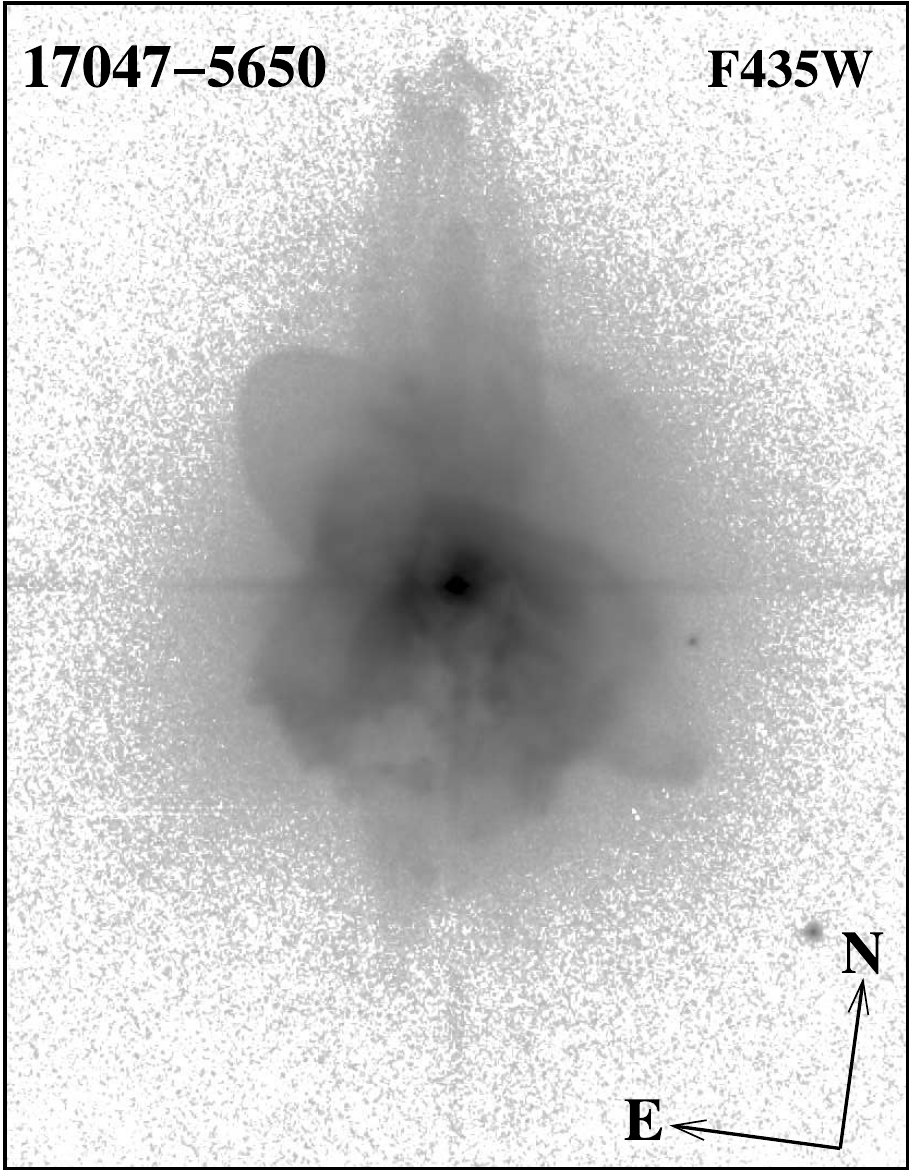}}
%\plotone{f1.eps}
\vskip 0.2in
\caption{HST image ({\it log stretch}) of the young planetary 
nebula IRAS\,17047-5650 ($10\farcs38\times13\farcs38$).}
\label{fig10}
\end{figure}

\begin{figure}[htb]
\resizebox{0.5\textwidth}{!}{\includegraphics{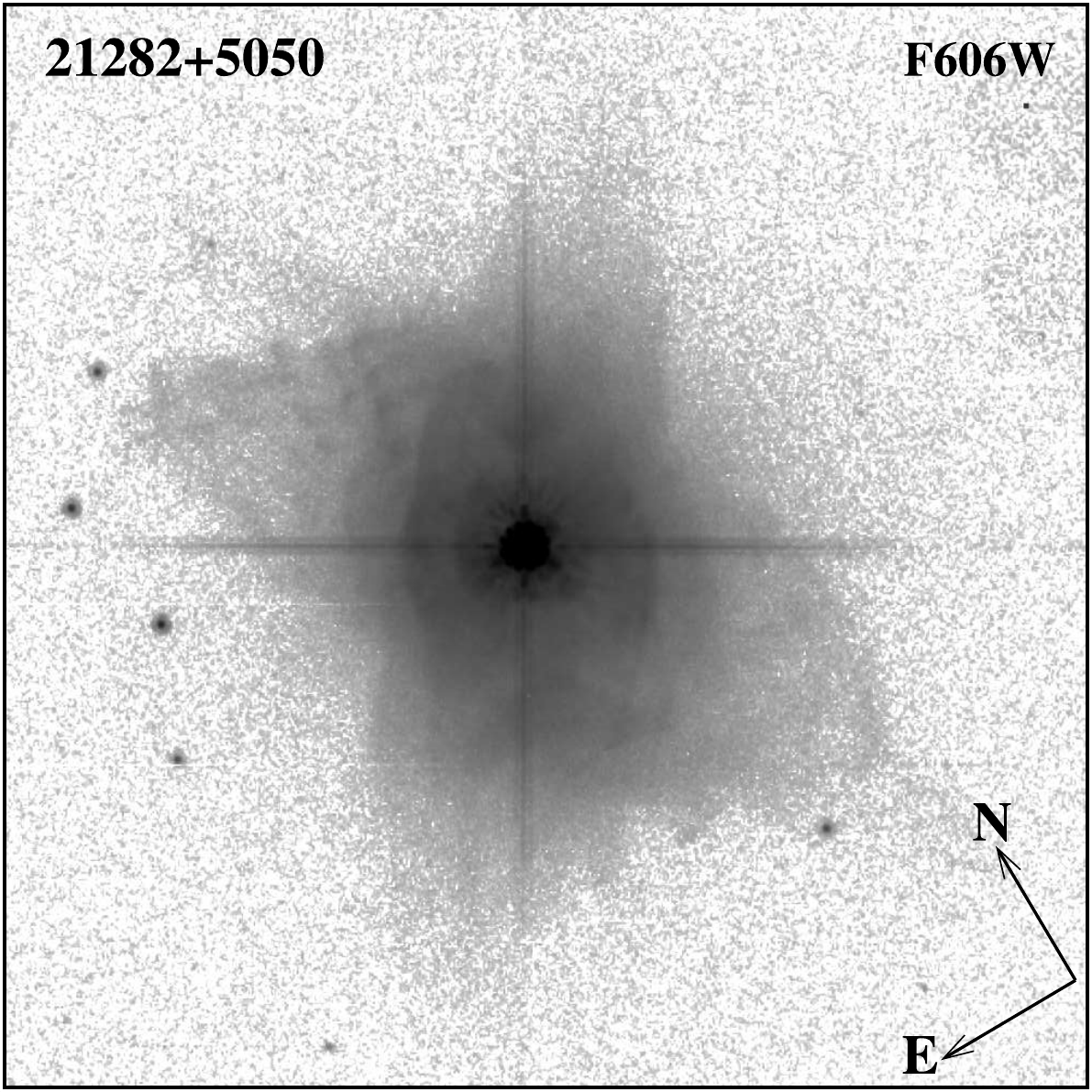}}
%\plotone{f1.eps}
\vskip 0.2in
\caption{HST image ({\it log stretch}) of the young planetary 
nebula IRAS\,21282+5050 (12\farcs5$\times$12\farcs5).}
\label{fig10b}
\end{figure}

\begin{figure}[htb]
\resizebox{0.5\textwidth}{!}{\includegraphics{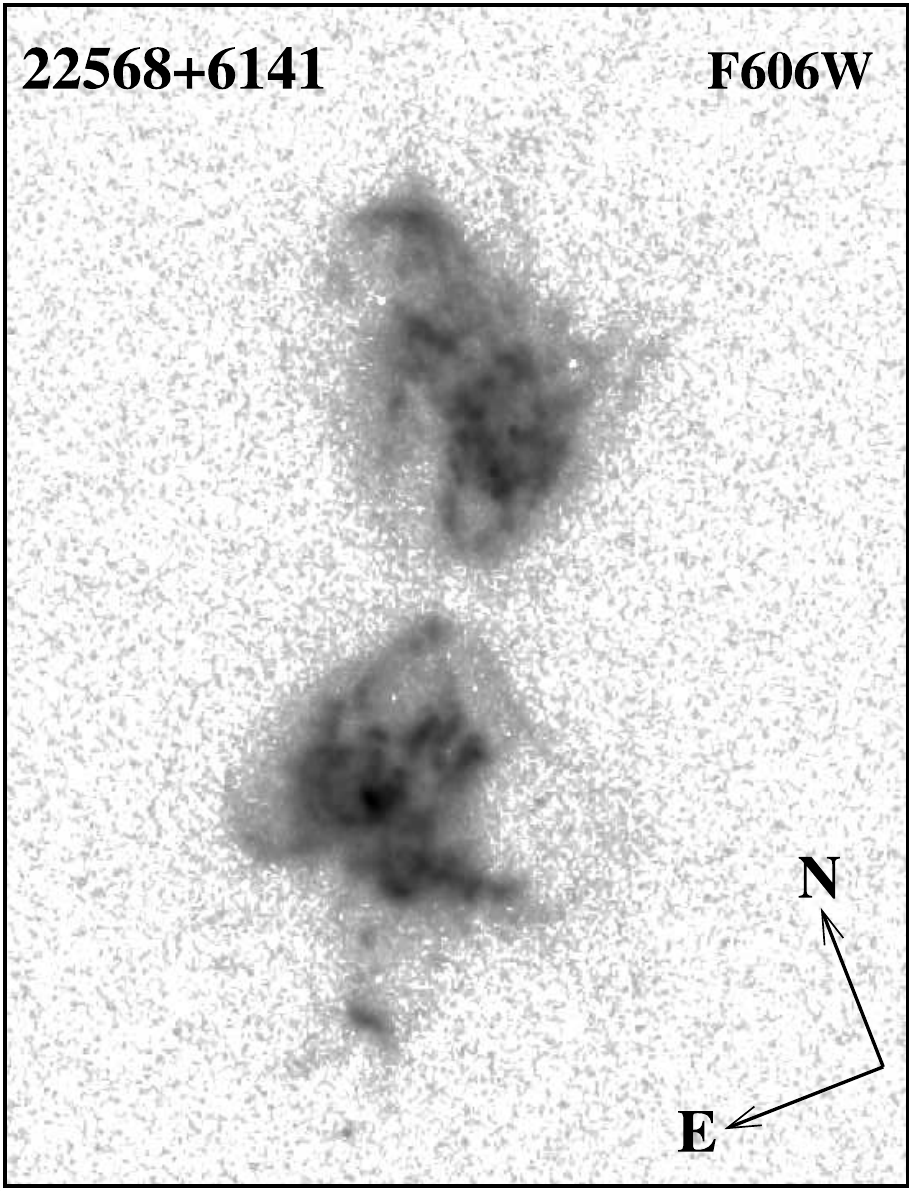}}
%\plotone{f1.eps}
\vskip 0.2in
\caption{HST image ({\it log stretch}) of the young planetary 
nebula IRAS\,22568+6141 ($6\farcs88\times9\farcs0$).}
\label{fig11}
\end{figure}

\noindent {\it IRAS\,17347-3139}
% 18.99 100.3 125.4 <249L Medm-Bip   22,77
% af/2x338 (606W), bd/2x338 (814W), 180x180, 0.5 WFC pix
%pg 22, OH maser, ISO, MSX+IRAS, 12/25=18.99/100.3
%ORIENTAT=92.23728945162559 / position angle of image y axis (deg. e of n)
%PHOTFLAM=7.7792272E-20 / inverse sensitivity, ergs/cm2/Ang/electron
% L=619(d/1kpc)^2=6000, so d=3.11 kpc, t=1971 yr, 1yr=3.156e7 sec
%  dust model L=619, integrated SED=595 Lsun
%  0.33E-03     0.19E-05     0.12E-08  80.5 180.0 511.0 Std.Dev.  0.3448E-01
%   Md=(6000/619)*0.33E-03=3.2e-3
% rPAGB=1.88, tPAGB=185.3
The F814W image shows a bipolar nebula with collimated lobes separated by a
dense waist; almost no nebulosity is visible in the F606W
image. The lobes show significant differences from each other in their
structure (Fig.\ref{fig4b}). We also obtained NICMOS images with NIC1 in the F110W (not
shown) and F160W
filters (Fig.\ref{fig4b}). The shape of the extended nebula appears quite similar in both
near-IR filters
compared to that in the F814W filter. The two lobes are rather asymmetric in shape and size,
with the NW lobe being significantly longer in extent. The latter also shows a very bright,
collimated feature in its innermost region (bottom left panel, Fig.\ref{fig4b}) located
mid-way between the lateral boundaries of the lobe. This collimated feature opens into a
Y-shape; the right hand fork of this Y-structure is significantly longer that the left-hand
one, and extends along the body of the lobe along a curved trajectory which, when
extrapolated, culminates at the bright, bow-shaped tip of this lobe. These collimated
structures are indicative of the presence of a highly-collimated jet whose axis has
precessed while it has sculpted out the NW lobe. A faint elliptical halo can be seen around
the lobes in the F160W image. Using azimuthal averaging, we can trace this halo out to a
radius of $\sim$2$''$. Although OH maser emission is detected, the line profile is weak and
complex (Zijlstra et al. 1989), and does not allow us to determine an AGB expansion
velocity. Therefore, assuming a typical AGB outflow velocity of
V$_{exp}$=15\,\kms, we find $t_{AGB}$=1950($D_L$/3.1 kpc) yr. 

The detection of radio continuum from this object (Gregorio-Monsalvo et
al. 2004) suggests that IRAS\,17347-3139 may be approaching the young planetary nebula
phase, or may already have become a very young planetary nebula (note that two other well
known pre-planetary nebulae, M\,1-92 and AFGL\,618, also show radio continuum emission). 
Gregorio-Monsalvo et al. also find H$_2$O maser emission, with a double-peaked profile
having a total width of about 9\,\kms, which they associate with a central toroid.

\noindent {\it IRAS\,17440-3310}
% 2q_aq_606/2x400, 5q_9q_814/2x50, 400x400, 0.5 HRC pix
% 4.05L 20.7 30.14 156.5L Medm-Bip   pg 70
% rPAGB=3.3/2=1.65 arccec 
% L=106(d/1kpc)^2=6000, so d=7.56 kpc, t=9844.3 yr, 1yr=3.156e7 sec
%  0.97E-04  0.38E-07  0.75E-12 79.2 138.7 1406.2 Std.Dev.  0.3178E-01
%   Md=(6000/106)*0.97E-04=5.49e-3
% Sevenster survey (ATCA), OH356.562-02.527; 17 47 22.733 -33 11 08.98, so O-rich:
% OH Vels -106.4  -77.2  -91.8, Ve=14.6
% rPAGB=1.4, tPAGB=335.4
% ESI survey, F5-9
This object has point-symmetrically shaped lobes, and is surrounded by an extended round
halo (Fig.\ref{fig5a}). In the F606W image, a few circular arcs can be seen in
the halo at low contrast. Using azimuthal averaging, we can trace the halo out 
to a radius of $\sim$4$''$; the radial distribution follows a power-law, r$^{-\beta}$ with
$\beta\sim$3. Using an AGB envelope expansion velocity of 14.6\kms~derived from OH maser
data,
we find that the AGB envelope was characterised by a 
roughly constant mass-loss rate over a period of $t_{AGB}$=9850($D_L$/7.6 kpc) yr.

\noindent {\it IRAS\,17543-3102}
% L=114(d/1kpc)^2=6000, so d=7.25 kpc, t=4594 yr, 1yr=3.156e7 sec
% rPAGB=1".35=1.468e17, tPAGB=310yr 
%  0.68E-04     0.10E-06     0.43E-10  83.0 190.0 650.0 Std.Dev.  0.7259E-01
%   Md=(6000/114)*0.68E-04=3.58e-3
% Sevenster survey (ATCA),  OH359.499 -03.302 so O-rich:
%  OH Vel 121.4 km/s (single-peaked)
This object shows a complex shape, with a central bright nebulous region, and two ansae
located along a SE-NW axis. The bright region is located closer to the SE ansa. Each of the
ansae are aligned tangentially to the long axis of the nebula, defined by the vector
joining the mid-points of the ansae (Fig.\ref{fig5ab}). A linear stretch of the F606W
image, with the central region saturated to bring out faint structures, shows that the ansae
define the ends of elongated, lobe-like structures. Two additional bright knotty structures 
can be seen, one located NE of the SE ansa, and another located NW of the central region
but closer to the center than the NW ansa. Using azimuthal averaging, a faint halo can be
traced out to a radius of $\sim$2$''$. Because of the presence of a large number of stars
in the vicinity of this object, including two rather bright ones within 1\farcs3 and
1\farcs8 of the center of the nebula, we cannot check whether or not the halo is round. The
radial distribution follows a power-law, r$^{-\beta}$ with $\beta\sim$2. The AGB expansion
velocity is unknown, since the OH maser emission line from this object
shows a single peak. Therefore, assuming a typical AGB outflow velocity of
V$_{exp}$=15\,\kms, we find $t_{AGB}$=4600($D_L$/7.3 kpc) yr.

\noindent {\it IRAS\,18276-1431}
% 22.65 132.0 120.0 38.6  Medm-Bip R1225=0.17  pg 3
% 2m8m_606/2x100sec, 4m6m_814/2x60 130x130 0.5 PC pix
% Sevenster survey  OH017.684-02.032 010694 18 30 30.676 -14 28 57.78
%  OH Vels +49.9  +72.6  +61.2; Ve=11.4
%  OH/H2O masers, ISO, MSX+IRAS 
%  give details of AGB envelope age
This object (with a K2/K3 spectral type central star: SIMBAD database) has a bipolar shape
with a central dense, roughly edge-on waist
(Fig.\ref{fig5b}) and an extended halo. The two lobes shows different shapes,
and a pair of ``searchlight beams" emanate from the fainter one. If the halo is due to a
spherical AGB mass-loss envelope, the elliptical shape indicates that the AGB envelope is
being indirectly illuminated by the lobes, and not by a central source, or that the light
from the central source cannot get out to the larger distances at low latitudes. A detailed
study of this object, including near-infrared imaging using Adaptive Optics from the ground,
has been presented by S\'anchez Contreras et al. (2007), and we have used their results in
Table\,\ref{t3}, scaled appropriately for the small difference in distance values (3\,kpc in
this paper, 2.7\,kpc in theirs). IRAS\,18276 shows OH maser emission and is therefore
O-rich.

\noindent {\it IRAS\,18420-0512}
% iq_qq_606.ps/2x400sec lq_pq_814.ps/2x50sec 400x400 0.5 HRC pix
% 1.03  26.72 26.22 237.1L Medm-Bip R1225=0.039 pg 65
% Sevenster survey OH027.577-00.853; 18 44 41.660 -05 09 17.0; so O-rich
%  OH Vels +93.1 +118.0 +105.6; Ve= 12.4
% L=139(d/1kpc)^2=6000, so d=6.6 kpc, t-agb=7600 yr, Md=2.6e-3
% 0.61E-04, 0.16E-06, 0.31E-10  88.0 144.8 850.0 Std.Dev.0.1399E-01
%  Md=(6000/139)*0.61E-04=2.63e-3
% rPAGB=0.8, tPAGB=167.3
This object shows a central bipolar shape embedded inside an extended halo; the
central star is visible (Fig.\ref{fig6a}). One, possibly two, secondary lobes 
with collimated shapes emanate from the central region. Faint, roughly circular, arcs can
be seen 
in the halo. The red and blue maser peaks in the OH line profile are
located at 116.0 and 93.1\kms, giving an expansion velocity of 12.4\kms. 
%The source has a 6
%cm VLA continuum flux of 82$\mu$Jy (Herman, Baud ...). 
Using azimuthal averaging, the halo
can be traced out to a radius of $\sim$3$''$; the radial distribution follows a power-law, 
r$^{-\beta}$ with $\beta\sim$3, signifying that the AGB envelope was characterised by a 
roughly constant mass-loss rate over a period of $t_{AGB}$=7600($D_L$/6.6 kpc) yr.

\noindent {\it IRAS\,19024+0044}
% 2.86 48.83  42.53  15.7 Large-Bip  pgs 67,86
Since a detailed multiwavelength study of this object, which has a multipolar morphology
(Fig.\ref{fig6ab}) and an early-G type central star,
has been reported by Sahai et al. (2005), we briefly summarise the main characteristics
here. The object shows at least six elongated lobes, two with limb-brightened
point-symmetric ansae at their tips, and a faint round halo characterised with
$\beta\sim$3.2, surrounds the lobes. Since Sahai et al. (2005) assumed a radial-velocity
distance of 3.5\,kpc resulting in a luminosity of 2600\ls, we have scaled up
the distance-dependent values given in Table\,\ref{t1} appropriately, from the values given
by
Sahai et al. (2005). IRAS\,19024 shows OH maser emission and is therefore O-rich.

\noindent {\it IRAS\,19134+2131}
% an_aw_606/2x338s ao_au_814/2x338s 80x80 0.5 WFC pix
%  5.06 15.56 8.56 3.95L  Smal-Bip  R1225=0.33 pg 32
% L=85.8(d/1kpc)^2=6000, so d=8.36 kpc, Md=1.54e-3
%  0.22E-04, 0.99E-06, 0.18E-09  83.3 159.7 473.4 Std.Dev. 0.4121E-01
% latest from SED, L=99.3, Fbol=3.2E-09 so d=7.8 kpc
%  0.38E-04   0.12E-05  0.18E-09 69.4 159.7 473.4 Std.Dev. 0.3474E-01
%   Md=(6000/85.8)0.38e-4=2.66e-3
% rPAGB=0.12, tPAGB=31.8
The object belongs to the class of ``Water-Fountain" nebulae -- preplanetary
nebulae with very high-speed jets seen in H$_2$O maser emission. I\,19134
appears quite compact in the optical images, but is clearly elongated along PA=
$\sim94\arcdeg$ (Fig.\ref{fig6b}), which is consistent with the orientation of the bipolar
H$_2$O jet (Imai et al. 2004, 2007). In spite of the limited resolution, there appear to be
two lobes separated by an intensity minimum, hence we tentatively classify this as a
bipolar object with closed lobes, and a waist. No significant halo was detected for this
source. 
%{\it add in NICMOS data?}

\noindent {\it IRAS\,19292+1806}
% also OH53.6-0.2, Ve=29.23/2.0, Vlsr=10.47 (Herman & Habing 1985), tAGB=4232
% detected at OVRO
% 5e_606/2x338sec 6c_814/2x338sec 140x140 0.5 WFC pix
% 5.01L 47.99 28.83: 318.7L Medm-Bip R1225=<0.10 pg 29
% L=220(d/1kpc)^2=6000, so d=5.2 kpc, t-agb=xx yr, Md=1.28e-3
%  0.47E-04, 0.98E-07, 0.19E-09 103.2 178.2 511.8 Std.Dev. 0.4177E-01
%  latest  from dust fit L=207;  from SED, L=193.9, Fbol=6.2e-9 (use IRAS 60 mu flux)
%  0.40E-04  0.17E-06  0.19E-09 103.2 178.2 511.8 Std.Dev. 0.3657E-01
%    Md=(6000/220)0.40E-04=1.09e-3
% rPAGB=0.9, tPAGB=148.3
This object shows a bipolar shape (Fig.\ref{fig7a}). A faint elliptical halo can be seen in
the F814W image. If the halo is due to a spherical AGB mass-loss envelope, the elliptical
shape indicates that the AGB envelope is being indirectly illuminated by the lobes, or that
light from the
central source cannot get out to the larger distances at low latitudes. By azimuthal
averaging, this halo can be traced out to a radius of
2\farcs5. Using an AGB envelope expansion velocity of 10.5\kms~derived from OH maser data,
we
find that $t_{AGB}$=4250($D_L$/5.2\,kpc)\,yr.

\noindent {\it IRAS\,19306+1407}
%11_61_606/2x300sec, 31_41_814/2x50sec 540 x 540 0.5 HRC pix
% 3.58 58.65 31.83 10.03: Larg-Bip R1225=0.061   pg 54
% B0Ie (ESI)
% OVRO Ve=14 km/s (check Vlsr=91 km/s)
% 2MASS: 20 01 59.52,+32 47 32.9: 8.021,0.020;7.103,0.017;6.580,0.023
% L=249(d/1kpc)^2=6000, so d=4.91 kpc, t-agb=6667.6 yr, Md=1.33e-3
% 0.55E-04     0.82E-07     0.58E-09  99.2 156.2 544.0 Std.Dev.  0.6575E-01
% original orient=104.74 (y axis PA)
% rPAGB=3.2, tPAGB=
This object shows a bipolar shape, with roughly cylindrical lobes (Fig.\ref{fig7b}). The
central star (spectral type B0:e, SIMBAD database) is directly visible, and the lobes are
surrounded by a tenuous round halo. By
azimuthally averaging, the halo can be traced out to a radius of $\sim$4$''$; the radial
distribution follows a power-law, r$^{-\beta}$ with $\beta\sim$3.3. No OH maser data are
available for this source. This source most likely has a mixed chemistry (i.e. shows both
carbon- and oxygen-rich dust features): Hrivnak et al. (2000) find PAH features in its ISO
spectra, and Hodge et al. (2003) who silicate features at 11, 19, and 23\micron. We have
determined an
AGB expansion velocity of 14\kms~from our OVRO CO J=1--0 survey (S{\'a}nchez Contreras \&
Sahai 2003) for this object, signifying that the AGB
envelope was characterised by a roughly constant mass-loss rate over a period
$t_{AGB}$=6650($D_L$/4.9\,kpc)\,yr. Lowe \& Gledhill (2006) find a dust mass of
$8.9\pm5\times 10^{-4}$\ms~using a distance of 2.7\,kpc from detailed modelling. Their
value, when scaled by the square of the ratio of our larger distance to theirs, i.e.,
$(4.9/2.7)^2$, gives a dust mass of $2.9\pm1.6\times 10^{-3}$\ms, which is larger than, but
consistent within uncertainties, with our value of $1.3\times
10^{-3}$\ms~(Table\,\ref{t3}).

\noindent {\it IRAS\,19475+3119}
% 11.ps/435W/75sec 41.ps/606W/7.4sec 600x600 HRC pix
% 0.54  37.99 55.83 14.76 Larg-Bip R1225=0.014  40
% L=324.3(d/1kpc), so L=7790 at d=4.9 kpc; but Sahai et al 07  give L=8300 Lsun
%  hence almost 513 Lsun extra from DUSTY model
%  from dust model L=214.9
% 0.16E-03     0.38E-05     0.14E-10  77.3 105.6 1159.8 Std.Dev.  0.5825E-01
A detailed multiwavelength study of this object, including HST imaging,   
has been reported by Sahai et al. (2007); and an interferometric study of millimeter-wave CO
line emission has been carried out by S\'anchez Contreras et al. (2006). Hence we briefly
summarise the main characteristics
here. IRAS\,19475 (F3Ib central star: SIMBAD database) has a quadrupolar shape
(Fig.\ref{fig8a}) showing two bipolar elongated lobes
emanating from the center of the nebula. One of the bipolar lobe pairs clearly
shows detailed point-symmetric structure with respect to the central star.
A faint, surface-brightness-limited, diffuse halo surrounds the lobes.
Since Sahai et al. (2007) assumed a far kinematic distance of 4.9\,kpc, at which this object
has a luminosity of 8300\ls, we have scaled the values given by Sahai et al. (2007) for the
distance-dependent parameters given in Table\,\ref{t1} appropriately. For the dust-mass, we
use the
study of Sarkar \& Sahai (2006), and exclude the additional mass of large, cold grains
derived from submillimeter-wave/millimeter-wave data for this object by Sahai et al.
(2007). The ISO spectra of IRAS\,19475 show that it is O-rich (Sarkar \& Sahai 2006).
% dust mass values -- how to scale?
%  From interferometric mapping of millimeter-wave molecular-line emission from this object,
% S\'anchez Contreras et al. (2006) derive values of $t_{PAGB}$=1660\,yr and
% $t_{AGB}$=4400-5250\,yr, scaled to our distance of 4.2\,kpc. Compared to our values for
% these parameters, their $t_{PAGB}$ agrees, but their $t_{AGB}$ estimate
% is significantly smaller because the AGB mass-loss envelope cannot be detected to as large
% a radius in molecular-line emission as in the HST images.
% D=4.8kpc,jet speed=53 km/s at max jet length (=3.2e17=4.4 asec)

\noindent {\it IRAS\,20000+3239}
% G069.6793+01.1610, 
%  (20 01 59.5, +32 47 33) 6.153e+00,4; 1.653e+01,4; 2.138e+01, 4;4.424e+01,4
% 2MASS 20 01 59.52	+32 47 32.9, 8.021,0.020;7.103,0.017;6.580,0.023
% 15.03 70.97 29.99 43.1L Medm-Ell R1225=0.21  pg 50
% 11_81_f606.ps/2x150sec 41_71_814.ps/2x35sec 200x200 HRC pix
% Ve=12.3 km/s
% L=477(d/1kpc)^2=6000, so d=3.54 kpc, t-agb=6851 yr, Md=8.79e-4
% Dust model gives L=434
% 0.52E-04     0.80E-07     0.24E-09 103.2 270.0 850.0 Std.Dev.  0.5693E-01
% rPAGB=1, tPAGB=112.1 yr
This object is roughly elliptical in shape, and is surrounded by a prominent round halo
(Fig.\ref{fig8b}). The radial brightness distribution in the halo can be roughly described
by a segmented power-law, with
an inner region extending to a radius of about $\sim$2$''$ described with a power-law index
$\beta\sim$2.3, and an outer region which can be traced to a radius of about 5$''$ and a
power-law index $\beta\sim$4.2. The age of the AGB envelope is 6850($D_L$/3.5\,kpc) yr.
The central star is classified as G8Ia, and the ISO/SWS spectra show it to be a C-rich
object (Hrivnak et al. 2000).

\noindent {\it IRAS\,22036+5306}
% 8.43  46.28 107.2 50.7  Larg-Bip   1   E,W
Since a detailed multiwavelength study of this O-rich PPN, which has an overall bipolar 
morphology (and an F5 [or earlier] central star), has been reported by Sahai et al. (2003),
we briefly summarise the main
characteristics here. The ACS images of this extended bipolar nebula (Fig.\ref{fig8c}) show
the main structures imaged with WFPC2 previously (Sahai et al. 2003) -- namely, the bipolar
lobes with bright regions S1 and S2, the central ring-like dense waist ($R_b$ and $R_d$),
and the knotty linear structures labelled $J_E$ and $J_W$ in Sahai et al. (2003). In
addition, the F814W image resolves the central star from surrounding bright nebulosity,
and shows the presence of two ansae just beyond the tips of the main lobes as well as a
faint round halo. The radial brightness 
distribution follows a segmented power-law, with
an inner region described with a power-law index
$\beta\sim$2.3 and extending to a radius of about $\sim$3$''$, and an outer region described
by a power-law index $\beta\sim$4.4, which can be traced to a radius of about 6\farcs5.
Sahai et al. (2003) argue for a distance of 2\,kpc for this object, at which its luminosity
is 2300\ls, and the total (dust+gas) mass of the dominant shell component used to model the
SED, about
4.7\ms. Given this large mass, and a further large (but uncertain) mass contribution from
an additional cold component estimated recently from IRAS\,22036+5306's submillimeter and
millimeter-wave continuum fluxes (Sahai et al. 2006a), it is unlikely that the object can be
at the larger distance which corresponds to a luminosity of 6000\ls. Hence for this object,
we
have retained a value of 2\,kpc for its distance. We derive $t_{AGB}$=8850($D_L$/2\,kpc)\,yr
assuming an AGB expansion velocity of 7\,\kms~(derived from the central narrow component of
the
$^{13}$CO line shown by Sahai et al. 2006).

\noindent {\it IRAS\,22223+4327}
% 21_435/50sec 51_606/4sec 320x320 HRC pix
% 2.12  37.12 22.4  9.54 Medm-Bip  R1225=0.057 pg 35
% L=319(d/1kpc)=6000, so d=4.3 (1yr=3.156e7 sec)
%  dust model L=211.8
% 0.42E-04     0.22E-06     0.10E-09  99.2 108.5 850.0 Std.Dev.  0.5665E-01
%  Md=7.9e-4
% CO Vexp=14 km/s, t=10985 yr, tPAGB=204.4
This object has a central round core, with two small lobes protruding on the southern side,
and a
single small lobe protruding on the northern side (Fig.\,\ref{fig9a}). A prominent, smooth
halo
surrounds the central aspherical nebula. Faint, thin, partial shells can be seen within the
halo in the F435W image, and to a lesser extent, in the F606W image. These structures
cannot be fitted with concentric circles around the central star. Radial streaks due to the
PSF of the central star are also present. The central star is classified as G0Ia in the
SIMBAD database. Using
azimuthally averaging, one can trace the halo out to a radius of $\sim$7\farcs5; the radial
distribution follows a power-law, r$^{-\beta}$ with $\beta\sim$3.8, signifying that the AGB
envelope is characterised by an increasing mass-loss rate over a period of 11000($D_L$/4.3
kpc)
yr. The circumstellar envelope is classified as C-rich since it shows millimeter-wave HCN
emission (Omont et al. 1993).  
The CO profile
shows weak line wings extending over 50\kms, indicating the presence of a fast post-AGB
outflow.

\noindent {\it IRAS\,23304+6147}
% 11_usat/606W/150sec 200x200 HRC pix
% 11.36 59.07 26.6 30.89L Medm-Bip R1225=0.19 pg 37
% CO 2-1 JCMT Vexp=15.5 km/s, Vr=-15.9 km/s 
% \bibitem[]{} Woodsworth, A.~W., Kwok, S., \& Chan, S.~J.\ 1990, \aap, 228, 503
% L=346(d/1kpc)=6000, so d=4.16, tAGB=7015.8
% dust model L=294
% 0.47E-04     0.22E-06     0.84E-10  99.2 225.0 850.0 Std.Dev.  0.5237E-01
%  Md=8.15e-4
The object probably has a quadrupolar shape, although the two pairs of lobes are not as well
separated, as, e.g, in IRAS\,19475+3119. We tentatively classify it as M. The lobes are 
surrounded by a prominent round halo
(Fig.\ref{fig9b}). The radial distribution in the halo can be roughly described as a
segmented
power-law, with an inner region extending to a radius of about $\sim$2\farcs5 described
with a power-law index $\beta\sim$2.5, and an outer region which can be traced to a radius
of $\sim$5\farcs5 and a power-law index $\beta\sim$3.8. The age of the AGB envelope is
7000($D_L$/4.2\,kpc) yr. The central star is visible, and is classified as G2Ia in the
SIMBAD
database. It has molecular carbon absorption features and is therefore C-rich (Hrivnak
1995). This object is thus remarkably similar to IRAS\,20000+3239 in the properties of its
post-AGB and AGB ejecta, except for its primary morphological classification, which is M,
whereas IRAS\,20000+3239 is E.

%\subsection{Unresolved Objects and Non-Detections}
% do not discuss in this paper

% inclined bipolar should show evidence for 2 lobes (i.e. on the other), even
% if waist not sufficiently edge-on to be seen
% elliptical: non-spherical and does not have 2 bifurcated lobes

\subsection{Related Objects: Nascent Preplanetary Nebulae \& Young Planetary Nebulae}
\label{other}
%\noindent {\it IRAS\,01037+1219 (WX Psc)}
%1155  967.6  215.2 72.08 Smal-Bip  pg 26
We included a few objects in our survey, which, from the evolutionary standpoint, bracket
the pre-planetary evolutionary phase, the main focus of our surveys. Thus, we included two
very late AGB stars, one which is O-rich (IRAS\,01037+1219: also WX Psc, IRC+10011), and
another, which is
C-rich (IRAS\,23166+1655: AFGL 3068). Our observations of IRAS\,01037+1219 show a
central,
asymmetrical, elongated (oval-shaped) nebulosity in the F814W image. The object appears much 
fainter and more compact in the F606W image (Fig.\ref{figm1}). The peak brightness region is
also elongated
and non-stellar (Fig.\ref{figm1a}). We believe that this object has just begun the
transition to the PPN phase and we label it as a ``nascent PPN", or nPPN. The distance and 
dust-shell related parameters (Table\,\ref{t3}) for IRAS\,01037 are taken directly from the
detailed modelling study by Vinkovic et al. (2004), and have not been re-scaled to our
standard luminosity value of 6000\ls.

Other well-studied objects which belong to the nPPN class include the carbon stars
IRC\,10216 (Skinner, Meixner \& Bobrowsky 1998), CIT-6 (Schmidt et al. 2002) and V Hydrae
(Sahai et al. 2003), all of which show strongly aspherical structures at their centers; a
brief summary of an HST survey directed at this class is given by Sahai et al. (2006b). A
detailed discussion of this class of objects is deferred to a future paper.

IRAS\,23166+1655
does not show any bright central nebulosity or a central star; however, there is a
substantial diffuse round halo with a large number of roughly circular arcs centered on the
2MASS coordinates of this object. The arcs and their interpretation have been discussed by
Mauron \& Huggins (2006) and Morris et al. (2006). Since there is no central organized
structure, this object is not relevant for our morphological classification system. 

%17047: Mc*,ps(m),h [Southern counterpart to long N lobe: appears open, or perhaps we cannot
%  see the faint end?]
%21282: Mc,o*,ps(m) [c: NE-SW pair, o: are the NW, SE lobe pair open, or we can't see faint
%  ends?]
%22568: Bcw,an,ps(an?,s) [N ansa not seen, but likely to be there]
%19255: Bcw,an,ps(an,s),h(e) [can't see end of S lobe, so don't know if its o or c]
We also included four young planetary nebulae, IRAS\,17047-5650 (better known as
CPD-56\arcdeg8032), 
IRAS\,21282+5050, IRAS\,22568+6141 (Garcia-Lario et al. 1991), and IRAS\,19255+2123 (better
known as K\,3-35),  in our survey.
IRAS\,17047-5650
is a well-known dusty planetary nebula with a late WC central star (Crowther, de Marco \&
Barlow 1998). Its F435W image
(Fig.\ref{fig10}) shows a multipolar object, with one well-defined set of
diametrically-opposed bipolar lobes oriented roughly along a NE-SW axis; in a second pair
oriented N-S, the northern lobe is observable to its tip, but the southern counterpart
appears to be surface-brightness limited along its length, and its tip is not seen. There
are complex brightness variations which likely represent additional geometrical structures,
e.g., one may infer the presence of a shell oriented roughly along a NW-SE axis. The
central star is visible in the image. 

The PN IRAS\,21282+5050 (Crowther et al. 1998) shows a
morphology
(Fig.\ref{fig10b}) qualitatively very similar to that of CPD-56\arcdeg8032 -- it has two
pairs of diametrically-opposed lobes, and a central roughly rhomboidal-shaped shell.
Our F606W image of IRAS\,22568+6141 (Fig.\ref{fig11}) shows a bipolar nebula with highly
structured lobes; no central star can be seen. A compact bright ansa can be seen just south
of the southeastern lobe, and it appears to be connected to this lobe via a narrow,
jet-like feature. A counterpart to this elongated feature is seen marginally just north of
the northwestern lobe. The overall shape of the lobes is roughly point-symmetric. 

\begin{figure}[htb]
\resizebox{0.5\textwidth}{!}{\includegraphics{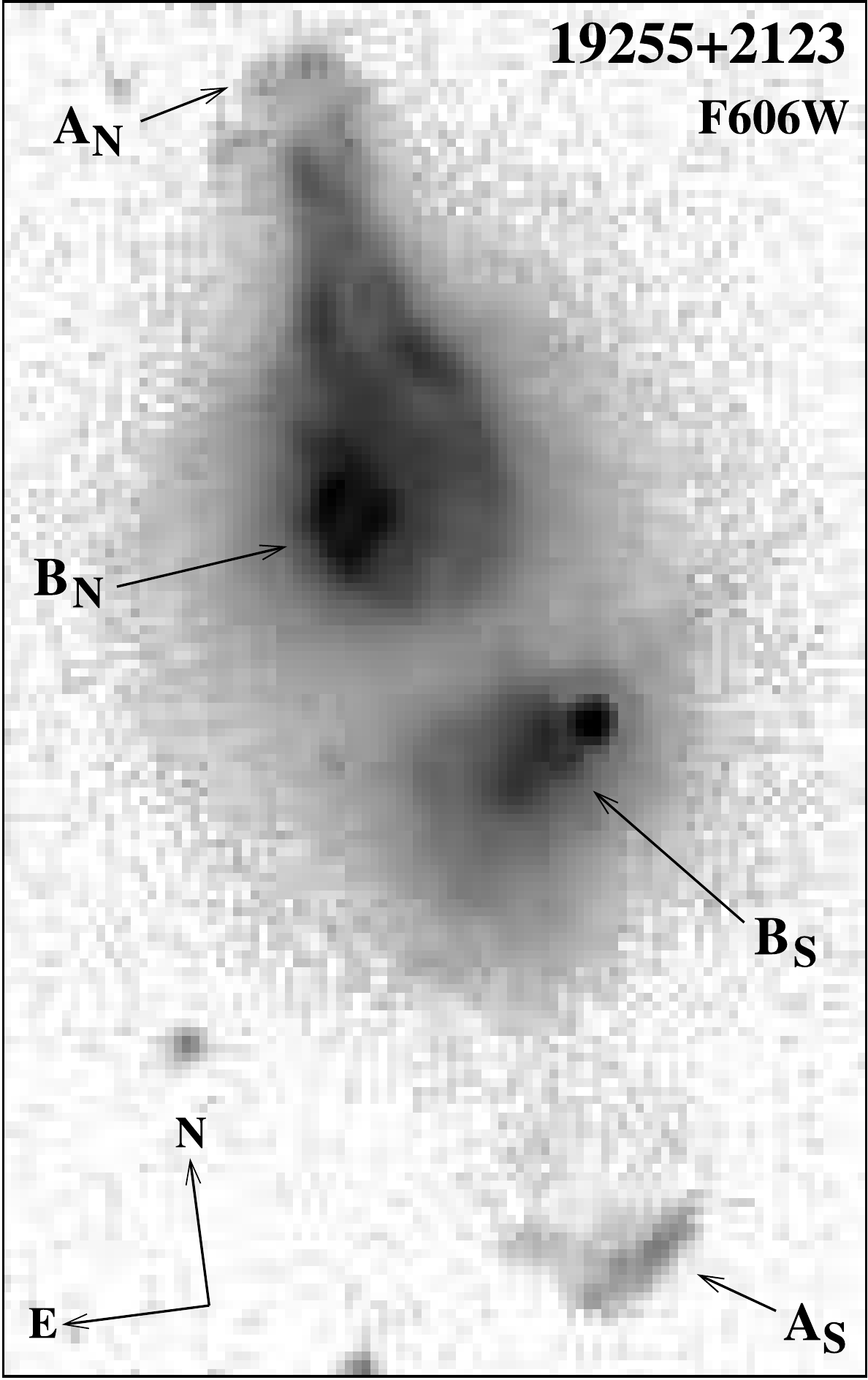}}
%\plotone{f1.eps}
\vskip 0.2in
\caption{HST image ({\it log stretch}) of the young planetary 
nebula IRAS\,19255+2123 ($4\farcs60\times7\farcs34$).}
\label{figk3-35}
\end{figure}

IRAS\,19255+2123 is a well-studied dusty planetary nebula noted for being the first PN in
which H$_2$O maser emission was found (Miranda et al. 2001); such PNs are exceedingly rare 
presumably due to the very short lifetime of H$_2$O masers during the PN phase. Its F606W
image shows a bipolar nebula (Fig.\,\ref{figk3-35}); the north lobe, which is the brighter
of the two, shows signficant brightness structure, including a bright region on its eastern
periphery ($B_N$), that has a bright point-symmetric counterpart in the southern lobe
($B_S$). A bright ansa is seen at the tip of the southern lobe ($A_S$), with a possible
counterpart at the tip of the northern lobe ($A_N$). A dense dusty waist separates the two
lobes, and a diffuse halo surrounds the lobes at low and intermediate latitudes.

\subsection{IRAS\,05506+2414: An Unrelated Object?}
\begin{figure}[htbp]
\vskip 1.0cm
\resizebox{1.0\textwidth}{!}{\includegraphics{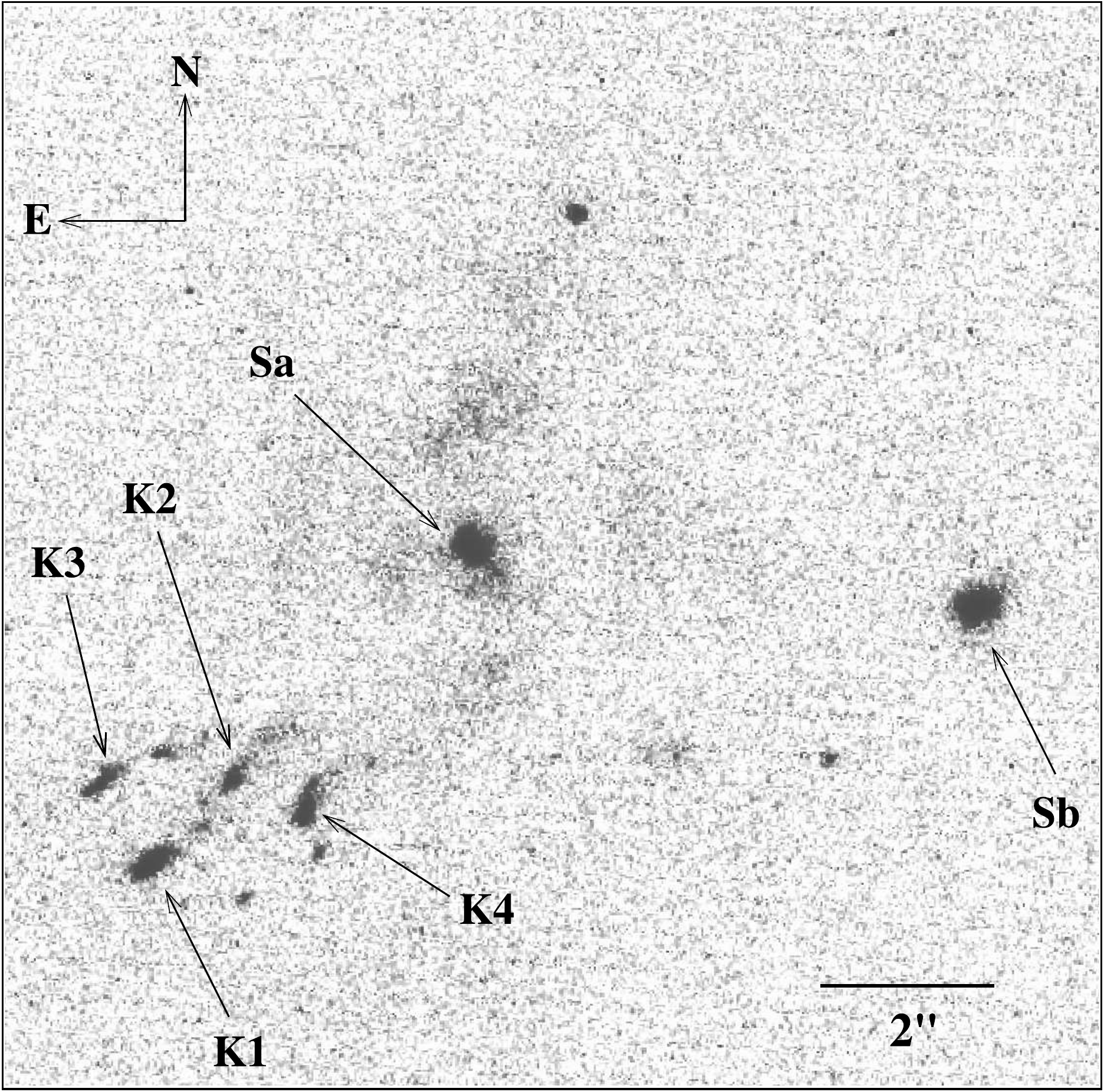}}
\vskip 0.5cm
\caption{HST (F606W) image of IRAS\,05506+2414.}
\label{ohir28}
\end{figure}

The morphology of this object is unlike that of any PPN or PN, although it met the selection
criteria for our survey. The HST image (Fig.\,\ref{ohir28}) shows a bright
compact knot ({\it Sa}), and a fan-like spray of elongated nebulous
features (e.g. knots {\it K1--4}; not all the knots seen in the image have been labelled)
which
are separated from the former, but appear to emanate from it. A second
compact knot ({\it Sb}) is seen to the west of the main source. We find that 
the MSX and 2MASS sources identified with the IRAS source are located  
at the position of knot {\it Sa}, indicating that it is associated with 
the stellar source powering an outflow that produces the knots {\it K1--4}\footnote{These
knots are most likely Herbig-Haro objects}. A 1612 MHz OH
maser source was discovered towards 
IRAS\,05506 with the Arecibo dish 
(Chengalur et al. 1993) -- the profile is not the typical
double-peaked profile seen towards most dying stars. 
The main lines at 1665,67 MHz were detected by Lewis (1997)
with peak fluxes of 300 and 150 mJy using Arecibo; the emission covers about
15\,\kms~in the 1612 MHz line and $\sim$40-45\,\kms~in the main lines.

If IRAS\,05506 is not associated with an evolved star, but instead is a young stellar object
(YSO), then the OH maser observations argue for it not being a low-mass YSO,
since there are no known OH masers toward low-mass YSOs. So if it is a
YSO, it must be related to a high-mass star-formation region (SFR), suggesting the
presence of an ultra-compact HII region with detectable continuum
emission. However, the NVSS VLA all sky survey at 20\,cm shows no
continuum source towards IRAS\,05506 down to an rms noise of $\sim$400
$\mu$Jy/beam. Could IRAS\,05506 then really be a PPN, but with a morphology which has never
been seen before
in PPNs or PNs? While this an improbable hypothesis, it is not implausible,
considering that not so long ago, Sahai \& Nyman (2000)
discovered a planetary nebula with a bipolar, knotty jet, bearing an uncanny
resemblance to jets seen in low-mass YSOs.
A detailed discussion of the nature of IRAS\,05506 is outside the scope of this paper; 
but will be presented, together with new optical,  millimeter-wave and radio data in a
forthcoming paper (Sahai et al. 2007, in preparation).

%The J2000 coordinates of {\it Sa} are 05 53 43.55, 24 14 44.0.
%(MSX: 05 53 43.6 +24 14 44, and 2MASS: 05 53 43.56 +24 14 44.7).

%\noindent {\it IRAS\,23166+1655 (AFGL 3068)}
% 706.7 775.6 248.5 73.7  Circ-Arc   28

%\subsection{Young Planetary Nebulae}

%\noindent {\it IRAS\,16268-4556}
% 6.51 49.55 47.64 266.1L Larg-Bip   42  --  yPN?

%\noindent {\it IRAS\,14562-5406 (He2-113, late WC central star)}
% 92.41 310.50 176.60 71.30 Large-Bip 55 --  yPN(He2-113)

%\noindent {\it IRAS\,17047-5650 (CPD-56\arcdeg8032, late WC central star)} DONE
% 143.70 257.20 199.10 91.69 Medm-Bip 19 --  yPN(CPD-56d8032), with late WC 
%central star

%\noindent {\it IRAS\,19234+1627}
% 1.65L 7.51 15.58 75.13L Medm-Ell   59 low-interest elliptical PN

%\noindent {\it IRAS\,21282+5050, late WC central star, but reclassified}
% 50.9  74.36 33.43 14.99 Medm-Bip   31  yPN -- with late WC central star

%\noindent {\it IRAS\,22568+6141} DONE
% 1.84  15.87 20.82 46.54L Medm-Bip  38

\section{Discussion}
\label{discus}
The 
morphological scheme we have presented in this paper is based on the PPNs in our survey and
on additional
well-studied PPNs outside our sample. Our PPN survey sample is a well-defined sample, but it
is not a complete sample. We have listed in Table 4 all PPNs with
high-resolution images in the literature outside our sample and we find that all of these
can be accomodated within our classification scheme. Thus, Tables 3 and 4
list all known galactic PPNs with existing high-resolution images -- a total of 53 objects.
Many of the objects in Table 4 are taken from the PPN survey by Ueta et al. (2000). These
authors classify PPNs as either SOLE (objects with
bright central star embedded in a faint extended nebulosity) or DUPLEX (objects with
bipolar shapes with a completely or partially obscured central star), and suggest that
the {\it axisymmetry} in PPNs is created by an equatorially enhanced superwind at the end
of the AGB phase. In view of the many very strong departures from axisymmetry in a very
significant fraction of the full sample of PPNs, we do not think that the Ueta et al.
classification provides a sufficiently complete framework for classifying the
morphologies of PPNs.

\subsection{Pre-Planetary and Young Planetary Nebulae: Morphological Similarities}
\label{discus-m}
The wide variety of morphologies which we have found are qualitatively similar to those
found for young planetary nebulae in a previous HST survey (e.g., Sahai \& Trauger 1998,
Sahai 2000, Sahai 2003) and other ground-based studies.
In particular, we find multipolar
objects like IRAS\,19024+0044, which closely resemble the ``Starfish Twins" -- two young PNs
with multiple lobes (Sahai 2000). The quadrupolar
object, IRAS\,19475+3119, probably belongs to the class of quadrupolar planetary nebulae
first identified by Manchado, Stanghellini \& Guerrero (1991). In the ``bipolar" class, the
cylindrical lobe shape together with the associated very dense waist found in
IRAS\,13428-6232, is 
very similar to that of the well-studied PN IC\,4406 (e.g., Sahai et al. 1991). The
(bipolar) hourglass-shaped PPN,  
IRAS\,15452-5459, resembles hourglass-shaped PN like MyCn\,18 (Sahai et al. 1999c) and Hb12
(Sahai \& Trauger 1998). 

There are a few PPNs
which are best described as elongated (class E). If the elongated morphology of these
objects is because they have prolate ellipsoidal shapes, then we expect that they will
evolve into the category of PNs which have been classified as elliptical in morphological 
schemes for PNs (e.g., Corradi \& Schwarz 1995). Alternatively, these may be intrinsically
bipolar but with (i) the polar axis
sufficiently tilted towards us so that it is difficult to see the pinching-in of the lobes
at their bases, or (ii) insufficiently resolved so that the pinching-in of the lobes at the
base is not visible. A good example of possibility (i) is provided by IRAS\,18420-0512
(Fig.\,\ref{fig6a}),
where the
northwest lobe shows some evidence for being pinched-in on its northeastern periphery. As
in the case of our surveys of young PNs with HST (Sahai \& Trauger 1998, Sahai 2003), we
found no round PPNs.

The similarity in morphologies between all PPNs observed with high spatial resolution and
young PNs 
strongly indicates that (i) the former are the progenitors of aspherical planetary
nebulae, and (ii) the onset of aspherical structure begins during the PPN phase (or
earlier). Thus, the primary shaping of a PN clearly does not occur during the PN phase via
the fast radiative wind of the hot central star, but significantly earlier in its
evolution.

The morphological system for PPNs which we have presented can be adapted for young PNs
directly.
The broad-band optical (and near-infrared) images of PPNs show light scattered off dust
grains in the circumstellar material. The interaction of the fast collimated wind from the
central star of the PPN with the slowly-expanding, spherical, AGB mass-loss envelope,
produces lobes with dense walls which consist mostly of swept-up AGB wind material. The
walls are overdense with respect to both the interiors and the exteriors of the lobes. The
shapes of the nebular lobes, which are the brightest regions of PPNs, thus represent the
shapes of the dense walls of the lobes, since the interiors of the lobes are tenuous as
demonstrated by the frequent presence of limb-brightening in the lobes. This basic
structure of the lobes in PPNs is reproduced in numerical simulations by Lee \& Sahai
(2003).

When the central star becomes hot enough to substantially ionise the circumstellar material,
the nebular walls then appear as bright features in H$\alpha$ and forbidden-line emission
(e.g. [NII]$\lambda\lambda$6548,6583). Note that because the optical-line emission is
proportional to $n_e^2\,l$, and thus, $n^2\,l$ (where $n_e$ and $n$ are the electron and
total density and $l$ is the path length), whereas scattered-light intensity is
proportional to $n\,l$, generally the brightness contrast between the nebular walls and its
surroundings is likely to be higher for PNs than for PPNs. The lobe morphology, both in the
PPN and PN phase, is a direct indicator of the geometrical shapes of the dense lobe walls,
and therefore of the wind-wind interaction process which created them.

The only distinction which needs to be kept in mind when modifying the PPN classification
scheme for PNs is related to the appearance of the waist region. Since most surveys of PNs
have been carried out in emission-line filters, the waist quite often (but not always)
appears as a bright
feature, rather than a dark feature. Hence, unlike the case of PPNs, in our classification
of young PNs, we will not define the waist in terms of a minimum in intensity along the
long axis of the nebula. Indeed, the waist is often the brightest structural component of a
bipolar or multipolar planetary nebula, which is understandable if we assume that the waist
region has much more mass than the lobes. Note also that if the waist region is in
expansion, then it will continue to flow radially outward from the star as a PPN evolves
into a PN. The central regions of PNs with waists are thus, in general, expected to be more
exposed and visible than those of the PPNs from which they evolve. Of course, very young
PNs such as IRAS\,22568+6141 (Fig.\ref{fig11}) and K\,3-35 (Fig.\,\ref{figk3-35}) still have
optically-thick waists fully or partially
obscuring their central stars. A detailed morphological classification of the young PNs
from the HST surveys described in Sahai \& Trauger (1998) and Sahai (2003), which are
presumably most closely related to PPNs in an evolutionary sense, will be presented in a
future paper (Sahai \& Morris 2007, in preparation). 

\subsection{Morphology and Nebular Chemistry}
\label{chem}
The ``total" sample of PPNs with resolved morphologies listed in Tables 3 \& 4 (our survey
plus previously observed
ones) includes objects with both oxygen-rich and carbon-rich 
chemistries, as well as a few which display mixed chemistry (i.e., show features
representative of both carbon-rich and oxygen-rich chemistry).  Although
O-rich objects outnumber the C-rich ones, there is a substantial fraction of
the latter (about 25\%). Since the ratio of dusty O-rich stars to dusty C-rich stars, as
estimated from a sample drawn from the IRAS PSC, is about 9 (Thronson et al. 1987), and
these objects are presumably the progenitors of dusty PPNs, C-rich PPNs are certainly not
under-represented in our total sample of PPNs relative to their parent population. Within
the total PPN sample, there does not appear to be any significant
correlation of morphology with nebular chemistry.

\subsection{Lobes}
\label{discus-l}
The variety of lobe shapes and structures seen in young PNs led to a new model by Sahai \&
Trauger (1998) for PN shaping, in which fast collimated outflows, operating during the PPN
or very late AGB phase, are the primary agents in producing aspherical planetary nebulae. 
Hence, our morphological classification system puts considerable emphasis on including 
descriptors which are related to the shapes, structures and symmetries of the lobes. 

An important example of this emphasis is the fairly detailed characterisation of
point-symmetry in our classification scheme, based on the recognition that various
sub-types of point-symmetry impose fairly strong constraints on specific formation
mechanisms. This has been discussed in detail by Sahai et al. (2005) in a study of the PPN,
IRAS\,19024+0044. Briefly, the M and ps(m) classification of this object suggests a
precessing {\it bipolar} jet, in which the fast outflow could be in the
form of discrete blobs ejected in discontinuous episodes, or a high-velocity jet 
that discontinuously changes direction, leading to the sculpting of 
multiple lobes in different directions within the ambient AGB envelope.
The presence of point-symmetric ansae at 
the tips of one of the lobe pairs, represented by the ps(an) descriptor, requires that
the collimated outflow
which produced these lobes must have changed its axis during its operation,
although by much less than the angular separation between lobes. Thus the physical 
mechanism(s) which produces the collimated outflows must be able to accomodate two different
time-scales, one associated with generating the point-symmetrical ansae within a lobe-pair,
and one associated with generating the point-symmetrically distributed multiple lobes. 

The $o$ and $c$ secondary classifiers (related to whether lobes are open or closed at their
ends, respectively) provide another example of descriptors that raise some important
questions related to the dynamical interactions believed to shape PPNs. Generally, post-AGB
collimated winds or jets interacting with AGB circumstellar envelopes (e.g., Lee \& Sahai
2003) will produce closed lobes as the fast wind sweeps up the ambient CSE material in
front of it. If the density of the ambient material falls sufficiently rapidly with radius
(e.g., as r$^{-3}$ or more steeply, perhaps even in a discontinous manner), and/or the fast
wind turns off, reducing the ram pressure compressing the dense material at the tip of the
lobe, the compressed layer will expand due to thermal pressure\footnote{note, however, that
since this material cools rapidly, the consequently relatively low sound speed compared to
the much higher expansion speed, will limit the efficacy of this effect}. Secondly,
fragmentation of the compresed layer due to hydrodynamic instabilities may also occur. As a
result of the above two effects, the end of the lobe could dissipate and become
undetectable, leading to the open-lobe appearance. The bipolar PPN Hen\,3-401, which shows
lobes with tattered ends (Sahai et al. 1999c), is a possible example of fragmentation of
the lobe-ends due to instabilities. But hourglass-shaped bipolar PPNs such as IRAS\,13557
(Fig.\,\ref{fig2a}) and IRAS\,15452 (Fig.\,\ref{fig3}) which have wide-open ends, show no
obvious evidence of fragmented clumpy material. Furthermore, if $c$ lobes evolve into $o$
lobes over time then we would expect the former to have systematically smaller values of
$t_{PAGB}$ than the latter, but this does not seem to be the case from an inspection of our
results in Table 3 (note, however, that our sample is still too small to make this a robust
conclusion). Perhaps magnetic fields stabilize the lobe walls against fragmentation in some
objects and not others, depending on the field strength and configuration. To summarise, a
theoretical investigation of issues related to the formation of open and closed lobes
using, e.g., numerical hydrodynamic simulations, is thus quite important for our general
understanding of the formation of aspherical structure in PPNs. 

\subsection{Halos}
\label{discus-h}
The prominent halos surrounding the central aspherical shapes seen in many of our objects
are direct signatures of the undisturbed circumstellar envelopes of the progenitor AGB
stars. The halos generally have round or elongated shapes. It is quite plausible that
objects with elongated haloes have AGB envelopes which are intrinsically round, but appear
elongated because they are either being indirectly illuminated by the lobes, or the light
from the central source cannot get out to the larger distances at low latitudes. The
majority of these have surface-brightness distributions consistent with a constant
mass-loss rate with a constant expansion velocity (e.g., IRAS\,17253-2831,
IRAS\,17440-3310, IRAS\,18420-0512, IRAS\,19024+0044). But there are also examples of
objects with varying mass-loss rates, with instances of both where the mass-loss rate was
higher (IRAS\,23304+6147), and lower (e.g., IRAS\,19475+3119), in the past. 

Three objects (IRAS\,20000+3239, 22036+5306 \& 23304+6147) show halos which require a
2-piece segmented power-law to describe
the surface brightness. This implies somewhat discontinuous changes in the AGB mass-loss
rate, if we assume that the envelope expansion velocity has remained constant\footnote{It
is unlikely that radial extinction of the illuminating light from the center plays a
significant role in establishing a steep power-law index $\beta$ for the surface brightness
in the outer region, because if radial extinction was important, it would have an even more 
dramatic effect on steepening the power-law in the inner region}. Comparing the three
objects, we find the interesting result that in all of these, the power-law in the outer
(inner) region is steeper (shallower) than that for a constant mass-loss rate. The
implication is that for these objects, the mass-loss rate went through a maximum, first
increasing over a period of $\sim$3800-4800\,yr, and then decreasing.

\subsection{Waists}
\label{discus-hw}
A significant fraction of preplanetary nebulae appear to have an optically thick 
concentration of dust in their equatorial plane, i.e., a waist. We refer to this dust
distribution as a disk, setting aside the  question of whether it is a bound, orbiting
disk, or an equatorial concentration of dust in an outflowing wind.  We have noted in our 
classification scheme that, in quite a few PPNs, the disk  has a sharp radial
edge, that is, a discontinuity in the extinction  projected against the extended emission
from the reflection nebula\footnote{Because the waist regions are best seen when 
they are edge-on or close to edge-on, in which case one cannot determine from the images
whether or not the disk is truncated at an outer radius, there is
an observational bias against seeing waists with such outer edges}. The curvature of this
discontinuity is typically convex
relative to the nebula's center of symmetry, consistent with its being caused by radial
structure in an axisymmetric disk.

This result carries interesting implications about the geometry of the disk.  For the simple
case of a bipolar nebula, for example, a  sharp edge projected against the far lobe is a
strong indication that  the disk is density bounded at its outside edge.   The sharp edge
to  the waist in that case is most likely due to the abrupt radial  termination of the
disk.  If the  surface density of dust in the disk were to fall off monotonically  without
an abrupt falloff (e.g., as $1/r$ for a disk that is radially  outflowing at constant
velocity and constant mass loss rate), then  the edge would be relatively diffuse.  These
remarks apply to a situation in which the disk scale height increases with radius as would
be expected for an outflowing disk.  Examples  of sharp edges are
found in the well-studied PPNs such as AFGL 2688 (Sahai et al. 1998a), and in  our sample
objects IRAS\,19292+1806 (Fig.\ref{fig7a}) and IRAS\,22036+5306 (Fig.\ref{fig8c}).

Determination of which lobe of a bipolar nebula is the far lobe can  be done by appeal to
the kinematics; the rear lobe is the redshifted  lobe.  It is usually, though not always,
the dimmer or more absorbed  lobe because of extinction by the disk, but this is not a
reliable  indicator because the lobes may be intrinsically asymmetric (e.g.,
OH\,231.8+4.2).

We note that it is possible to have a sharp edge against the near lobe of a bipolar PPN if
several conditions are met: 1) the disk has  a sharp surface boundary, 2) the scale height
of the disk density  distribution rises linearly or less rapidly than linearly with 
radius, and 3) the observer's line of sight is approximately tangent  to that surface.  The
last condition implies that the observer be  viewing the system from near the equatorial
plane, which, in any  case, is the best vantage point from which to be able to identify 
such a system as a bipolar or multipolar nebula.

It is also possible for the disk geometry to be arranged so that it  gives a sharp edge
against the far lobe even if the disk does not  have a sharp outer boundary.  The
conditions for that, however, are  quite restrictive: 1) a sharp disk surface boundary, 2) a
disk scale  height that reaches a maximum at some radius at which the disk is  still
optically thick, and 3) the observer's line of sight be tangent  to the disk at the point
where the disk reaches its maximum scale  height.  While this is possible, it is not well
motivated physically,  so we regard any object that shows a sharp radial edge projected 
against the far lobe as a case in which the disk has a sharp outer  boundary.

The inferred sharp outer boundary to the disks in a number of systems  can most easily be
explained by an abrupt transition in  mass loss geometry from spherically
symmetric to an axisymmetric  geometry having some combination of a latitude-dependent mass
loss  rate and a latitude-dependent outflow velocity.  Models for such a transition can be
based on the presence of a close binary companion  (e.g., Morris 1987, 1990) or on the
emergence of a strong stellar magnetic field  after mass loss has stripped away enough of
the overlying stellar  atmosphere (Blackman et al. 2001a,b).  These issues will be discussed
in a separate paper.

\begin{figure}[htbp]
\vskip 1.0cm
\resizebox{1.0\textwidth}{!}{\includegraphics{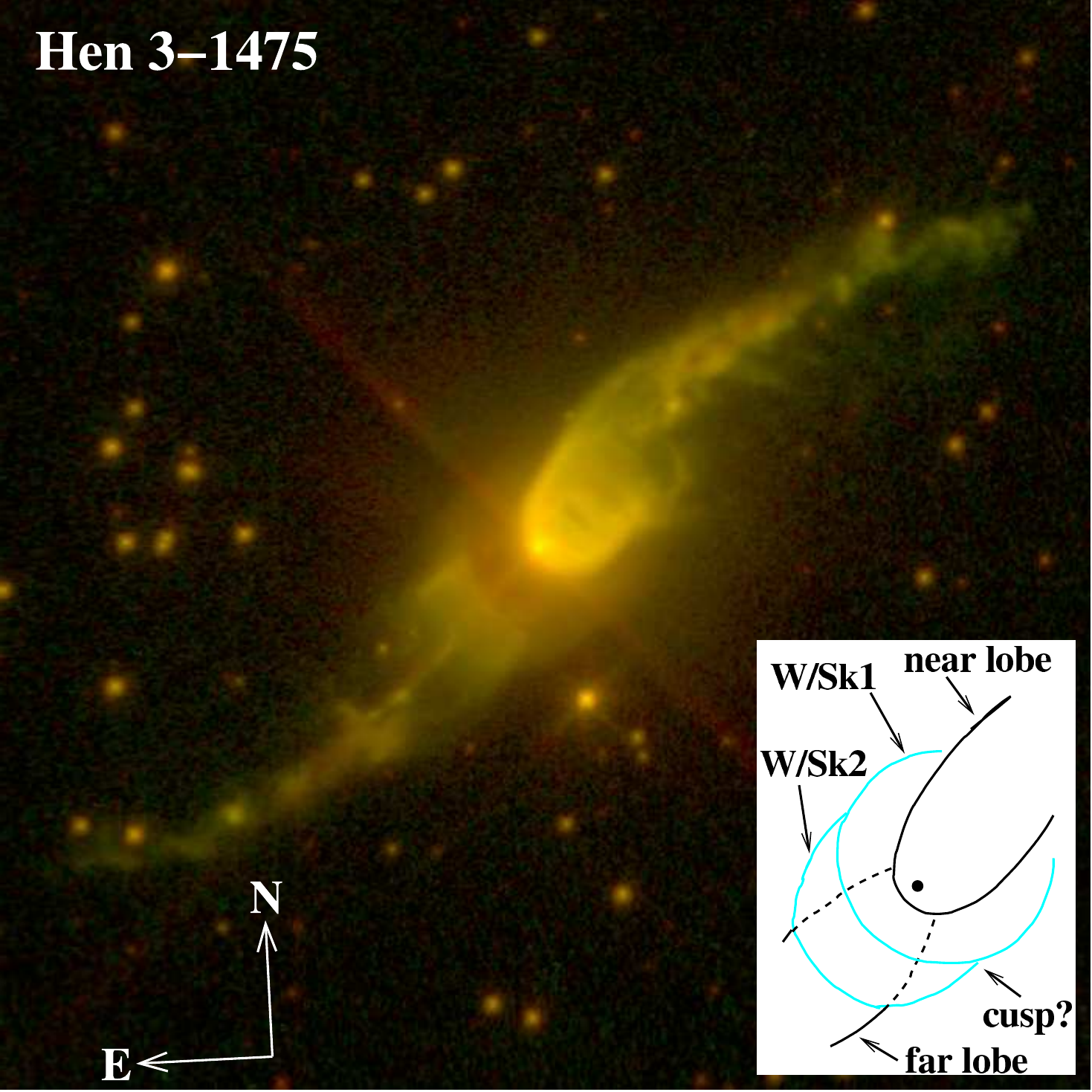}}
\vskip 0.5cm
\caption{Color-composite ({\it green}: F555W, {\it red}: F814W) HST image of the 
pre-planetary nebula Hen\,3-1475 ($18\farcs2\times18\farcs2$). Inset is a schematic of the
central region, showing the outlines of the main structural features -- background/shadowed
features are shown as dashed curves. The cyan curves show the waist/skirt features whereas
the black curves show the bipolar lobes.}
\label{he3-1475}
\end{figure}

It is important to note that waists with sharp outer edges may not be distinguishable from
skirt structures if the nebular axis is sufficiently inclined towards us. An example of
this is provided by Hen\,3-1475 (Fig.\,\ref{he3-1475}). A color image combining images
taken through the F555W and F814W filters with WFPC2 clearly demarcates the waist/skirt
structure -- most prominently defined by a relatively red, sharp-edged elliptical band
(feature labelled $W/Sk1$ in inset,
Fig.\,\ref{he3-1475}) which 
cuts sharply across the south-east or far lobe. South-east of $W/Sk1$, one can see 
a bluer structure whose outer edge is a partial ellipse ($W/Sk2$ in inset) which joins 
$W/Sk1$ near the equatorial plane. An apparent cusp can be seen on the south-west side of
the nebula where $W/Sk1$ and $W/Sk2$ meet, suggesting a skirt structure for these features.
However, since this cusp feature is not well-defined, and its counterpart on the north-east
side is not apparent, we cannot definitively argue for the presence of a skirt in
Hen\,3-1475. 

\begin{figure}[htbp]
\vskip 1.0cm
\resizebox{1.0\textwidth}{!}{\includegraphics{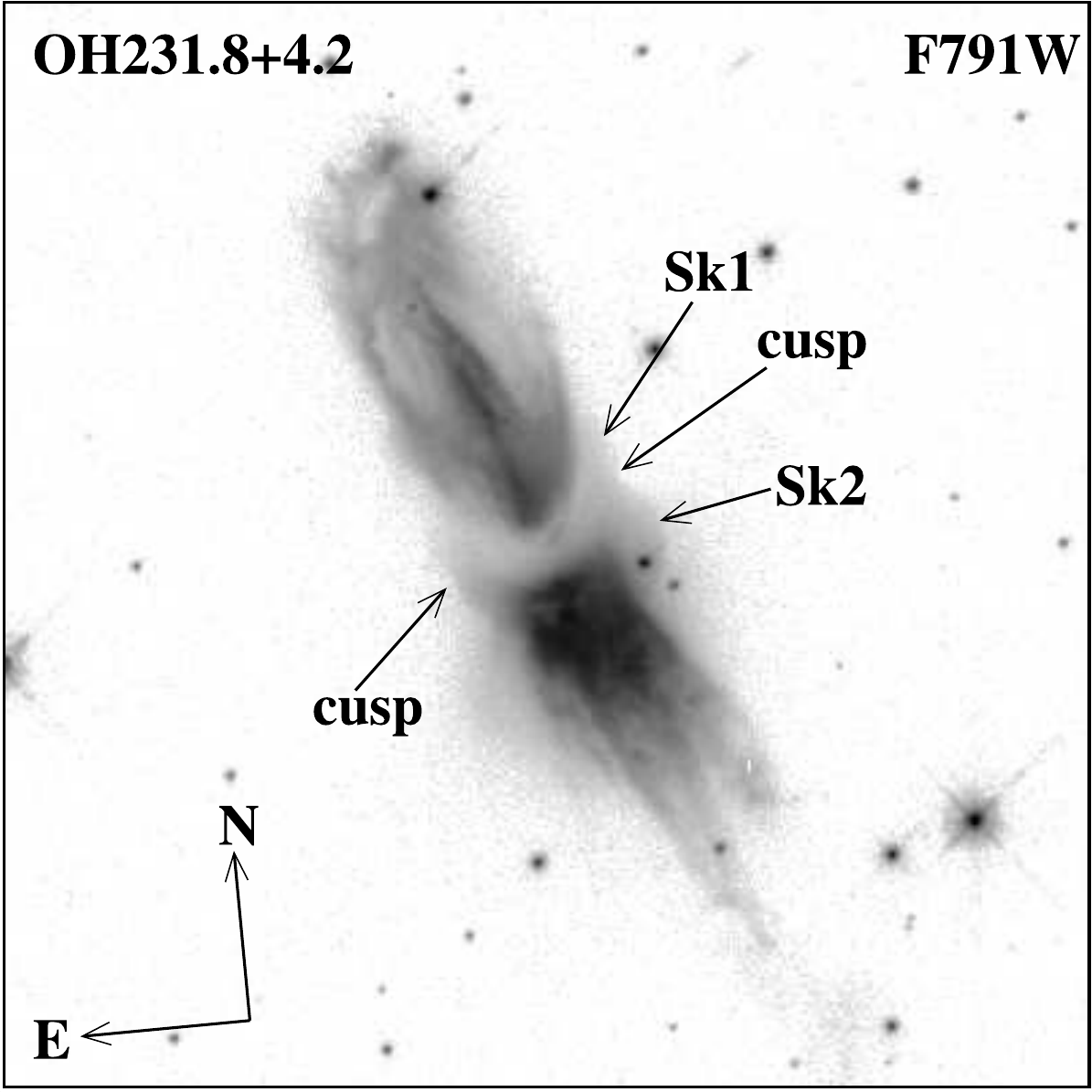}}
\vskip 0.5cm
\caption{HST (F791W) image of the pre-planetary nebula OH231.8+4.2
($39\farcs8\times39\farcs8$).}
\label{oh231.8}
\end{figure}

The central obscuring structure in
OH231.8+4.2 (Fig.\,\ref{oh231.8}) is quite similar to that of Hen\,3-1475, with its
near-side seen as an obscuring
elliptical band, partially cutting across the south-west nebular lobe and partially seen as
a bright structure projected against the sky background (feature labelled $Sk1$). Its
far-side counterpart
is seen most clearly as a bright structure against the sky background on the eastern and
western flanks of the bright south-eastern nebular lobe ($Sk2$). The skirt-shape can be 
inferred from the cusp resulting from the intersection of the projected edges of $Sk1$ and
$Sk2$ on the north-western side of the nebula; a similar cusp can be seen somewhat less
clearly on the south-eastern side. A quantitative analysis of the colors in the central
region of these objects may help in better elucidating the structural relationship between
the waist and skirt features.

%hen3-1475  /u4/sahai/data/jaz1/sahai/optdata/he3-1475/3t9t_6tct.epsi (f555w,f814w)

\section{Conclusions}
\label{conclude}
Using the Hubble Space Telescope, we have carried out a survey of a well-defined set of
candidate preplanetary nebulae. We present images of 22 new PPN and 1 nascent PPN with
well-resolved geometrical structures. Combining this sample of PPNs with previously studied
objects in this class, we have devised a comprehensive new morphological classification
system for this class of objects. We summarize our major findings below:

(1) We have found a wide variety of aspherical morphologies which include bipolar and
multipolar shapes, many of which display point-symmetries by virtue of their shapes and/or
structures. This variety of morphologies is qualitatively similar to those found for young
planetary nebulae in previous surveys. As in the case of our surveys of young planetary
nebulae (PNs), we found no round PPNs.

(2)  We also find prominent halos,
surrounding the central aspherical shapes, in many of our objects -- these 
are direct signatures of the undisturbed circumstellar envelopes of the progenitor AGB
stars. The majority of these have surface-brightness distributions consistent with
a constant mass-loss rate at a constant outflow velocity. But 3 objects require a 
segmented power-law to describe
the surface brightness indicating that in these objects, the AGB mass-loss rate went through
a maximum, first
increasing over a period of $\sim$3800-4800\,yr, and then decreasing.

(3) We have estimated the physical sizes of the aspherical nebulae and the halos. Distances
have been derived by measuring the bolometric flux of each object from its SED, and using a
fixed luminosity of 6000\ls. A time-scale for the duration of the post-AGB mass-loss 
($t_{PAGB}$) and a lower limit to the time-scale for AGB mass-loss ($t_{AGB}$), have been
derived from the expansion ages of the aspherical nebulae and the haloes, respectively, by
making simplifying assumptions about the expansion speeds of the nebular material. A rather
wide range of time-scales is found in both cases: $t_{PAGB}$ lies in the range
(30--1660)\,yr, with a median age of 180\,yr, and $t_{AGB}$ lies in the range
(300--14700)\,yr, with a median age of 6650\,yr.

(4) We have computed dust masses by fitting the SED of each object from the near-infrared
(K-band) to the far-infrared
(as defined by MSX and IRAS fluxes) using a 
simple multi-component model. The mass of the coolest 
component dominates the mass budget for each object. The dust masses cover the range
$(0.53-23.5)\times10^{-3}$\ms; the corresponding total masses, assuming a typical
gas-to-dust ratio of 200, are $(0.1-4.7)$\ms.

(5) No obvious correlations are discernible between the morphologies and the post-AGB ages,
dust masses, temperatures, or nebular chemistry.

(6) The similarities in morphologies between our survey objects and
young PNs 
supports the view that the former are the progenitors of aspherical planetary
nebulae, and the onset of aspherical structure begins during the PPN phase (or
earlier). Thus, the primary shaping of a PN clearly does not occur during the PN phase via
the fast radiative wind of the hot central star, but significantly earlier in its
evolution.

\acknowledgments
RS and MM thank NASA for partially funding this work by a
NASA LTSA award (no. 399-20-40-06); RS also received partial support for
this work from HST/GO awards (nos.
GO-09463.01-A and GO-09801.01-A) from the Space Telescope Science Institute
(operated by the Association of Universities for Research in Astronomy,
under NASA contract NAS5-26555). CSC is partially funded for this work by National
Science Foundation grant 9981546 to Owens Valley Radio Observatory; the Spanish
MCyT under project AYA2003-2785; and the Astrocam project (Ref: S-0505
ESP-0237).

%\section{Conclusions}

%The quadrupolar and elliptical objects do not show dense waists, whereas all the remaining
%(multipolar, hourglass, bipolar do).
%The most prominent halos are found in the carbon-rich objects. The radial surface
%brightness of the halos indicates a $r^{-\alpha}$ density law, where $\alpha$ is xx,
%indicating a xx mass-loss rate with time in the past >>xx years of the objects history.

%\acknowledgments
%We thank ... RS and MM thank NASA were partially funded for this work by a
%NASA LTSA grant (no. 399-30-61-00-00).

\clearpage
%11385-5517 92.6 138.3 193.0 104.0   Smal Irr   pg 74, no NIC
%11385-5517/47a_606w.epsi, 189_435w.epsi

%13428-6232 21.88 337.4 394.6 185.8L vLarg Bip  pg 68, no NIC
%13428-6232/11_41_606.epsi, 21_31_814.epsi

%13557-6442 33.73 63.58 24.92 13.4L  Smal Bip  pg 76, no NIC
%13557-6442/11_41_606.epsi, 21_31_814.epsi HRC

%15405-4945 2.38L 26.51 82.99 34.22: Medm Bip  pg 48, NIC pg 82
%15405-4945/hm_606.epsi, ik_814.epsi

%15452-5459 87.05 242.7 273.6 401.6L Larg Hrlg  pg 43, NIC pg 78
%15452-5459/v1_606.epsi, wz_814.epsi, v1_cen.pdfi, wz_cen.pdfi

%17253-2831/9m8m_bw.epsi, 5m6m_814.epsi pg 2 no NIC

%17347-3139/af.epsi, bd.epsi, pg 22, NIC1 pg 77
% f160_inset.ps/  f160_inset_r.eps LINEAR STRETCH, different for inner patch
 
%17440-3310/2q_aq_606.epsi, 5q_9q_814.epsi, pg 70, no NIC

%18276-1431/2m8m_606.epsi, 4m6m_814.epsi, pg 3, no NIC

\clearpage
%18420-0512/iq_qq_606.epsi, lq_pq_814.epsi, pg 65, no NIC

% 606_11_51.epsi 814_31_41.epsi 500 x 500 1/2 HRC pix=6.25 asec

%19134+2131/an_aw_606.epsi, ao_au_814.epsi, pg 32, pg 84 NIC (not useful? what about AO)

%
%19292+1806/5e_606.epsi, 6c_814.epsi, pg 29, no NIC

%19306+1407/11_61_606.epsi, 31_41_814.epsi, pg 54

%19475+3119/11.epsi, 41.epsi, no NIC

%20000+3239/11_606b.epsi, 41_814b.epsi, pg 50, no NIC

%pg 106

%22223+4327/11_81.epsi, 41_71.epsi, pg 35, no NIC

%23304+6147 11.36 59.07 26.6 30.89L Medm Bip  pg 37, no NIC
%23304+6147/11_usat.epsi, 41_71_usat_814.epsi

%Nascent PPN
%
%/data/sahai/ohirdata/hst/i01037+1219/RI_shrp.ps
% 01037+1219 1155  967.6  215.2 72.08 Smal Bip  pg 26, pg 98 NIC (useful?)
% nPPN, IRC+10011

%
%/data/sahai/ohirdata/hst/i01037+1219/RI_shrp.ps

%
%/data/sahai/ohirdata/hst/

\clearpage

%hen3-1475  /u4/sahai/data/jaz1/sahai/optdata/he3-1475/3t9t_6tct.epsi (f555w,f814w)
%/data/sahai/ohirdata/ohir-bppn-29.eps

%  /u4/sah1/sahai/optdata/wfpc2/oh231.8/oh231.8-f791w-cen2.ps best at showing skirt
%   log stretch 400x400 WF3 ([272:671, 110:509] patch, 791cf), -0.1 to 3.5, sky=0.15

%/data/sahai/ohirdata/hst/
%Point Source: 03206+6521 95.76 133.7 37.54 9.9    Comp --    46
%09425-6040 26.82 55.53 21.18 5.05:  Stellar   34

\clearpage
%%%%%%%%%%%%%%%%%%%%%%%% TABLE %%%%%%%%%%%%%%%%%%%%%%%%%%%%%%%%
%\tiny
%\scriptsize
%\footnotesize
%\small
%\normalsize
%\large
%\Large
%\LARGE
%\huge
%\Huge

\begin{footnotesize}
\begin{table}
\begin{center}
\caption{Survey Targets\label{t1}}
\begin{tabular}{p{0.9in}p{0.9in}p{0.9in}p{0.7in}p{0.7in}}
\tableline\tableline
IRAS     & RA (2000) & DEC(2000)   & Chem. & \\
name     & {\it hhmmss.ss} & {\it ddmmss.s} & & \\
\tableline
\multicolumn{3}{l} {Pre-Planetary Nebulae} & & \\
$01037+1219$\tablenotemark{a} & 010625.98 & $+123553.0$ & O &\\
$11385-5517$ & 114058.81 & $-553425.8$ & O &\\
$13428-6232$ & 134620.54 & $-624759.6$ & -- &\\
$13557-6442$ & 135928.92 & $-645718.3$ & O &\\
$15405-4945$ & 154411.06 & $-495522.1$ & O &\\
$15452-5459$ & 154911.38 & $-550851.6$ & O &\\
$15553-5230$ & 155911.18 & $-523841.6$ & -- &\\
$16559-2957$ & 165908.22 & $-300140.3$ & O &\\
$17253-2831$ & 172832.96 & $-283325.8$ & O &\\
%$17340-3757$ & 173729.72 & $-375922.3$ & O &\\
$17347-3139$ & 173800.61 & $-314055.2$ & O &\\
$17440-3310$ & 174722.72 & $-331109.3$ & O &\\
$17543-3102$ & 175733.61 & $-310303.5$ & O &\\
$18276-1431$ & 183030.70 & $-142857.0$ & O &\\
$18420-0512$ & 184441.65 & $-050917.0$ & O &\\
$19024+0044$ & 190502.06 & $+004850.9$ & O &\\
$19134+2131$ & 191535.21 & $+213634.0$ & O &\\
$19292+1806$ & 193125.37 & $+181310.3$ & O &\\
$19306+1407$ & 193255.08 & $+141336.9$ & C+O &\\
$19475+3119$ & 194929.56 & $+312716.3$ & O &\\
$20000+3239$ & 200159.52 & $+324732.9$ & C &\\
%$20068+4051$ & & & &\\
$22036+5306$ & 220530.29 & $+532132.8$ & O &\\
$22223+4327$ & 222431.43 & $+434310.9$ & C &\\
$23304+6147$ & 233244.79 & $+620349.1$ & C &\\
%$23541+7031$ & & & &\\
\tableline
\multicolumn{2}{l} {Other Objects}  & & & \\
$05506+2414$\tablenotemark{b} & 055343.56 & $+241444.7$ & & \\
$17047-5650$\tablenotemark{c} & 170900.93 & $-565448.0$ & & \\
$19255+2123$\tablenotemark{c} & 192744.02 & $+213003.4$ & & \\
$21282+5050$\tablenotemark{c} & 212958.47 & $+510400.3$ & & \\
$22568+6141$\tablenotemark{c} & 225854.89 & $+615758.0$ & & \\

\tableline
\end{tabular}
\tablenotetext{a} {nascent pre-planetary nebula}
\tablenotetext{b} {most likely a young stellar object}
\tablenotetext{c} {planetary nebula}
\end{center}
\end{table}
\end{footnotesize}
\clearpage

\begin{table}
\begin{center}
\caption{Morphological Classification Codes\label{t2}}
\begin{tabular}{p{0.9in}p{4in}}
\tableline\tableline
\multicolumn{2}{c}{PRIMARY CLASSIFICATION -- {\it Nebular Shape}} \\
{\bf B} & Bipolar \\
{\bf M} & Multipolar \\
{\bf E} & Elongated \\
{\bf I} & Irregular \\
\tableline
\multicolumn{2}{c}{SECONDARY CLASSIFICATIONS} \\
\multicolumn{2}{l}{{\it lobe shape:}} \\
{\bf o} & lobes open at ends\\
{\bf c} & lobes closed at ends \\
\\
\multicolumn{2}{l}{{\it obscuring waist:}} \\
{\bf w} & central obscuring waist \\
{\bf w(b)}  & obscuring waist has sharp radial boundary \\
\\
\multicolumn{2}{l}{{\it central star:}} \\
{\bf $\star$} & central star evident in optical images \\
\\
\multicolumn{2}{l}{{\it other nebular characteristics:}} \\
{\bf an}  & ansae present \\
{\bf ml}  & minor lobes are present \\
{\bf sk}  & a skirt-like structure present around the primary lobes \\
\\
\multicolumn{2}{l}{{\it point symmetry:}} \\
{\bf ps(m)}  & two or more pairs of diametrically-opposed lobes \\
{\bf ps(an)} & diametrically-opposed ansae present \\
{\bf ps(s)}  & overall geometric shape of lobes is point-symmetric \\
\\
\multicolumn{2}{l}{{\it halo:}} \\
{\bf h}    & halo emission is present (relatively low-surface brightness diffuse region
around primary nebular structure)\\
{\bf h(e)} & halo has elongated shape \\
{\bf h(i)} & halo has indeterminate shape \\
{\bf h(a)} & halo has centro-symmetric arc-like features \\
{\bf h(sb)}  & searchlight-beams are present \\
\tableline
\end{tabular}
\end{center}
\end{table}
\clearpage

\begin{footnotesize}
\begin{table}
\begin{center}
\caption{Properties of Survey Pre-Planetary Nebulae\label{t3}}
\begin{tabular}{p{0.9in}p{0.4in}p{1.45in}p{0.3in}p{0.3in}p{0.3in}p{0.3in}p{0.4in}p{0.5in}p{0.3in}p{0.3in}}
\tableline\tableline
IRAS & f25/f12 &  Morphology & $r_{PAGB}$ & $t_{PAGB}$ & $r_{AGB}$ & $V_{exp}$ &
$t_{AGB}$ & T$_d$ & M$_d$ & Dist. \\
%            & &    &    & &                        &  &  &  & \\
name & &      & (\arcsec) & (yr) & (\arcsec) & \kms & (yr) & (K) &
$10^{-3}$ \ms & (kpc) \\
%% Morphology
\tableline
$01037+1219$\tablenotemark{a} & 0.83 & E & ...  & ... & 35 & 20 & 5500 & ... & 1.3 & 0.65 \\ 
$11385-5517$    & 1.5    & I   & 4.6  & 180 & (4.6) & (15) & 1800 &  74 & 1.3 & 1.2 \\
$13428-6232$    & 15.4   & Bow & 24.7 & 1470 & (24.7) & (15) & 14700 & 88 & 2.9 & 1.9 \\ 
$13557-6442$    & 1.9    & Bow,h& 0.8 & 85  & 3.0 & (15) & 3500 & 103 & 0.53 & 3.3 \\ 
$15405-4945$ & $>$11 & Bcw,ps(s),h(e) & 1.0 & 155 & 3.0 & 13.5 & 5200 & 72 & 7.0 & 4.9\\ 
$15452-5459$    & 2.9  & Bow  & 11.7 & 625  & (11.7) & 9.5 & 9850 & 87 & 1.8 & 1.7\\ 
$15553-5230$    & 4.9  & Bow  &  1.1 & 145  & (1.1)  & (15)& 1430 &104 & 1.1 & 4.3\\
$16559-2957$ & 3.5 & Bw,ps(s),h & 0.35 & 60 & 3.5 & 14.8 & 5850& 118 & 0.38 & 5.2\\
$17253-2831$ & 15.1& Ec*,ps(s),h & 0.66 & 160 & 2.7 & 9.1 & 10800 & 98 & 1.5 & 7.7\\ 
%$17340-3757$ & 2.5 &  Bcw &  &   & & & &  & & \\
$17347-3139$    & 5.3  & Bcw,h(e) & 1.9 & 185 & 2.0 & (15) & 1950 & 81 & 3.2 & 3.1\\
$17440-3310$  & $>$5.1 & Bc*,ps(s),h(a) & 1.4 & 335 & 4 & 14.6 & 9850 & 79 & 5.5 & 7.6 \\
$17543-3102$  & 7.9 & Bc,an,h(i) & 1.35 & 310 & 2 & (15) & 4600 & 83 & 3.6 & 7.3 \\
$18276-1431$\tablenotemark{b} & 5.8 & Bcw,h(e,a,sb) & 0.5 & 45 & 2.8 & 17.0 & 2300 & 50-105
& 4.75 & 3\\ 
$18420-0512$    & 25.9   & Ec*,h(a) & 0.8 & 170 & 3 & 12.4 & 7600 & 88 & 2.6 & 6.6 \\
$19024+0044$ & 17.1 & Mcw(b),an,ps(m,an),h & 1.8 & 440 & 2 & 13.4 & 3700 & 109 & 1.2 & 5.1\\
$19134+2131$    & 3.1    & Bcw & 0.12 & 30 & (0.12) & (15) & 300 & 69 & 2.7 & 8.4 \\
$19292+1806$  & $>$9.6   & Bcw(b),h(e) & 0.9 & 150 & 2.5 & 14.6 & 4250 & 103 & 1.1 & 5.2 \\
$19306+1407$    & 16.4   & Bo*,h & 3.2 & 500 & 4 & 14 & 6650 & 99 & 1.3 & 4.9 \\
$19475+3119$\tablenotemark{c}  & 70.4 & Mc*,ps(m,s),h & 5.2 & 1660  & 6 & 15 & 8000 &
46-94 & 4.3 & 4.2\\ 
$20000+3239$    & 4.7    & Ec*,h & 1.0 & 110  & 5 & 12.3 & 6850 & 103 & 0.88 & 3.5 \\
%$20068+4051$    & 7.7    & B &  &   & &  & & \\
$22036+5306$ & 5.5 & Bcw(b),an,ps(s,an),h & 4.0 & 255 & 6.5 & 7 & 8850 & 35-50 & 23.5 & 2 \\
$22223+4327$    & 17.5   & Ec*,h(a) & 1.5 & 204 & 7.5 & 14 & 11000 & 99 & 0.79 & 4.3 \\
$23304+6147$    & 5.2    & Mc*,ps(m),h(a) & 1.1 & 145 & 5.5 & 15.5 & 7000 & 99 & 0.82 & 4.2
\\
%$23541+7031$    & 1.0    & B &  &   & &  & & \\
\end{tabular}
\tablenotetext{a}{values of $r_{AGB}$, $t_{AGB}$, M$_d$, Dist from Vinkovic et al. 2004}
\tablenotetext{b}{values of parameters from S\'anchez-Contreras et al. (2007), scaled
appropriately for L=6000\ls~as necessary}
\tablenotetext{c}{values of parameters from Sahai et al. (2007) and S\'anchez Contreras et
al. 2006, scaled appropriately for L=6000\ls~as necessary}
\tablecomments{Numbers in parenthesis imply {\it assumed} values}
\end{center}
\end{table}
\end{footnotesize}

%\begin{tiny}
\vskip -0.5in
\begin{table}
\scriptsize
\begin{center}
\caption{Properties of Previously Observed, Well-Resolved Pre-Planetary Nebulae\label{t4}}
\begin{tabular}{p{1.3in}p{0.4in}p{0.6in}p{0.4in}p{1.2in}}
\tableline\tableline
IRAS & f25/f12 & Imaging & Chem.\tablenotemark{a} & Morphology \\
name &         & Refs.   &  \\
\tableline
{\it Well-Studied PPNs} & & & & \\
AFGL\,618\tablenotemark{b}  & 2.3    & 1  & C & Mcw,ml,h(e,a) \\
Red Rectangle   & 1.1    & 2  & C+O & Bow \\
OH\,$231.8+4.2$ & 11.9   & 3  & O & Bcw,sk  \\
$07131-0147$    & 1.6    & 4  & O & Bow*  \\
Frosty Leo Neb. & $>$17  & 5  & O & Bcw(b)*,an,ml,ps(m,an,s) \\
Roberts\,22     & 5.5    & 6  & C+O & Bcw,ml,ps(s),h(e)  \\
Hen\,3-401      & 9.3    & 7  & C & Bow*,sk       \\
Boomerang Neb.  & 1.3    & 8  & ... & Bow* \\
$16342-3814$    & 12.3   & 9  & O & Bcw,ps(s)   \\   
Hen\,3-1475     & 4.0    & 10 & O & Bow(b)*,an,sk,ps(an,s),h  \\ 
M\,1-92         & 3.4    & 11 & O & Bcw(b)*,an   \\ 
AFGL\,2688      & ...    & 12 & C & Bcw(b),ml,h(e,a,sb) \\
\tableline
{\it Less Well-Studied PPNs} & & & & \\
$02229+6208$   & 3.1    &  10  & C & E         \\
$04296+3429$   & 3.6    &  13  & C & Bcw(b),h  \\
$05341+0852$   & 2.2    &  10  & C & Ec*,ps(s),h \\
$06530-0213$   & 4.5    &  10  & C & Bc*,ml,ps(s),h(e)  \\
$07134+1005$   & 4.8    &  10  & C & Ec*,h(e) \\
$07430+1115$   & 3.9    &  10  & C & E*,h   \\
$08005-2356$   & 2.9    &  10  & O & Bc*   \\
$16594-4656$   & 6.6    &  14  & C & Mcw*,an,ps(m,an),h(a)  \\
$17106-3046$   & 15.5   &  15  & O & Bw(b)   \\
$17150-3224$   & 5.6    &  16  & O & Bcw,h(a,sb)  \\
$17245-3951$   & 13.3   &  14  & O & Bcw,h(sb)   \\
$17436+5003$   & 30.1   &  10  & O & Ec*   \\
$17441-2411$   & 4.5    &  17  & C & Bcw,h(e,a)   \\
$18095+2704$   & 2.8    &  10  & O & Bc,h    \\
$19374+2359$   & 4.2    &  10  & O & B,h(e)   \\
$19477+2401$   & 4.9    &  14  & C & E,h(e)   \\
$20028+3910$   & 5.0    & 18   & C & Bcw(b),h(e,a,sb) \\
$22272+5435$   & 4.1    & 10   & C & Ec*,h(a) \\
$22574+6609$   & 3.3    & 14   & C & Bwh(i) \\
$23321+6545$   & 6.3    & 10   & C & Ec \\
\end{tabular}
%% Any table notes must follow the \end{tabular} command.
%\clearpage
\tablenotetext{a}{Information about the nebular chemistry has been taken either directly (in
most cases) from the imaging reference paper, or from papers cited in the latter}
\tablenotetext{b}{The ``a" descriptor is based on an unpublished HST/ACS F606W image
provided by Dr. Bruce Balick (Huehnerhoff, Baerny, \& Balick, in preparation) using data
from GO 9430/PI S. Trammell}
\tablerefs{{\it References for HST Images} 1: Lee \& Sahai 2003; 2: Cohen et al.
2004, 3: Bujarrabal et al. 2002, 4: Scarrott et al. 1990 ({\it ground-based image only}), 
5: Sahai et al. 2000a, 6: Sahai et al. 1999d, 7: Sahai et al. 1999a, 8: Sahai et al. 2000b, 
9: Sahai et al. 1999b, 10: Ueta et al. 2000, 11: Bujarrabal et al. 1998, 12: Sahai et al.
1998a, 13: Sahai 1999, 14: Su et al. 2001, 15: Kwok et al. 2000, 16: Kwok, Su \& Hrivnak
1998, 17: Su et al. 1998, 18: Hrivnak,
Kwok, \& Su 2001, 19: Su, Hrivnak, \& Kwok 2001
}
\end{center}
\end{table}
%\end{tiny}

%figures
%oh231.8    /u4/sahai/data/jaz1/sahai/nicmos/oh231.8/oh231_f160w.ps
%  /u4/sah1/sahai/optdata/wfpc2/oh231.8/oh231.8-f791w-cen2.ps best at showing skirt
%   log stretch 400x400 ([272:671, 110:509] patch from 791cf)image, -0.1 to 3.5, sky=0.15
%hen3-1475  /u4/sahai/data/jaz1/sahai/optdata/he3-1475/3t9t_6tct.epsi (f555w,f814w)

%19477+2401   11.2   54.9   27.1     38    Comp       B ????
%/u4/sah1/sahai/optdata/wfpc2/iras20028

%Handling overlong column cells
%For columns with overly long sentences in a given cell, you can force them to wrap with the
%"p" specifier in the tabular environment. For example,
%\begin{tabular}{|l|l|p{8cm}|}
%The first two columns here are the usual left-justified entries, there are vertical lines
%about all columns, but the 3rd column is forced to be 8cm wide, so it can't overfill the
%page. Additionally, any text that would be longer than 8cm is wrapped nicely in the cell.
%%%%%%%%%%%%%%%%%%%%%%%% TABLE %%%%%%%%%%%%%%%%%%%%%%%%%%%%%%%%

\end{document}